\pdfoutput=1

\documentclass[12pt,a4paper]{article}

\usepackage{ifthen} 
\newboolean{pdflatex}
\setboolean{pdflatex}{true} 

\newboolean{articletitles}
\setboolean{articletitles}{true} 

\newboolean{uprightparticles}
\setboolean{uprightparticles}{false} 

\newboolean{inbibliography}
\setboolean{inbibliography}{false} 

\def\paperauthors{LHCb collaboration} 
\def\paperasciititle{Test of Lepton Flavor Universality by the
measurement of the B0 -> D*- tau+ nu/_tau
      branching fraction
      using three-prong $\tau$ decays} 
\def\papertitle{Test of Lepton Flavor Universality by the measurement of the $B^0 \to D^{*-} \tau^+ \nu_{\tau}$
      branching fraction
      using three-prong $\tau$ decays} 
\def\paperkeywords{{High Energy Physics}, {LHCb}} 
\def\papercopyright{CERN on behalf of the LHCb collaboration}
\def\paperlicence{CC-BY-4.0}
\def\paperlicenceurl{https://creativecommons.org/licenses/by/4.0/}

\usepackage[top=1in, bottom=1.25in, left=1in, right=1in]{geometry}

\usepackage{microtype}
\usepackage{lineno}  
\usepackage{xspace} 
\usepackage{caption} 

\usepackage{graphicx}  
\usepackage{color}
\usepackage{colortbl}
\graphicspath{{./figs/}} 

\usepackage{amsmath} 
\usepackage{amssymb}
\usepackage{amsfonts}
\usepackage{upgreek} 

\newcommand*\patchAmsMathEnvironmentForLineno[1]{%
\expandafter\let\csname old#1\expandafter\endcsname\csname #1\endcsname
\expandafter\let\csname oldend#1\expandafter\endcsname\csname
end#1\endcsname
 \renewenvironment{#1}%
   {\linenomath\csname old#1\endcsname}%
   {\csname oldend#1\endcsname\endlinenomath}%
}
\newcommand*\patchBothAmsMathEnvironmentsForLineno[1]{%
  \patchAmsMathEnvironmentForLineno{#1}%
  \patchAmsMathEnvironmentForLineno{#1*}%
}
\AtBeginDocument{%
\patchBothAmsMathEnvironmentsForLineno{equation}%
\patchBothAmsMathEnvironmentsForLineno{align}%
\patchBothAmsMathEnvironmentsForLineno{flalign}%
\patchBothAmsMathEnvironmentsForLineno{alignat}%
\patchBothAmsMathEnvironmentsForLineno{gather}%
\patchBothAmsMathEnvironmentsForLineno{multline}%
\patchBothAmsMathEnvironmentsForLineno{eqnarray}%
}

\usepackage{hyperxmp}

\usepackage[pdftex,
            pdfauthor={\paperauthors},
            pdftitle={\paperasciititle},
            pdfkeywords={\paperkeywords},
            pdfcopyright={Copyright (C) \papercopyright},
            pdflicenseurl={\paperlicenceurl}]{hyperref}

\usepackage[all]{hypcap} 

\usepackage{xspace} 
\usepackage{upgreek}

\def\lhcb {\mbox{LHCb}\xspace}

\def\babar  {\mbox{BaBar}\xspace}
\def\belle  {\mbox{Belle}\xspace}

\def\MagUp {\mbox{\em Mag\kern -0.05em Up}\xspace}

\ifthenelse{\boolean{uprightparticles}}%
{

 \def\Peta        {\ensuremath{\upeta}\xspace}

 \def\Pmu         {\ensuremath{\upmu}\xspace}                 
 \def\Pnu         {\ensuremath{\upnu}\xspace}                 
                  
 \def\Ppi         {\ensuremath{\uppi}\xspace}                 
                  
 \def\Prho        {\ensuremath{\uprho}\xspace}                 
                  
 \def\Ptau        {\ensuremath{\uptau}\xspace}

 \def\PDelta      {\ensuremath{\Delta}\xspace}                 
 \def\PXi      {\ensuremath{\Xi}\xspace}                 
 \def\PLambda      {\ensuremath{\Lambda}\xspace}                 
 \def\PSigma      {\ensuremath{\Sigma}\xspace}                 
 \def\POmega      {\ensuremath{\Omega}\xspace}                 
 \def\PUpsilon      {\ensuremath{\Upsilon}\xspace}                 
 
 \def\PB      {\ensuremath{\mathrm{B}}\xspace}                 
                  
 \def\PD      {\ensuremath{\mathrm{D}}\xspace}

 \def\PK      {\ensuremath{\mathrm{K}}\xspace}

 \def\Pb      {\ensuremath{\mathrm{b}}\xspace}                 
 \def\Pc      {\ensuremath{\mathrm{c}}\xspace}

 \def\Pi      {\ensuremath{\mathrm{i}}\xspace}

 \def\Ps      {\ensuremath{\mathrm{s}}\xspace}

}
{

 \def\Peta        {\ensuremath{\eta}\xspace}

 \def\Pmu         {\ensuremath{\mu}\xspace}                 
 \def\Pnu         {\ensuremath{\nu}\xspace}                 
                  
 \def\Ppi         {\ensuremath{\pi}\xspace}                 
                  
 \def\Prho        {\ensuremath{\rho}\xspace}                 
                  
 \def\Ptau        {\ensuremath{\tau}\xspace}

 \mathchardef\PDelta="7101
 \mathchardef\PXi="7104
 \mathchardef\PLambda="7103
 \mathchardef\PSigma="7106
 \mathchardef\POmega="710A
 \mathchardef\PUpsilon="7107
                  
 \def\PB      {\ensuremath{B}\xspace}                 
                  
 \def\PD      {\ensuremath{D}\xspace}

 \def\PK      {\ensuremath{K}\xspace}

 \def\Pb      {\ensuremath{b}\xspace}                 
 \def\Pc      {\ensuremath{c}\xspace}

 \def\Pi      {\ensuremath{i}\xspace}

 \def\Ps      {\ensuremath{s}\xspace}

}

\makeatletter
\ifcase \@ptsize \relax
  \newcommand{\miniscule}{\@setfontsize\miniscule{4}{5}}
\or
  \newcommand{\miniscule}{\@setfontsize\miniscule{5}{6}}
\or
  \newcommand{\miniscule}{\@setfontsize\miniscule{5}{6}}
\fi
\makeatother

\DeclareRobustCommand{\optbar}[1]{\shortstack{{\miniscule (\rule[.5ex]{1.25em}{.18mm})}
  \\ [-.7ex] $#1$}}

\def\muon       {{\ensuremath{\Pmu}}\xspace}
\def\mup        {{\ensuremath{\Pmu^+}}\xspace}
\def\mun        {{\ensuremath{\Pmu^-}}\xspace} 

\def\tauon      {{\ensuremath{\Ptau}}\xspace}
\def\taup       {{\ensuremath{\Ptau^+}}\xspace}
\def\taum       {{\ensuremath{\Ptau^-}}\xspace}

\def\neu        {{\ensuremath{\Pnu}}\xspace}
\def\neub       {{\ensuremath{\overline{\Pnu}}}\xspace}
\def\neue       {{\ensuremath{\neu_e}}\xspace}

\def\neum       {{\ensuremath{\neu_\mu}}\xspace}
\def\neumb      {{\ensuremath{\neub_\mu}}\xspace}
\def\neut       {{\ensuremath{\neu_\tau}}\xspace}
\def\neutb      {{\ensuremath{\neub_\tau}}\xspace}

\def\squark    {{\ensuremath{\Ps}}\xspace}

\def\cquark    {{\ensuremath{\Pc}}\xspace}

\def\bquark    {{\ensuremath{\Pb}}\xspace}
\def\bquarkbar {{\ensuremath{\overline \bquark}}\xspace}

\def\pion   {{\ensuremath{\Ppi}}\xspace}
\def\piz    {{\ensuremath{\pion^0}}\xspace}

\def\pip    {{\ensuremath{\pion^+}}\xspace}
\def\pim    {{\ensuremath{\pion^-}}\xspace}

\def\rhomeson {{\ensuremath{\Prho}}\xspace}
\def\rhoz     {{\ensuremath{\rhomeson^0}}\xspace}
\def\rhop     {{\ensuremath{\rhomeson^+}}\xspace}

\def\kaon    {{\ensuremath{\PK}}\xspace}
  \def\Kbar    {{\kern 0.2em\overline{\kern -0.2em \PK}{}}\xspace}

\def\KorKbar    {\kern 0.18em\optbar{\kern -0.18em K}{}\xspace}
\def\Kz      {{\ensuremath{\kaon^0}}\xspace}

\def\Kp      {{\ensuremath{\kaon^+}}\xspace}
\def\Km      {{\ensuremath{\kaon^-}}\xspace}

\def\KS      {{\ensuremath{\kaon^0_{\mathrm{ \scriptscriptstyle S}}}}\xspace}

\newcommand{\etapr}{\ensuremath{\Peta^{\prime}}\xspace}

  \def\Dbar    {{\kern 0.2em\overline{\kern -0.2em \PD}{}}\xspace}
\def\D       {{\ensuremath{\PD}}\xspace}

\def\DorDbar    {\kern 0.18em\optbar{\kern -0.18em D}{}\xspace}
\def\Dz      {{\ensuremath{\D^0}}\xspace}
\def\Dzb     {{\ensuremath{\Dbar{}^0}}\xspace}
\def\Dp      {{\ensuremath{\D^+}}\xspace}
\def\Dm      {{\ensuremath{\D^-}}\xspace}

\def\Dstar   {{\ensuremath{\D^*}}\xspace}

\def\Dstarm  {{\ensuremath{\D^{*-}}}\xspace}

\def\Ds      {{\ensuremath{\D^+_\squark}}\xspace}
\def\Dsp     {{\ensuremath{\D^+_\squark}}\xspace}

\def\Dssp    {{\ensuremath{\D^{*+}_\squark}}\xspace}

\def\B       {{\ensuremath{\PB}}\xspace}
\def\Bbar    {{\ensuremath{\kern 0.18em\overline{\kern -0.18em \PB}{}}}\xspace}

\def\BorBbar    {\kern 0.18em\optbar{\kern -0.18em B}{}\xspace}
\def\Bz      {{\ensuremath{\B^0}}\xspace}

\def\Bu      {{\ensuremath{\B^+}}\xspace}
\def\Bub     {{\ensuremath{\B^-}}\xspace}
\def\Bp      {{\ensuremath{\Bu}}\xspace}
\def\Bm      {{\ensuremath{\Bub}}\xspace}

\def\Bd      {{\ensuremath{\B^0}}\xspace}
\def\Bs      {{\ensuremath{\B^0_\squark}}\xspace}

\def\Bc      {{\ensuremath{\B_\cquark^+}}\xspace}

  \def\Y#1S{\ensuremath{\PUpsilon{(#1S)}}\xspace}

\def\FourS {{\Y4S}}

\def\Lz          {{\ensuremath{\PLambda}}\xspace}
\def\Lbar        {{\ensuremath{\kern 0.1em\overline{\kern -0.1em\PLambda}}}\xspace}
\def\LorLbar    {\kern 0.18em\optbar{\kern -0.18em \PLambda}{}\xspace}

\def\Lb      {{\ensuremath{\Lz^0_\bquark}}\xspace}

\def\BF         {{\ensuremath{\mathcal{B}}}\xspace}

\def\BR         {\BF}
\newcommand{\decay}[2]{\ensuremath{#1\!\to #2}\xspace}         

\def\to                 {\ensuremath{\rightarrow}\xspace}

\def\qsq       {{\ensuremath{q^2}}\xspace}

\def\AT#1     {\ensuremath{A_{\mathrm{T}}^{#1}}\xspace}           

\def\C#1      {\ensuremath{\mathcal{C}_{#1}}\xspace}                       
\def\Cp#1     {\ensuremath{\mathcal{C}_{#1}^{'}}\xspace}                    
\def\Ceff#1   {\ensuremath{\mathcal{C}_{#1}^{\mathrm{(eff)}}}\xspace}        
\def\Cpeff#1  {\ensuremath{\mathcal{C}_{#1}^{'\mathrm{(eff)}}}\xspace}       
\def\Ope#1    {\ensuremath{\mathcal{O}_{#1}}\xspace}                       
\def\Opep#1   {\ensuremath{\mathcal{O}_{#1}^{'}}\xspace}                    

\newcommand{\tev}{\ifthenelse{\boolean{inbibliography}}{\ensuremath{~T\kern -0.05em eV}}{\ensuremath{\mathrm{\,Te\kern -0.1em V}}}\xspace}
\newcommand{\gev}{\ensuremath{\mathrm{\,Ge\kern -0.1em V}}\xspace}
\newcommand{\mev}{\ensuremath{\mathrm{\,Me\kern -0.1em V}}\xspace}
\newcommand{\kev}{\ensuremath{\mathrm{\,ke\kern -0.1em V}}\xspace}
\newcommand{\ev}{\ensuremath{\mathrm{\,e\kern -0.1em V}}\xspace}
\newcommand{\gevc}{\ensuremath{{\mathrm{\,Ge\kern -0.1em V\!/}c}}\xspace}
\newcommand{\mevc}{\ensuremath{{\mathrm{\,Me\kern -0.1em V\!/}c}}\xspace}
\newcommand{\gevcc}{\ensuremath{{\mathrm{\,Ge\kern -0.1em V\!/}c^2}}\xspace}
\newcommand{\gevgevcccc}{\ensuremath{{\mathrm{\,Ge\kern -0.1em V^2\!/}c^4}}\xspace}
\newcommand{\mevcc}{\ensuremath{{\mathrm{\,Me\kern -0.1em V\!/}c^2}}\xspace}

\def\mm   {\ensuremath{\mathrm{ \,mm}}\xspace}

\def\mum  {\ensuremath{{\,\upmu\mathrm{m}}}\xspace}

\def\invfb   {\ensuremath{\mbox{\,fb}^{-1}}\xspace}

\newcommand{\stat}{\ensuremath{\mathrm{\,(stat)}}\xspace}
\newcommand{\syst}{\ensuremath{\mathrm{\,(syst)}}\xspace}

\newcommand{\chisq}{\ensuremath{\chi^2}\xspace}

\def\gsim{{~\raise.15em\hbox{$>$}\kern-.85em
          \lower.35em\hbox{$\sim$}~}\xspace}
\def\lsim{{~\raise.15em\hbox{$<$}\kern-.85em
          \lower.35em\hbox{$\sim$}~}\xspace}

\def\sqs   {\ensuremath{\protect\sqrt{s}}\xspace}

\def\ptot       {\mbox{$p$}\xspace}
\def\pt         {\mbox{$p_{\mathrm{ T}}$}\xspace}

\def\evtgen     {\mbox{\textsc{EvtGen}}\xspace}

\def\geant      {\mbox{\textsc{Geant4}}\xspace}

\def\photos     {\mbox{\textsc{Photos}}\xspace}

\def\pythia     {\mbox{\textsc{Pythia}}\xspace}

\def\tell1  {TELL1\xspace}
\def\ukl1   {UKL1\xspace}

\newcommand{\ie}{\mbox{\itshape i.e.}\xspace}

\usepackage{cite} 
\usepackage{mciteplus}

\usepackage{longtable} 

\begin{document}

\renewcommand{\thefootnote}{\fnsymbol{footnote}}
\setcounter{footnote}{1}


\begin{titlepage}
\pagenumbering{roman}

\vspace*{-1.5cm}
\centerline{\large EUROPEAN ORGANIZATION FOR NUCLEAR RESEARCH (CERN)}
\vspace*{1.5cm}
\noindent
\begin{tabular*}{\linewidth}{lc@{\extracolsep{\fill}}r@{\extracolsep{0pt}}}
\ifthenelse{\boolean{pdflatex}}
{\vspace*{-2.7cm}\mbox{\!\!\!\includegraphics[width=.14\textwidth]{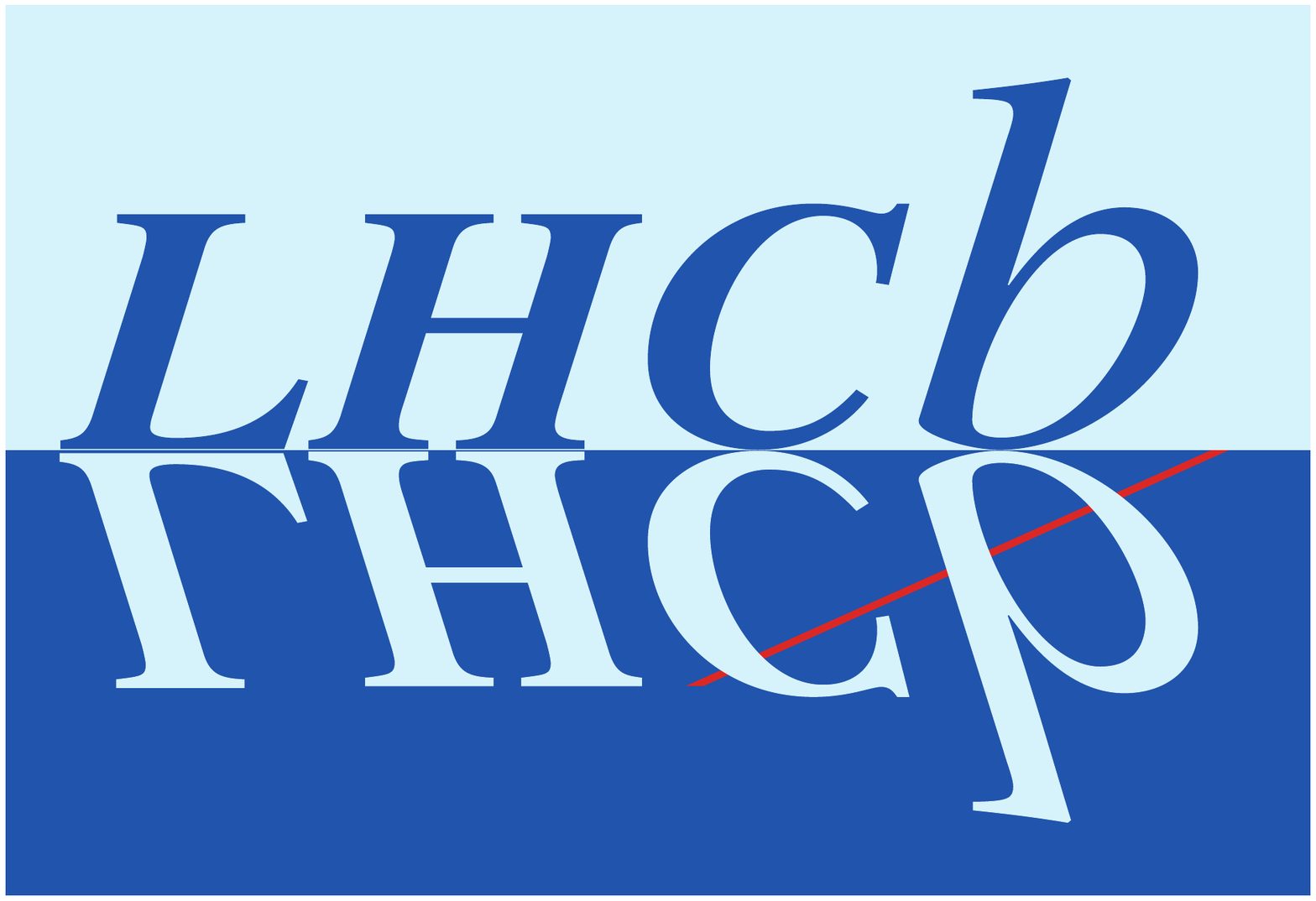}} & &}%
{\vspace*{-1.2cm}\mbox{\!\!\!\includegraphics[width=.12\textwidth]{lhcb-logo.eps}} & &}%
\\
 & & CERN-EP-2017-256 \\  
 & & LHCb-PAPER-2017-027 \\  
 & & 26 April 2018 \\ 
 & & \\
\end{tabular*}

\vspace*{2.5cm}

{\normalfont\bfseries\boldmath\huge
\begin{center}
  \papertitle
\end{center}
}

\vspace*{0.4cm}

\begin{center}
\paperauthors\footnote{Authors are listed at the end of this paper.}
\end{center}

\vspace{\fill}

\begin{abstract}
\vspace*{0.2cm}
  \noindent
 The ratio of
branching fractions \mbox{${\cal{R}}(D^{*-})\equiv {\cal{B}}(B^0 \to D^{*-} \tau^+
\nu_{\tau})/{\cal{B}}(B^0 \to D^{*-} \mu^+\nu_{\mu})$} is measured using a data
sample of proton-proton collisions collected with
the LHCb detector  at center-of-mass energies of 7 and 8\tev, corresponding to an integrated
luminosity of 3\invfb.
The \tauon lepton is reconstructed with three charged pions in the final state. A
novel method is used that exploits the different vertex topologies of signal and backgrounds
to isolate samples of semitauonic decays of $b$ hadrons with high purity.
Using the \Bz \to \Dstarm\pip\pim\pip decay as the normalization channel, the ratio
${\cal{B}}(B^0 \to D^{*-} \tau^+ \nu_{\tau})/{\cal{B}}(B^0 \to
D^{*-}\pip\pim\pip)$ is measured to be $1.97 \pm 0.13 \pm 0.18$, where the
first uncertainty is statistical and the second systematic.
An average of branching fraction measurements for the normalization channel is used to derive
\mbox{${\cal{B}}(B^0 \to D^{*-} \tau^+ \nu_{\tau}) = (1.42 \pm 0.094 \pm 0.129
\pm 0.054) \%$}, where the third uncertainty is due to the limited
knowledge of ${\cal{B}}(B^0\to
D^{*-}\pi^+\pi^-\pi^+)$. A test of lepton flavor universality is
performed using  the well-measured branching fraction ${\cal{B}}(B^0 \to
D^{*-} \mu^+\nu_{\mu})$ to compute
\mbox{${\cal{R}}(D^{*-}) = 0.291 \pm 0.019 \pm 0.026 \pm 0.013$},
where the third uncertainty originates from the uncertainties
on ${\cal{B}}(\Bz \to \Dstarm\pip\pim\pip)$ and
${\cal{B}}(B^0 \to D^{*-} \mu^+\nu_{\mu})$.
This measurement is in agreement with the Standard Model prediction and with previous
measurements.
\end{abstract}

\vspace*{1.0cm}

\begin{center}
  Published in Phys.~Rev.~D {\bf{97}} (2018) 072013
\end{center}

\vspace{\fill}

{\footnotesize
\centerline{\copyright~\papercopyright, licence \href{\paperlicenceurl}{\paperlicence}.}}
\vspace*{2mm}

\end{titlepage}


\newpage
\setcounter{page}{2}
\mbox{~}
%
%
%
%

\cleardoublepage


\renewcommand{\thefootnote}{\arabic{footnote}}
\setcounter{footnote}{0}

\newcommand{\extrn}{\ensuremath{\mathrm{\,(ext)}}\xspace}



\pagestyle{plain} 
\setcounter{page}{1}
\pagenumbering{arabic}


\section{Introduction}
\label{sec:Introduction}

In the Standard Model (SM) of particle physics, lepton flavor
universality (LFU) is an accidental symmetry  broken only by
the Yukawa interactions.
Differences between the expected branching fraction of semileptonic
decays into the three lepton families originate from
the different masses of the charged leptons. Further deviations from LFU would
be a signature of physics processes beyond the SM.

Measurements of the couplings of $Z$ and $W$ bosons to light leptons, mainly constrained by
LEP and SLC experiments, are compatible with LFU. Nevertheless, a
2.8 standard deviation difference exists between the measurement of the
branching fraction of the $W^+\to\tauon^+\neut$ decay with respect
to those of the branching fractions of $W^+\to\muon^+\neum$ and $W^+\to e^+\neue$
decays~\cite{Schael:2013ita}.

Since uncertainties due to hadronic effects cancel to a large
extent, the SM prediction for the ratios between branching fractions of semitauonic
decays of \PB mesons relative to decays involving lighter lepton
families, such as
\begin{eqnarray}
{\cal{R}}(D^{(*)-})&\equiv& {\cal{B}}(\Bz \to D^{(*)-} \taup
    \neut)/{\cal{B}}(\Bz \to D^{(*)-} \mup\neum),  \\
 {\cal{R}}(D^{(*)0})&\equiv& {\cal{B}}(\Bm \to D^{(*)0} \taum
    \neutb)/{\cal{B}}(\Bm \to D^{(*)0} \mun\neumb),
\end{eqnarray}
is known with an uncertainty at the percent
level~\cite{Fajfer:2012vx,Bigi:2016mdz,Bernlochner:2017jka,Jaiswal:2017rve}. For \Dstar decays, recent papers~\cite{Bigi:2017jbd,Jaiswal:2017rve} argue for larger uncertainties, up to 4\%.
 These decays therefore provide a sensitive probe of SM extensions
with flavor-dependent couplings, such as models with an extended Higgs sector~\cite{Tanaka:1994ay},
 with leptoquarks~\cite{Buchmuller:1986zs,Davidson:1993qk}, or with an extended gauge sector~\cite{Greljo:2016,Boucenna:2016,Bhattacharya:2017}.

The \decay{\PB}{\D^{(*)} \taup \neut} decays have recently been
subject to intense experimental scrutiny.
Measurements of
${\cal{R}}(D^{0,-})$ and ${\cal{R}}(D^{*-,0})$ and their averages ${\cal{R}}(D)$
and ${\cal{R}}(D^{*})$ have been reported by the
\babar~\cite{Lees:2012xj,Lees:2013uzd} and
\belle~\cite{Huschle:2015rga,Sato:2016svk} collaborations in final
states
involving
electrons or muons from the \tauon decay.
The \lhcb collaboration measured ${\cal{R}}(\D^{*})$~\cite{LHCb-PAPER-2015-025}
with results compatible with those from \babar,
while the result from the \belle collaboration
is compatible with the SM within 1 standard deviation.
The measurements from both the \babar and \belle collaborations were performed with events that
were ``tagged'' by fully reconstructing the decay of one of the two
\PB mesons from the \FourS\ decay to a fully hadronic final state (hadronic
tag); the other \PB meson was used to search for the signal.
In all of the above measurements, the decay of the \tauon lepton into a
muon, or an electron, and two neutrinos was exploited.
More recently, the \belle collaboration published a measurement~\cite{Sato:2016svk} with
events tagged  using semileptonic decays, compatible with the SM
within $1.6$ standard deviations.
A simultaneous measurement of ${\cal{R}}(\Dstar)$ and of the $\tau$
polarization, using hadronic
tagging and reconstruction of the
$\tau^-\to\pi^-\neut$ and $\tau^-\to\rho^-\neut$ decays, was published by the
\belle collaboration~\cite{Hirose:2016wfn,Hirose:2017dxl}.
The average of
all these ${\cal{R}}(\Dstar)$ measurements is in tension with
the SM expectation at $3.3$ standard deviations.
All these ${\cal{R}}(D^{(*)-,0})$ measurements yield values that are above the
SM predictions with a combined significance of
$3.9$ standard deviations~\cite{HFAG}.

This paper presents a measurement of ${\cal{B}}$(\decay{\Bd}{\Dstarm \taup
  \neut}), using for the first time the \tauon decay with three charged particles (three-prong)
in the final state, \ie \mbox{\decay{\taup}{\pip \pim \pip \neutb}} and
\mbox{\decay{\taup}{\pip \pim \pip \piz\neutb}},
denoted as {\it{signal}} in this paper. The \Dstarm meson is reconstructed through the
\mbox{\decay{\Dstarm}{\Dzb (\to K^+ \pim) \pim}} decay chain.\footnote{The inclusion of
charge-conjugate decay modes is implied in this paper.} The visible final
state consists of six charged tracks; neutral pions are
not reconstructed in this analysis. A data sample of proton-proton
collisions, corresponding to an integrated luminosity of $3$\invfb,
collected with the \lhcb detector at center-of-mass energies of \sqs=
7 and 8\tev is used. A shorter version of this paper can be found in Ref.~\cite{LHCb-PAPER-2017-017}. 

The three-prong \tauon decay modes have different features with respect to
leptonic \tauon decays, leading to measurements with a
better signal-to-background ratio and statistical significance.
The absence of charged leptons in the final state avoids
backgrounds originating from semileptonic decays of $b$ or $c$ hadrons.
The three-prong topology enables the precise reconstruction of  a \tauon decay vertex detached from the
\Bz decay vertex due to the non zero \tauon lifetime, thereby allowing the discrimination between signal
decays and the most abundant background due to
\decay{\PB}{\Dstarm 3\pion X} decays, where $X$ represents unreconstructed
particles and 3\pion$\equiv \pip\pim\pip$.\footnote {The notation $X$ is used
when unreconstructed particles are known to be present in the decay chain and
$(X)$ when unreconstructed particles may be present in the decay chain.} The
requirement of a 3\pion decay vertex detached from the \B vertex suppresses the
$\Dstarm 3\pion X$ background by three orders of magnitude, while retaining
about 40\% of the signal.
Moreover,  because only one neutrino is produced in the \tauon decay, the
measurements of the \Bz and \tauon lines of flight allow the determination of
the complete kinematics of the decay, up to two quadratic ambiguities, leading
to four solutions.

 After applying the $3\pi$ detached-vertex requirement, the dominant
background consists of \B decays with a \Dstarm and another charm
hadron in the final state, called {\it{double-charm}} hereafter. The largest component is due to
\decay{\B}{\Dstarm\Dsp(X)} decays. These decays have the same topology as the signal, as
the second charm hadron has a measurable lifetime and its decay vertex
is detached from the \B vertex.
The double-charm background is
suppressed by applying vetoes on the presence of additional particles
around the direction of the \tauon and \PB
  candidates, and exploiting the different resonant structure of the 3\pion system
  in \taup and \Dsp decays.

The signal yield, $N_{\mathrm{sig}}$, is
normalized to that of the exclusive $\Bz \to \Dstarm 3\pi$
decay, $N_{\mathrm{norm}}$, which has the same charged particles in the final state. This choice
minimizes experimental systematic uncertainties.
The measured quantity is
\begin{equation}{\cal{K}}(\Dstar^-)\equiv\frac{{\cal{B}}(B^0 \to D^{*-} \tau^+
    \nu_{\tau})
}{{\cal{B}}(B^0 \to
    D^{*-} 3\pion)} =
    \frac{N_{\mathrm{sig}}}{N_{\mathrm{norm}}}\frac{\varepsilon_{\mathrm{norm}}}{\varepsilon_{\mathrm{sig}}}\frac{1}{{\cal{B}}(\tau^+\to
3\pi\neutb)+{\cal{B}}(\tau^+\to
3\pi\piz\neutb)},
\label{eqn:kappa}
\end{equation}

\noindent where $\varepsilon_{\mathrm{sig}}$ and
$\varepsilon_{\mathrm{norm}}$ are the efficiencies for the signal and
normalization decay modes, respectively. More precisely, $\varepsilon_{\mathrm{sig}}$ is the weighted average efficiency  for the 3\pion and the 3\pion\piz modes, given their respective branching fractions. The absolute branching fraction is
obtained as
${\cal{B}}(\decay{\Bz}{\Dstarm\taup\neut})={\cal{K}}(\Dstar^-)\times{\cal{B}}(B^0 \to  D^{*-} 3\pion)$,
 where the branching fraction of the $\Bz\to\Dstarm{3\pi}$
decay is taken by averaging the measurements of
Refs.~\cite{LHCb-PAPER-2012-046,TheBABAR:2016vzj,Majumder:2004su}. 
A value for ${\cal{R}}(\Dstar^-)$ is then derived by using the branching
fraction of the \decay{\Bz}{\Dstarm\mup\neum}
decay from Ref.~\cite{HFAG}.

This paper is structured as follows. Descriptions of the \lhcb
detector, the data and simulation samples and the trigger
selection criteria are given in Sec.~\ref{sec:Detector}.
Signal selection and background suppression strategies are summarized in Sec.~\ref{sec:selection}.
Section~\ref{sec:double_charm} presents the study performed to
characterize double-charm backgrounds due to
\decay{\PB}{\Dstarm\Dsp(X)},
\decay{\PB}{\Dstarm\Dp(X)}
and \decay{\PB}{\Dstarm\Dz(X)} decays.
The strategy used to fit the signal yield and the corresponding results
are presented in Sec.~\ref{sec:signal_yield}. The
determination of the yield of the normalization mode is discussed
in Sec.~\ref{sec:normfit}. The determination of
${\cal{K}}(\Dstarm)$ is presented in Sec.~\ref{sec:kDstar} and
systematic uncertainties are discussed in
Sec.~\ref{sec:systematic}. Finally, overall results and conclusions are given in
Sec.~\ref{sec:results}.

\section{Detector and simulation}
\label{sec:Detector}

The \lhcb detector~\cite{Alves:2008zz,LHCb-DP-2014-002} is a single-arm forward
spectrometer covering the \mbox{pseudorapidity} range $2<\eta <5$,
designed for the study of particles containing \bquark or \cquark
quarks. The detector includes a high-precision tracking system
consisting of a silicon-strip vertex detector surrounding the $pp$
interaction region~\cite{LHCb-DP-2014-001}, a large-area silicon-strip detector located
upstream of a dipole magnet with a bending power of about
$4{\mathrm{\,Tm}}$, and three stations of silicon-strip detectors and straw
drift tubes~\cite{LHCb-DP-2013-003} placed downstream of the magnet.
The tracking system provides a measurement of momentum, \ptot, of charged particles with
a relative uncertainty that varies from 0.5\% at low momentum to 1.0\% at 200\gevc.
The minimum distance of a track to a primary vertex (PV), the impact parameter (IP),
is measured with a resolution of $(15+29/\pt)\mum$,
where \pt is the component of the momentum transverse to the beam, in\,\gevc.
Different types of charged hadrons are distinguished using information
from two ring-imaging Cherenkov detectors~\cite{LHCb-DP-2012-003}.
Photons, electrons and hadrons are identified by a calorimeter system consisting of
scintillating-pad and preshower detectors, an electromagnetic
calorimeter and a hadronic calorimeter. Muons are identified by a
system composed of alternating layers of iron and multiwire
proportional chambers~\cite{LHCb-DP-2012-002}.

Simulated samples of $pp$ collisions are generated using
\pythia~\cite{Sjostrand:2006za,*Sjostrand:2007gs}
with a specific \lhcb
configuration~\cite{LHCb-PROC-2010-056}.  Decays of hadronic particles
are described by \evtgen~\cite{Lange:2001uf}, in which final-state
radiation is generated using \photos~\cite{Golonka:2005pn}. The
{\mbox{\textsc{Tauola}}\xspace} package~\cite{Davidson:2010rw} is used
to simulate the decays of the \tauon lepton into the 3\pion\neutb and
3\pion\piz\neutb final states according to the resonance chiral Lagrangian model~\cite{Nugent:2013hxa} with a tuning
based on the results from the \babar collaboration~\cite{Nugent:2013ij}. The
interaction of the generated particles with the detector, and its response,
are implemented using the \geant
toolkit~\cite{Allison:2006ve, *Agostinelli:2002hh} as described in
Ref.~\cite{LHCb-PROC-2011-006}.
The signal decays are simulated using form factors that are derived
from heavy-quark effective theory~\cite{Caprini:1997mu}. The experimental
values of the corresponding parameters are taken from Ref.~\cite{HFAG}, except for an
unmeasured helicity-suppressed amplitude, which is taken
from Ref.~\cite{Korner:1989qb}.

The trigger~\cite{LHCb-DP-2012-004} consists of a
hardware stage, based on information from the calorimeter and muon
systems, followed by a software stage, in which all charged particles
with $\pt>500\,(300)\mevc$ are reconstructed for 7 TeV (8 TeV) data. At the hardware trigger stage, candidates are required to have a muon with high \pt or a
hadron, photon or electron with high transverse energy. 
The software trigger requires a two-, three-, or four-track secondary
vertex with significant displacement from any PV
consistent with the decay of a \bquark ~hadron, or
a two-track vertex with a significant displacement from any PV
consistent with
a $\Dzb\to K^+\pi^-$ decay.
In both cases, at least one
charged particle must have a transverse momentum $\pt > 1.7\gevc$ and
must be
inconsistent with originating from any PV. A multivariate algorithm~\cite{BBDT}
is used for the identification of secondary vertices consistent with
the decay of a \bquark ~hadron. Secondary vertices consistent with the
decay of a \Dzb meson must
satisfy additional selection criteria, based on the momenta and transverse momenta
of the \Dzb decay products ($p>5$\gevc and \pt$>800$\mevc),
and on the consistency, as a loose requirement, of the \Dzb momentum vector with the direction
formed by joining the PV and the \Bz vertex.

\section{Selection criteria and multivariate analysis}
\label{sec:selection}

The signal selection proceeds in two main steps. First, the dominant background,
consisting of candidates where the 3\pion system originates from the \Bz vertex,
called {\it{prompt}} hereafter, is suppressed by applying a $3\pi$ detached-vertex
requirement. Second, the double-charm background is suppressed using
a multivariate analysis (MVA). This is the only background with the same vertex
topology as the signal.

This section is  organized as follows. After a summary of the principles of the
signal selection in Sec.~\ref{ssec:sigvtxtopo}, the categorization of the
remaining background processes is given in Secs.~\ref{ssec:double_charm}
and~\ref{ssec:othbkg}. This categorization motivates (Sec.~\ref{ssec:list}) the
additional selection criteria that have to be applied to the  tracks
and vertices of the candidates in order to exploit the requirement of vertex detachment in its
full power. Section \ref{ssec:isol} describes the isolation tools used to take
advantage of the fact that, for the $\tau^+\to 3\pion\neutb$ channel, there is no other charged or neutral particle at the \Bz vertex
beside the reconstructed particles in the final state. Particle identification requirements are presented in
Sec.~\ref{ssec:misid}. The
selection used for the normalization channel is described in
Sec.~\ref{ssec:selnorm}. Section~\ref{ssec:reconstr} details the kinematic
techniques used to reconstruct the decay chains in the signal and background
hypotheses. Finally, the  MVA that is used to reduce the double-charm
backgrounds is presented in Sec.~\ref{ssec:BDT} and, in
Sec.~\ref{ssec:comp}, the background composition at various stages of the
selection process is illustrated.

\subsection{The detached-vertex topology}
\label{ssec:sigvtxtopo}

\begin{figure}[t]
	\centering
	\includegraphics[width=0.6\textwidth]{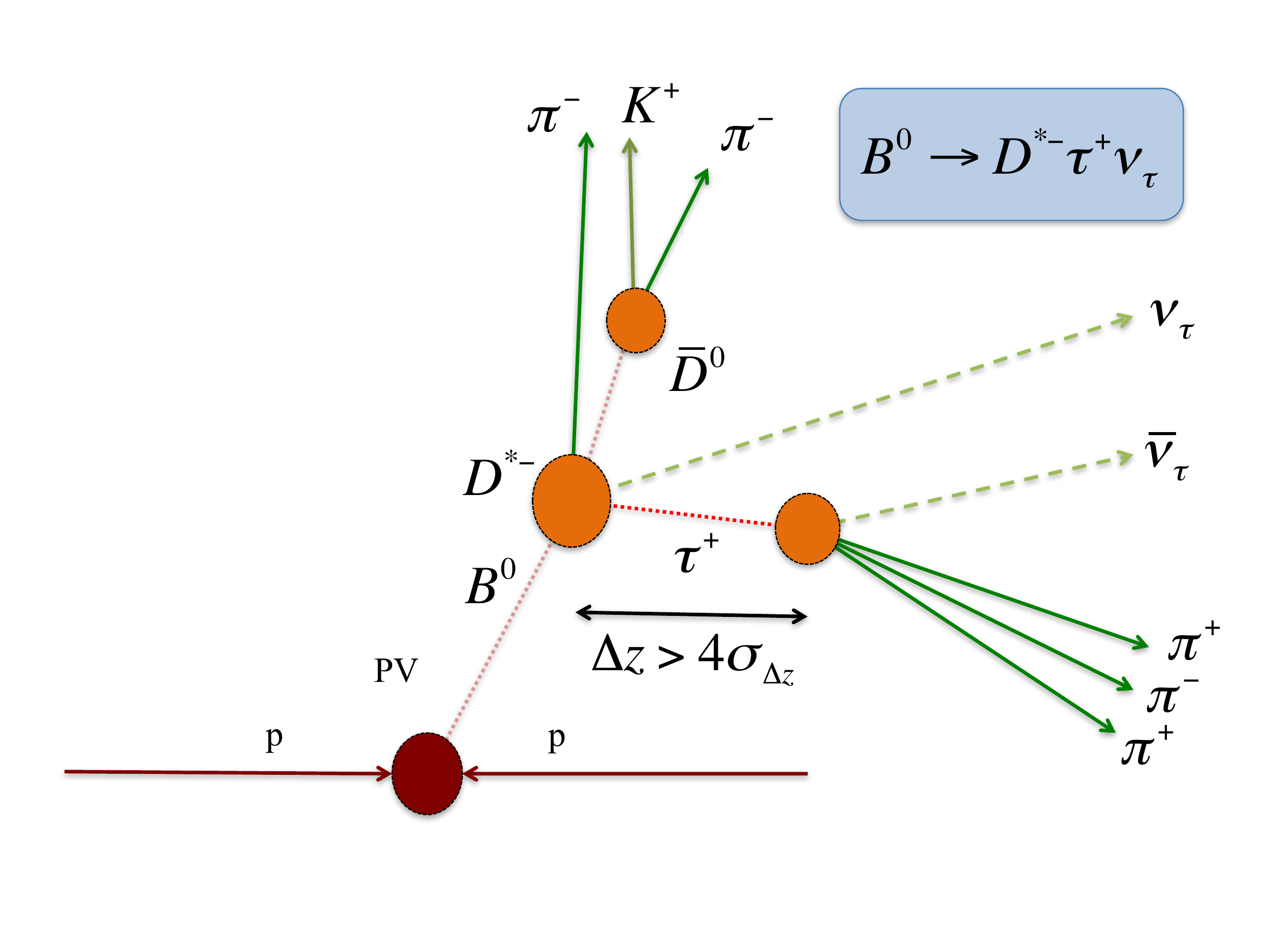}
	\caption{Topology of the signal decay. A requirement on the
          distance between the
          3\pion and the \Bz vertices along the beam direction to be greater than
          four times its uncertainty is applied.}
\label{fig:new_inversion}
\end{figure}

The signal final state consists of a \Dstarm meson, reconstructed in the
\decay{\Dstarm}{\Dzb\pim}, \decay{\Dzb}{\Kp\pim} decay
chain, associated with a 3\pion system.
 The selection of \Dstarm candidates starts by requiring \Dzb candidates with
masses between $1845$ and $1885$\mevcc, \pt larger than 1.6\gevc, combined with pions of
\pt larger than $0.11$\gevc such that the difference between the \Dstarm and the
\Dzb masses lies between $143$ and $148$\mevcc. The $\Dstarm 3\pion$ combination is very common in \PB meson
decays, with a signal-to-background ratio smaller than $1$\%. The dominant
background is prompt, {\it{i.e.}} consisting of candidates where the 3\pion
system is produced at the \Bz vertex. However, in the signal case,
because of the significant \tauon lifetime and boost along the forward
direction, the 3\pion system is detached from the \Bz vertex, as shown in
Fig.~\ref{fig:new_inversion}.
The requirement for the detached vertex is that the distance between the 3\pion
and the \Bz vertices along the beam direction, $\Delta z \equiv
z(3\pion)-z(\Bz)$, is greater than four times its uncertainty, $\sigma_{\Delta
z}$. This leads to  an improvement in the signal to noise ratio by a factor 160, as shown in Fig.~\ref{fig:plotz}. The remaining background consists
of two main categories: candidates with a true detached-vertex topology, and
candidates that appear to have such a detached-vertex topology.

\subsubsection{Background with detached-vertex topology}
\label{ssec:double_charm}
\begin{figure}[b]
	\centering
	\includegraphics[width=0.6\textwidth]{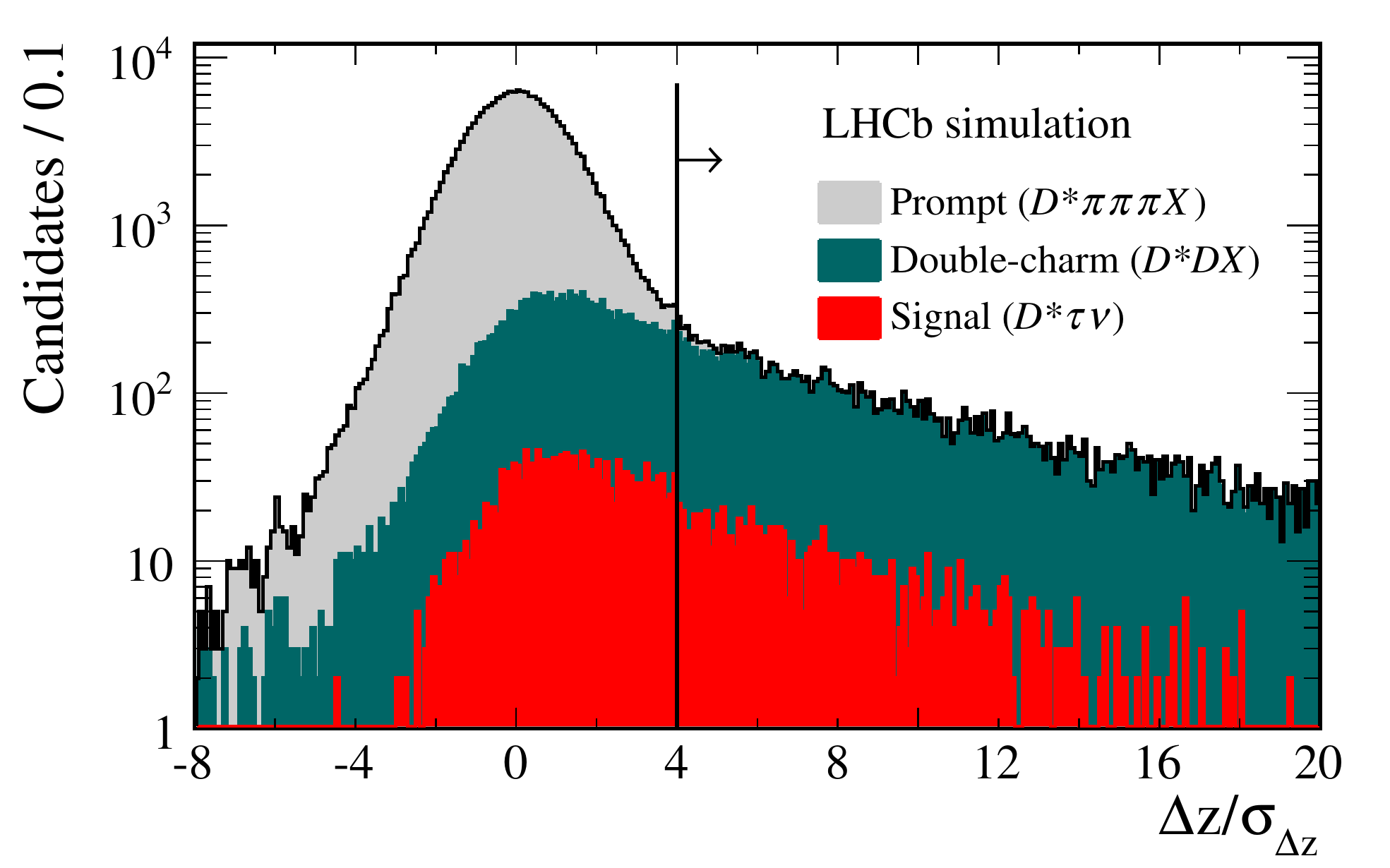}
	\caption{Distribution of the distance between the \Bz vertex and the
3\pion vertex along the beam direction, divided by its uncertainty, obtained
using simulation.
The vertical line shows the 4$\sigma$ requirement
used in the analysis to reject the prompt background component.}
\label{fig:plotz}
\end{figure}

\begin{figure}[htb]
 \centering
 \includegraphics[width=0.6\textwidth]{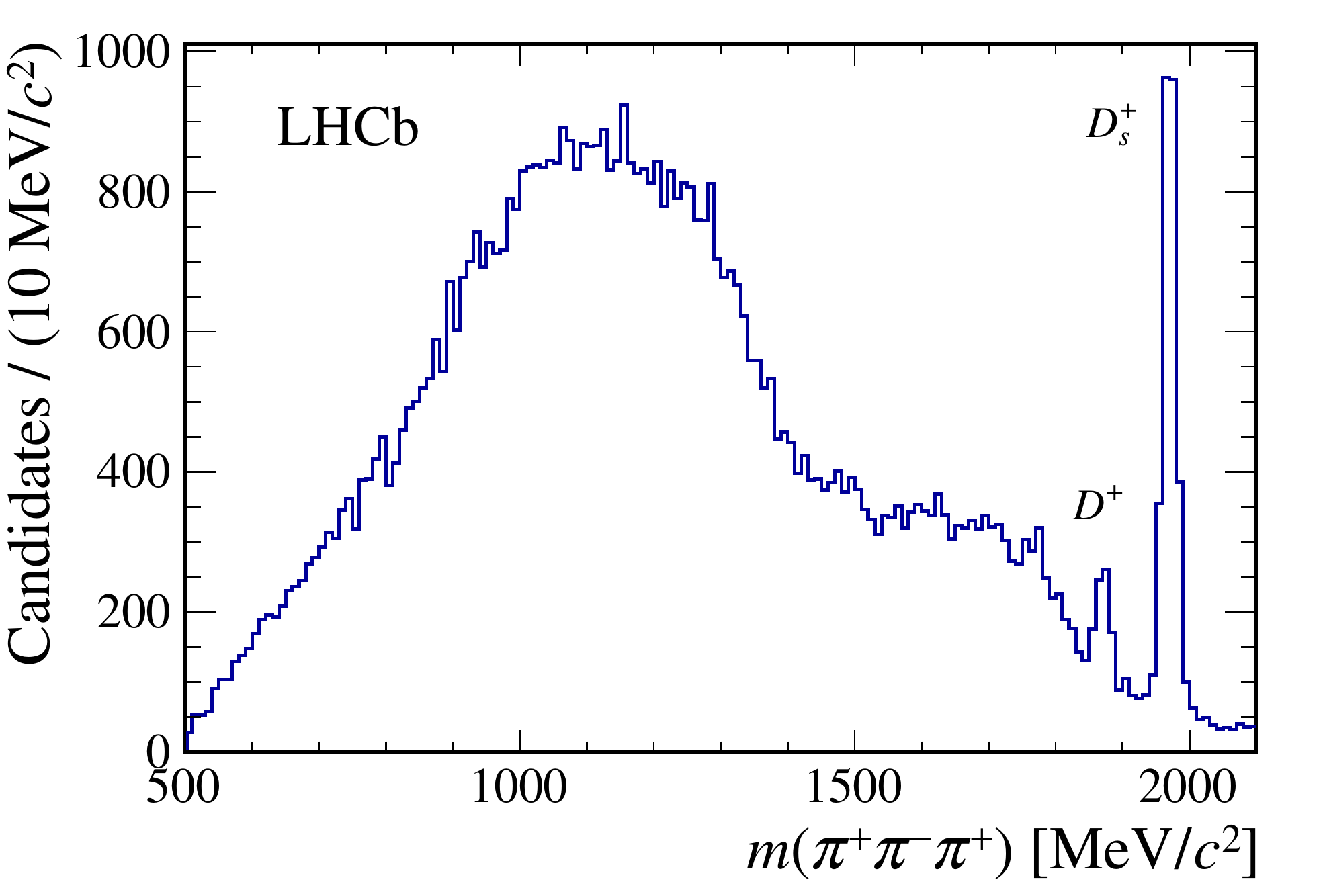}
\caption{\small Distribution of the $3\pi$ mass for candidates after the
detached-vertex requirement. The \Dp and \Dsp mass peaks are indicated.}
 \label{fig:control-ds}
\end{figure}

The double-charm \decay{\PB}{\Dstarm \D (X)} decays are the only other \PB decays
with the same vertex topology as the signal. Figure~\ref{fig:plotz} shows, on
simulated events, the
dominance of the double-charm background over the signal after the detached-vertex
requirement.
Figure \ref{fig:control-ds} shows the $3\pi$ mass data distribution
after the detached-vertex requirement, where peaking structures corresponding to the
\decay{\Dp}{3\pi} decay and \decay{\Dsp}{3\pi} decay -- a very important control channel for
this analysis -- are clearly visible.


\subsubsection{Background from other sources}
\label{ssec:othbkg}

Requirements additional to the detached vertex are needed to reject
spurious background sources with vertex topologies similar to the signal. The
various background sources are classified to distinguish candidates where the
3\pion system originates from a common vertex and those where one of the three
pions originates from a different vertex.

The background category, where the 3\pion system stems from a common vertex, is
further divided into two different classes depending on whether or not the \Dstarm and
3\pion system originate from the same  \bquark hadron.
In the first case, the $3\pi$ system either comes from the decay
of a $\tau$ lepton or a \Dz , \Dp, $D_s^+$ or $\Lz_c^+$ hadron.{~Candidates originating from $b$ baryons form only 2\% of this double-charm category.} In this case, the
candidate has the correct signal-like vertex topology.  Alternatively, it comes
from a misreconstructed prompt background candidate containing a \Bz , \Bp, \Bs
or $\Lz_b^0$ hadron. The detailed composition of
these different categories at the initial and at the final stage of the analysis is
described in
Sec.~\ref{ssec:comp}.
In the second case, the \Dstarm and the 3\pion systems  are not
daughters of the same \bquark hadron. The 3\pion system originates from one of
the following
sources: the other \bquark hadron present in the event ({\it B1B2} category); the decay of charm hadrons
produced at the PV ({\it{charm}} category); another
PV; or an interaction in the beam pipe or in the detector material.

The 3\pion background not originating  from the same vertex is dominated by candidates where two pions originate from the same
vertex whilst the third may come directly from the PV, from a different vertex
in the decay chain of the  same \bquark hadron, from the other \bquark hadron
produced at the PV, or from another PV.
Due to the combinatorial origin of this background, there is no strong
correlation between the charge of the 3\pion system and the $D^{*-}$
charge. This enables the normalization of the combinatorial background
with the wrong-sign data sample.

\subsubsection{Summary of the topological selection requirements}
\label{ssec:list}
The requirements applied to suppress combinatorial and charm backgrounds, in
addition to the detached-vertex criterion, are reported in
Table~\ref{tab:FirstCut}.
These include a good track quality and a minimum transverse momentum of 250\mevc
for each pion, a good vertex reconstruction quality for the 3\pion system and large $\chisq_{\text{IP}}$ with respect to any PV for each pion
of the 3\pion system and for the \Dzb candidate, where $\chisq_{\text{IP}}$ is defined as the difference in the vertex-fit \chisq of a given PV
reconstructed with and without the particle under
consideration. In addition, the 3\pion vertex must be
  detached from its primary vertex along the beam axis by at least 10 times the
corresponding uncertainty. The distance
from the 3\pion vertex position to the beam center in the plane transverse to
the beam direction, $r_{3\pion}$, must be outside
the beam envelope and inside the beam pipe to avoid 3\pion vertices coming from
 proton interactions or secondary interactions with the beam-pipe
material. The attached primary vertex to the \Dzb and 3\pion candidates must be the same. The number of candidates per event must be equal to one; this cut is the first rejection step against nonisolated candidates. Finally, the difference between the reconstructed \Dstarm and \Dzb masses must lie between 143 and 148 \mevcc.
\begin{table}[htbp]
  \centering
  \caption{List of the selection cuts. See text for further explanation.}
  \label{tab:FirstCut}
  \begin{tabular}{lrl}
    \hline
    \text{Variable} & \text{Requirement} & \text{Targeted background}\\ \hline
    $[z(3\pion) - z(\Bz)]/\sigma_{(z(3\pion) - z(\Bz))}$ & $>4$ & prompt \\
    \pt{(\pion)}, \pion from 3\pion    & $> 250$\mevc & all \\
    $3\pi$ vertex $\chisq$ & $< 10$ & combinatorial \\
    $\chisq_{\text{IP}}(\pi)$, $\pi$ from 3\pion  & $> 15$   & combinatorial \\
    $\chisq_{\text{IP}}(\Dzb)$ & $> 10$ &charm\\
    $[z(3\pion) - z({\text{PV}})]/\sigma_{(z(3\pion) - z({\text{PV}}))}$ & $>10$ & charm \\
    $r_{3\pion}$ & $\in [0.2,5.0]\mm$& spurious 3\pion \\
    $\text{PV}(\Dzb)$ & $=\text{PV}(3\pion)$ &charm/combinatorial\\
    number of \Bz candidates & $=1$&all \\
   $\Delta m \equiv m(\Dstarm) - m(\Dzb)$ & $\in [143,148]\mevcc$  &combinatorial \\ \hline
  \end{tabular}
\end{table}

\subsection{Isolation requirements}
\label{ssec:isol}
\subsubsection{Charged isolation}
A charged-isolation algorithm ensures that no extra tracks are compatible with
either the \Bz or 3\pion decay vertices. It is implemented by counting the
number of charged tracks having \pt larger than 250\mevc, $\chisq_{\text{IP}}$
with respect to the PV larger than 4, and $\chisq_{\text{IP}}(3\pion)$
and $\chisq_{\text{IP}}(\Bz)$, with respect to the vertex of the 3\pion
and \Bz candidates, respectively, smaller than 25.
The $\Dstarm 3\pion$ candidate is rejected if any such track is found.
As an example, the performance of the charged-isolation algorithm is determined on a simulated
sample of double-charm decays with a \Dz meson in the final state. In
cases where $\Bz\to\Dstarm\Dz\Kp(X)$, with $\Dz\to\Km 3\pi(X)$, two
charged kaons are present in the decay chain, one originating from the \Bz
vertex and the other from the \Dz vertex.
For these candidates, the rejection rate is  95\%. The
charged-isolation algorithm has a selection efficiency of 80\% on a data sample of
exclusive $\Bz\to\Dstarm 3\pi$ decays. This sample has no additional
charged tracks from the \Bz vertex
and has thus similar charged-isolation properties as the signal.
This value is in good
agreement with the efficiency determined from simulation.

\begin{figure}[htb]
\centering
 \includegraphics[width=0.6\textwidth]{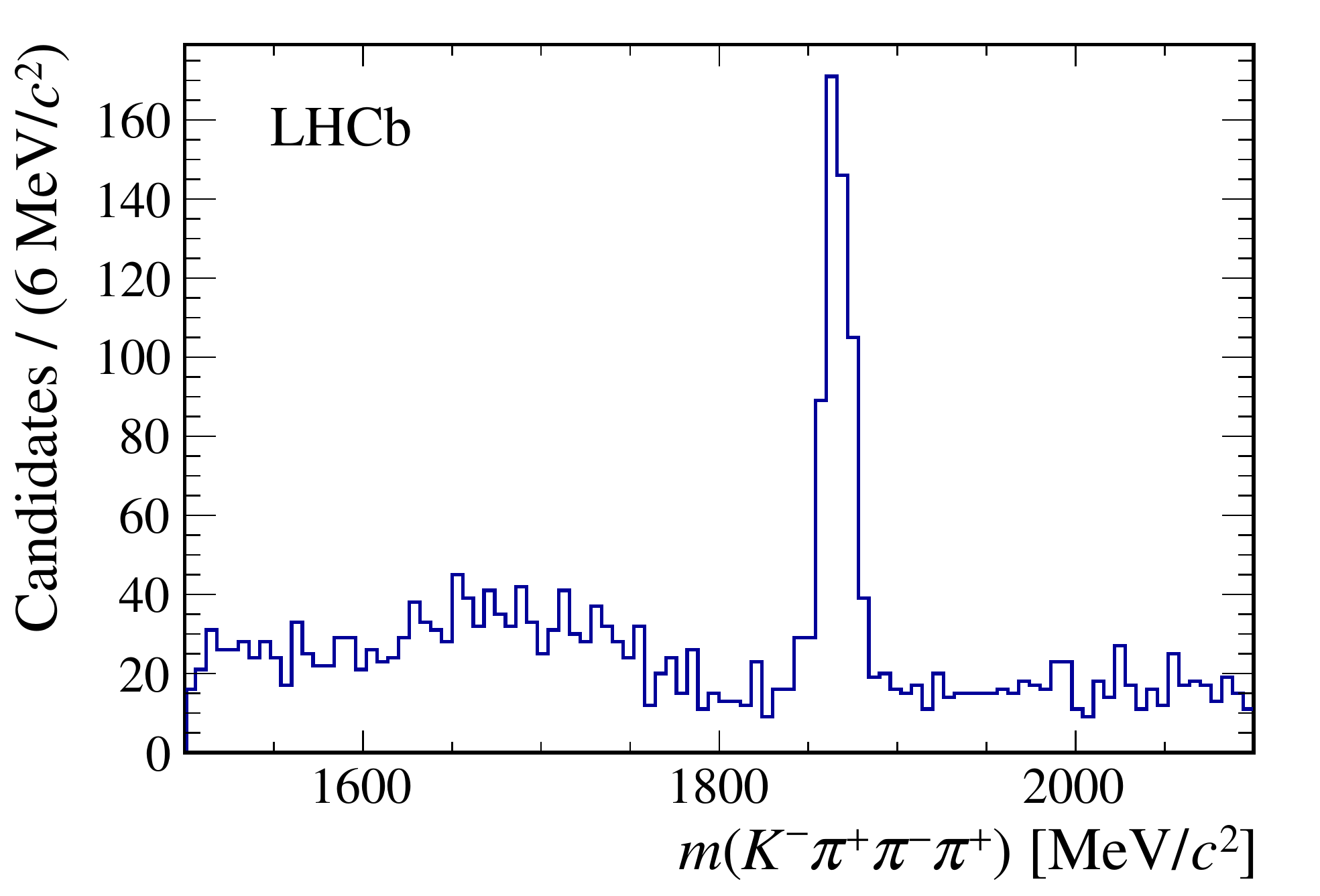}
\caption{\small Distribution of the $\Km 3\pi$ mass for $D^0$ candidates where a charged kaon has been associated to the 3\pion vertex.}
\label{fig:controlDz}
\end{figure}

Reversing the isolation requirement provides a sample of
candidates from the inclusive \Dz decay chain mentioned above, where a \Dz meson
decays into $\Km 3\pi$ and the charged kaon has been found as a nearby
track. Figure~\ref{fig:controlDz} shows the $\Km3\pion$ mass distribution
featuring a prominent \Dz peak.
This control sample is used to determine the
properties of the \decay{\B}{\Dstarm\Dz(X)} background in the signal fit.

\subsubsection{Neutral isolation}
\label{ssec:neutralCones}
\begin{figure}[htb]
\centering
 \includegraphics[width=0.6\textwidth]{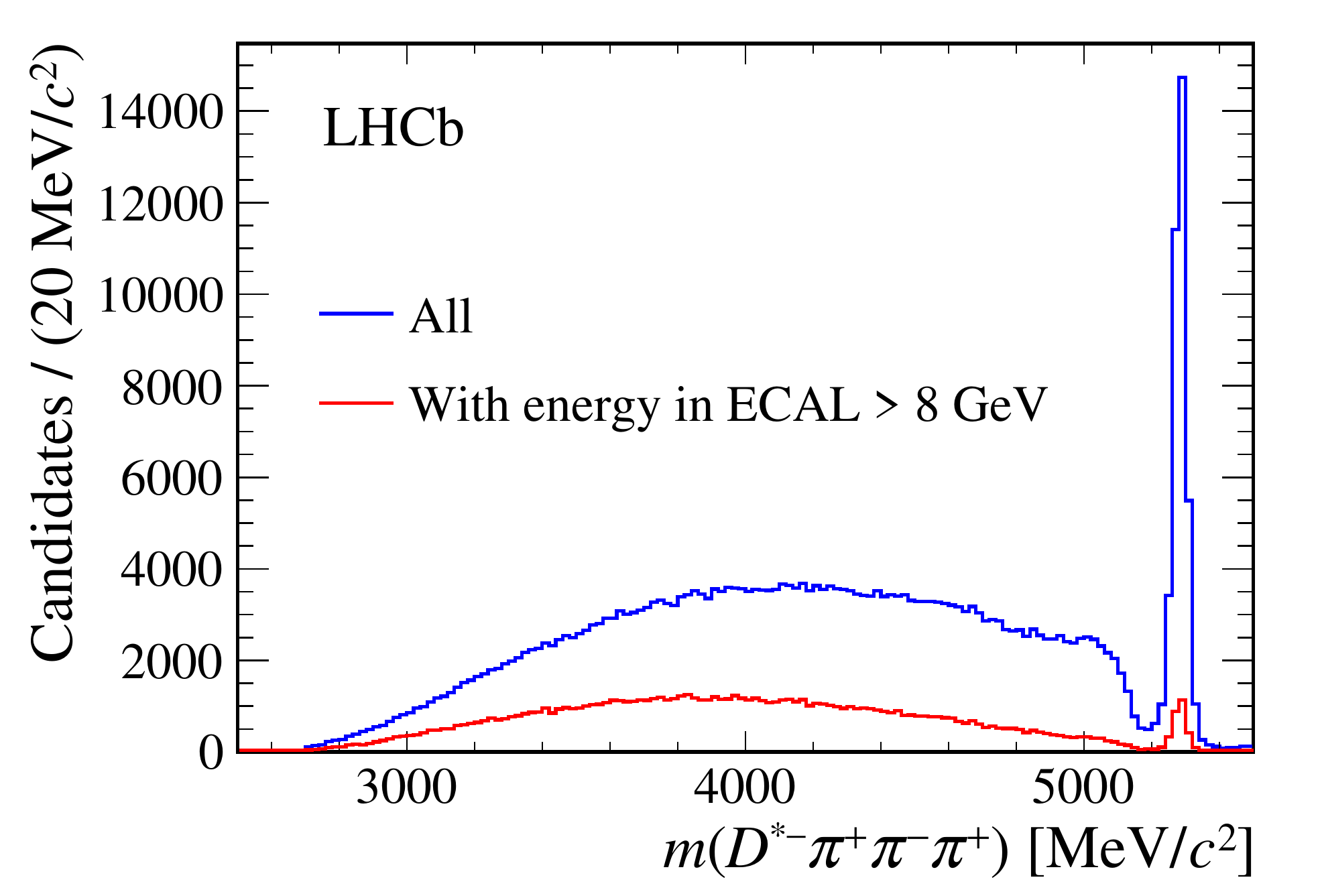}
\caption{Distribution of the $\Dstarm{3\pi}$ mass (blue) before
and (red) after a requirement of finding an energy of at least 8\gev in the
electromagnetic calorimeter around the 3\pion direction.}
\label{fig:neu}
\end{figure}

Background candidates from decays with additional neutral particles
are suppressed by using the energy deposited in the electromagnetic
calorimeter in a cone of $0.3$ units in $\Delta\eta-\Delta\phi$ around the
direction of the 3\pion system, where $\phi$ is the azimuthal angle in the plane
perpendicular to the beam axis. For this rejection
method to be effective, the amount of collected energy in the region of interest
must be small when no neutral particles are produced in the \Bz meson decay.
Candidates where the \Bz meson decays
to $\Dstarm 3\pi$, with $\Dstarm\to\Dzb\pim$, are used as a check.
Figure~\ref{fig:neu} compares the distributions of the $\Dstarm{3\pi}$
mass with  and without the requirement of an energy deposition of at
least 8\gev in the electromagnetic calorimeter around the 3\pion direction. Since it is known that no neutral
particle is emitted in this decay, the
inefficiency of this rejection method is estimated by the ratio of the yields of the
two spectra within $\pm 30$\mevcc around the \Bz mass, and it is found to be small enough to allow the use of this method.
  The energy deposited in the electromagnetic calorimeter around the 3\pion
direction is one of the input quantities to the
MVA described below, used to suppress inclusive \Dsp decays to $3\pion X$,
which contain photons and \piz mesons
in addition to the three
pions. Photons are also produced when \Ds excited states decay to the \Ds ground state.  The use of this variable
 has an impact on signal, since it vetoes the
\decay{\taup}{3\pi\piz\neutb} decay, whose efficiency is roughly one
half with respect to that of
the 3\pion mode, as can be seen later in Table~\ref{tab:efficiency-summary}.

\subsection{Particle identification requirements}
\label{ssec:misid}
\begin{figure}[t]
 \begin{center}
   \includegraphics*[width=0.6\textwidth]{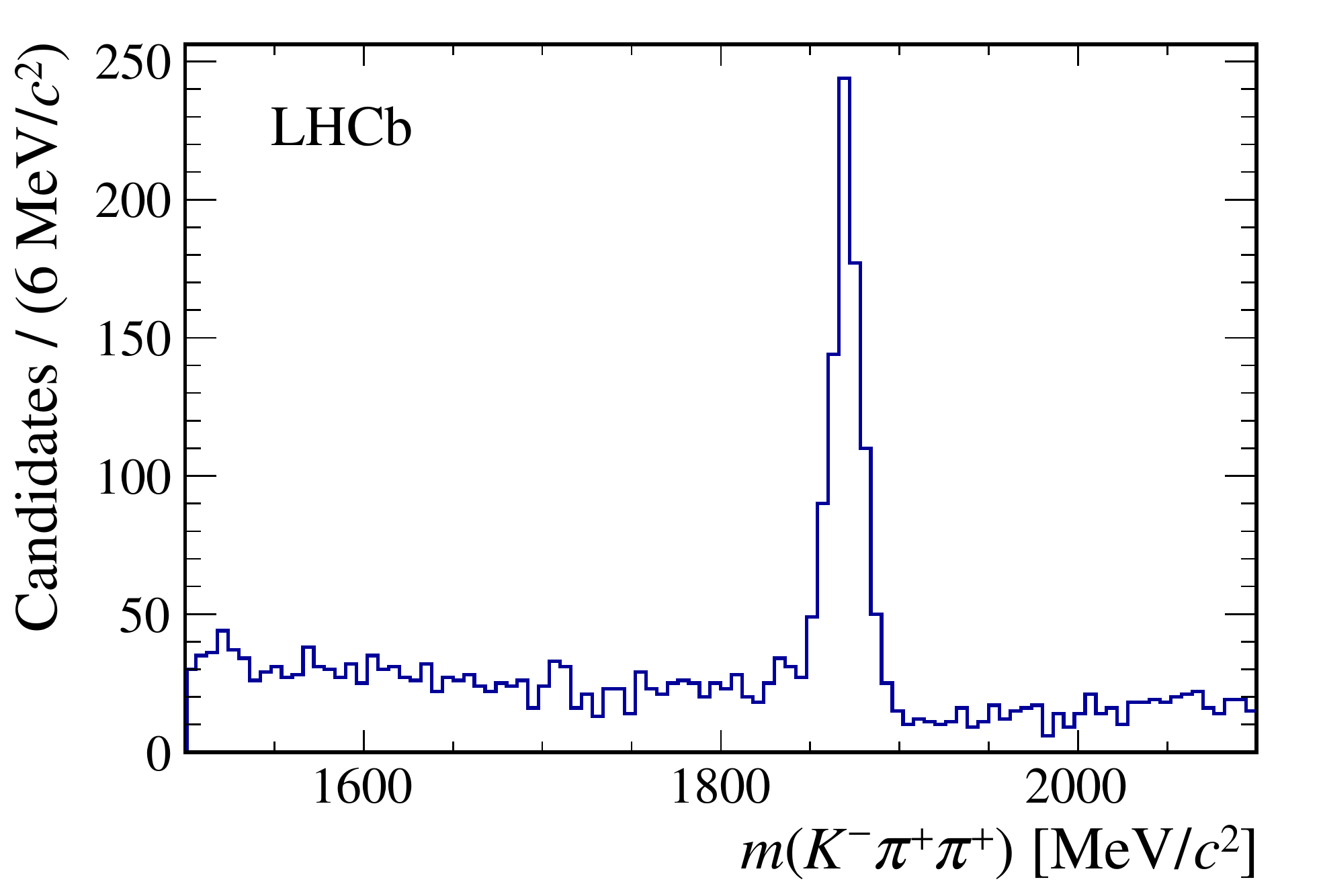}
 \end{center}
\caption{Distribution of the \Km\pip\pip mass for $D^+$ candidates passing
the signal selection, where the negative pion has been identified as a
kaon and assigned the kaon mass.}
\label{fig:kpipi}
\end{figure}

 In order to ensure that the tracks forming the 3\pion
   candidate are real pions, a positive pion identification is
   required and optimized taking into account the  efficiency and
   rejection performance of particle identification (PID) algorithms,
   and the observed kaon to pion ratio in the 3\pion candidates, as measured through the \Dm peak when giving a kaon mass to the negatively charged pion. As a result, the kaon identification probability is required to be less than 17\%. To keep
the \Dstarm reconstruction efficiency as high as possible, the requirement on the
kaon identification probability for the soft-momentum pion originating from the
\Dstarm decay is set to be less than $50$\%.   The
\mbox{\decay{\Dp}{\Km\pip\pip}} and
\decay{\Dp}{\Km\pip\pip\piz} decays have large branching fractions and
contribute to the $\PB\to\Dstarm\Dp(X)$ background, that is significant when the
kaon is misidentified as a pion.  A remaining  kaon contamination of about 5\%
in the final
sample is estimated by studying the \Km\pip\pip mass when assigning the
 kaon mass to the negative pion. Figure~\ref{fig:kpipi} shows the \Km\pip\pip
mass distribution for candidates that have passed all analysis
requirements, except that the \pim candidate must have a high kaon identification probability. A clear \Dp signal
of $740\pm 30$ candidates is visible, with little
combinatorial background. Therefore, an additional requirement on the kaon
identification probability of the \pim candidate is applied.
All of these PID requirements are chosen in order to
get the best discrimination between signal and background.  They form, together
with the topology selection and the isolation requirement defined above, the
final selection.

\subsection{Selection of the normalization channel}
\label{ssec:selnorm}
\begin{figure}[htb]
\centering
 \includegraphics[width=0.6\textwidth]{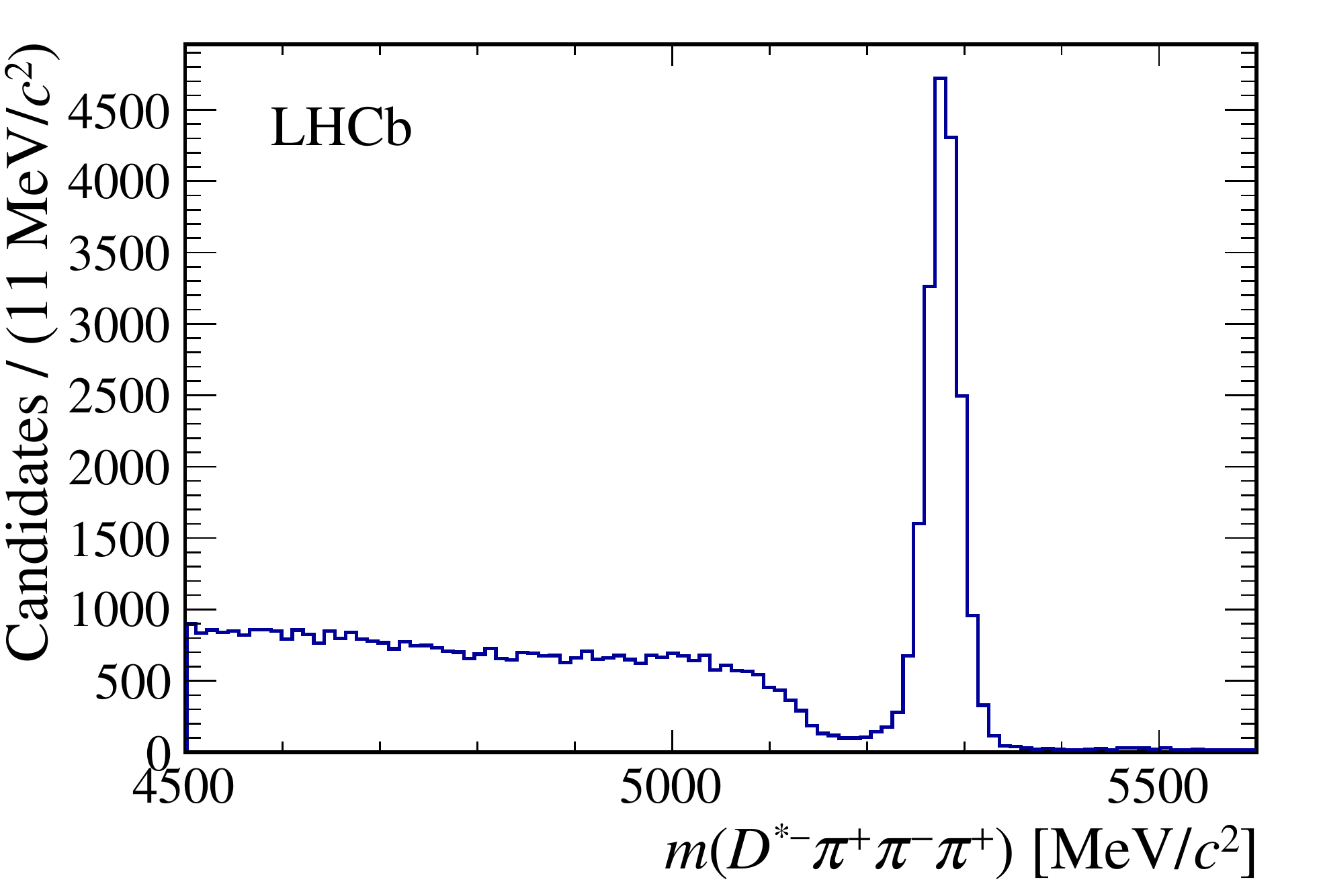}
\caption{Distribution of the $\Dstarm 3\pi$ mass for candidates passing the
selection.}
\label{fig:largebmass}
\end{figure}

The $\Bz \to \Dstarm 3\pi$ normalization channel is selected by requiring the \Dzb
vertex to be located at least
4$\sigma$ downstream of the 3\pion vertex along the beam direction, where
$\sigma$ is the distance between the $B^0$ and $\Dzb$ vertices divided by
their uncertainties added in quadrature. All other selection criteria are
identical to that of the signal case, except for the fact that no MVA
requirement is applied to the normalization channel. Figure~\ref{fig:largebmass}
shows the
$\Dstarm 3\pi$ mass spectrum after all these requirements.
Moreover, the high purity of this sample of
exclusive \Bz decays allows the validation of the selection efficiencies
derived using simulation.
 \subsection{Reconstruction of the decay kinematics}
\label{ssec:reconstr}
Due to the precise knowledge of the \Dzb, 3\pion and \Bz decay vertices, it is
possible to reconstruct the decay chains of both signal and background
processes, even in the presence of unreconstructed
particles, such as two neutrinos in the case of the signal, or neutral particles
originating at the 3\pion vertex in the case of double-charm background. The
relevant reconstruction techniques are detailed in the following.

\subsubsection{Reconstruction in the signal hypothesis}
The missing information due to the two neutrinos emitted in the signal decay chain can be recovered with the
measurements of the \Bz and $\tau$ line of flight (unit vectors joining the \Bz
vertex to the PV and the 3\pion vertex to the \Bz vertex,
respectively) together with the known \Bz and $\tau$ masses.
The reconstruction of the complete decay kinematics of both
the \Bz and $\tau$ decays is thus possible, up to two two-fold ambiguities.

\begin{figure}[t]
\begin{center}
    \includegraphics[width=0.59\linewidth]{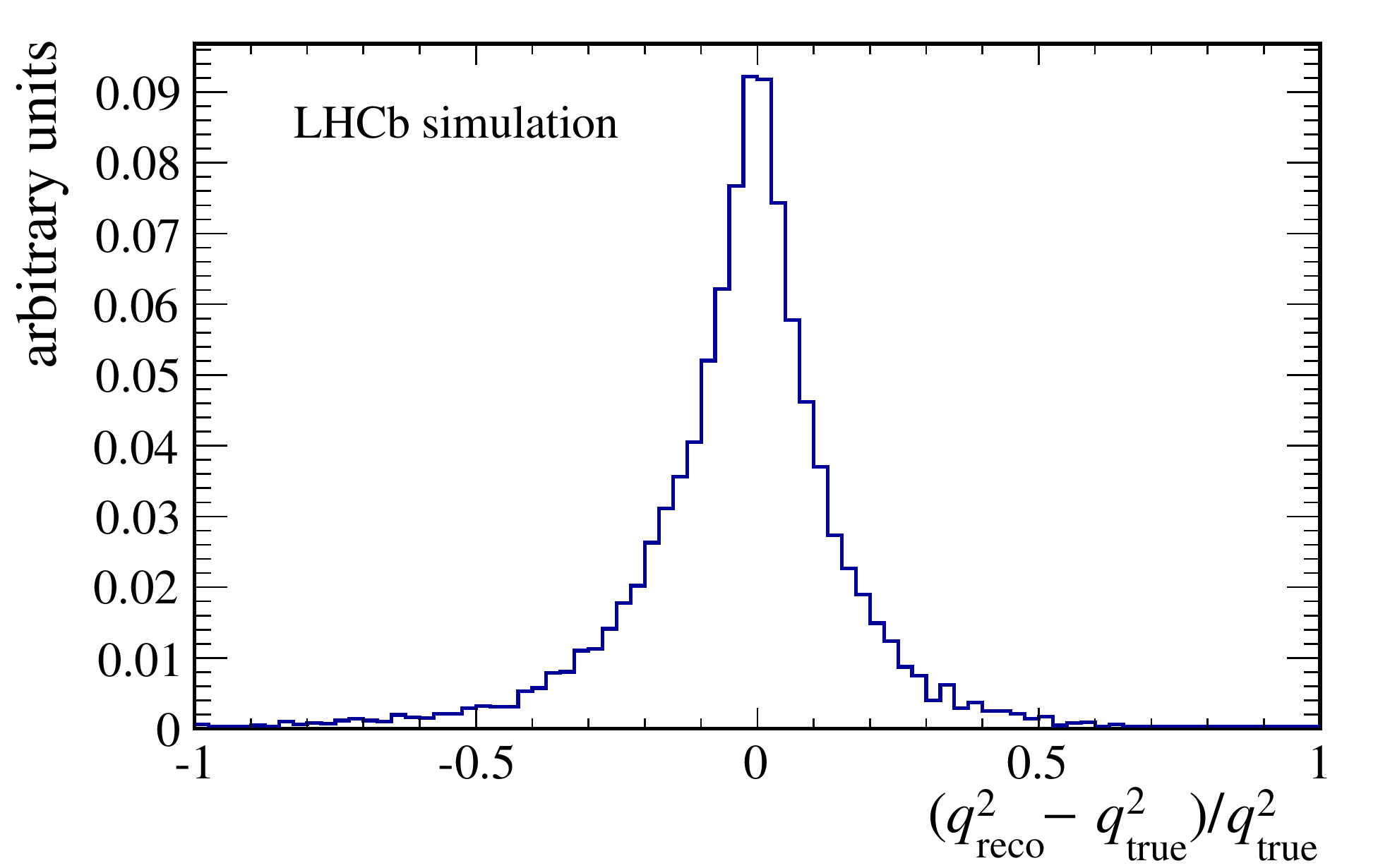}
  \end{center}
  \caption{
    \small 
    Difference between the reconstructed and true \qsq variables divided by the
true \qsq, observed in
the \mbox{\decay{\Bz}{\Dstarm\taup\neut}} simulated signal sample after partial reconstruction.}
  \label{fig:q2resol}
\end{figure}

The \tauon momentum in the laboratory frame is obtained as (in units where $c=1$)
\begin{equation}
  |\vec{p}_{\tau}| =
\frac{(m_{3\pi}^2+m_{\tau}^2)|\vec{p}_{3\pi}|\cos{\theta_{\tau,3\pi}}
    \pm E_{3\pi}\sqrt{(m_{\tau}^2-m_{3\pi}^2)^2-4
      m_{\tau}^2|\vec{p}_{3\pi}|^2\sin^2\theta_{\tau,3\pi}}}{2(E_{3\pi}^2-|\vec{p}_{3\pi}|^2\cos^2\theta_{\tau,3\pi})},
\label{eq:ptau}
\end{equation}

\noindent where $\theta_{\tau,3\pi}$ is the angle between the $3\pi$ system
three-momentum and the
$\tau$ line of flight; $m_{3\pi}$, $|\vec{p}_{3\pi}|$ and $E_{3\pi}$ are the
mass, three-momentum and energy of the $3\pi$ system, respectively; and
$m_{\tau}$ is the known $\tau$ mass.
Equation~\ref{eq:ptau} yields a single solution, in the limit where the opening angle between the $3\pi$
and the $\tau$ directions takes the maximum allowed value
\begin{equation}
  \theta^{\max}_{\tau,3\pi} = \arcsin
\left(\frac{m_{\tau}^2-m_{3\pi}^2}{2m_{\tau}|\vec{p}_{3\pi}|}\right).
\end{equation}
At this value, the argument of the square root in Eq.~\ref{eq:ptau} vanishes,
leading to only one solution, which is used as an estimate of the $\tau$ momentum.
The same procedure is applied to estimate the $B^0$ momentum

\begin{eqnarray}
  |\vec{p}_{B^0}| &=&
\frac{(m_{Y}^2+m_{B^0}^2)|\vec{p}_{Y}|\cos{\theta_{B^0,Y}} \pm
    E_{Y}\sqrt{(m_{B^0}^2-m_{Y}^2)^2-4
      m_{B^0}^2|\vec{p}_{Y}|^2\sin^2\theta_{B^0,Y}}}{2(E_{Y}^2-|\vec{p}_{Y}|^2\cos^2\theta_{B^0,Y})}
\label{eq:pB}
\end{eqnarray}

\noindent by defining
\begin{eqnarray}
  \theta^{\max}_{B^0,Y} &=& \arcsin
\left(\frac{m_{B^0}^2-m_{Y}^2}{2m_{B^0}|\vec{p}_{Y}|}\right),
\end{eqnarray}
\noindent  where $Y$ represents the $D^{*-}\tau$ system.
Here, the three-momentum and mass of the  $D^{*-}\tau$ system are calculated using the previously estimated $\tau$ momentum
\begin{eqnarray}
  \vec{p}_{Y} = {\vec{p}}_{D^{*-}} + {\vec{p}}_{\tau},\,\,\,\,\,\,E_Y = E_{D^{*-}} + E_\tau,
\end{eqnarray}
\noindent where ${\vec{p}}_{D^{*-}}$ and ${\vec{p}}_{\tau}$ are the three-momenta of the $D^{*-}$ and the
$\tau$ candidates, and $E_{D^{\ast-}}$ and $E_\tau$ their energies. Using this method,
the rest frame variables \mbox{$\qsq\equiv
(p_{\Bz}-p_{\D^{\ast-}})^2 = (p_{\tau}+p_{\neut})^2$} and the $\tau$ decay time,
$t_{\tau}$, are determined with sufficient accuracy
to retain their discriminating power against double-charm backgrounds, as
discussed in Sec.~\ref{sec:signal_yield}. Figure~\ref{fig:q2resol} shows the
difference between the reconstructed  and the true value of $\qsq$ divided by
the true \qsq on simulated events.
No significant bias is
observed and an average resolution of 1.2\gevgevcccc is obtained.
The relative \qsq resolution is $18\%$ full-width half-maximum.{~The slight asymmetry is due to the presence at low \qsq of a tail of reconstructed \qsq below the kinematical limit for true \qsq.}

\subsubsection{Reconstruction assuming a double-charm origin for the candidate}

A full kinematic reconstruction of the \B decay chain specifically adapted to
two-body double-charm \B decays provides additional discrimination.
After the detached-vertex requirement, the main source of background candidates is attributed to decays of the form
\decay{\PB}{\Dstarm\Dsp (X)}, with \decay{\Dsp}{3\pion N}, $N$ being a system
of unreconstructed neutral particles. For these decays, the
missing information is due to a neutral system of unknown mass originating from the
\Dsp decay vertex, {\em{i.e.}} four unknowns. The measurements of the \Bz and \Dsp lines of flight,
 { providing four constraints, }together with the known \Bz mass, are sufficient to
reconstruct the full decay kinematics

\begin{equation}
  \label{eq:pConserv}
  |\vec{p}_B|\hat{u}_B = |\vec{p}_\Dsp|\hat{u}_\Dsp + \vec{p}_\Dstarm.
\end{equation}
{This equation assumes the absence of any other particles in the \B decay. It is however also valid when an additional particle is aligned with the \Dsp momentum direction, as in the case of \Bz\to\Dstarm\Dssp where the soft photon emitted in the \Dssp decay has a very low  momentum in the direction transverse to that of the \Dsp momentum. It is also a good approximation for quasi-two-body \Bz decays to \Dstarm and higher excitations of the \Dsp meson.}
This equation can be solved with two mathematically equivalent ways, through a vectorial or scalar product methods, noted $v$ and $s$ respectively. This equivalence does not hold in the presence of extra particles. This difference is used to provide some further discrimination between signal and nonisolated backgrounds. The magnitudes of the momenta obtained for each method are:
\begin{subequations}
  \begin{gather}
    \label{eq:PBSV}
    P_{B,v} = \frac{| \vec{p}_\Dstarm \times \hat{u}_\Dsp |}{| \hat{u}_B \times \hat{u}_\Dsp |}, \\
    P_{B,s} = \frac{\vec{p}_\Dstarm \cdot \hat{u}_B - (\vec{p}_\Dstarm \cdot \hat{u}_\Dsp)(\hat{u}_B \cdot \hat{u}_\Dsp)}{1 - (\hat{u}_B \cdot \hat{u}_\Dsp)^2},
    \end{gather}
  \end{subequations}
for the \Bz momentum, and
\begin{subequations}
  \begin{gather}
    \label{eq:PDsSV}
    P_{D_s,v} = \frac{| \vec{p}_\Dstarm \times \hat{u}_B |}{| \hat{u}_\Dsp \times \hat{u}_B |}, \\
    P_{D_s,s} = \frac{(\vec{p}_\Dstarm \cdot \hat{u}_B)(\hat{u}_B \cdot \hat{u}_\Dsp) - \vec{p}_\Dstarm \cdot \hat{u}_\Dsp}{1 - (\hat{u}_B \cdot \hat{u}_\Dsp)^2},
    \end{gather}
  \end{subequations}
for the \Ds momentum.

\begin{figure}[t]
\begin{center}
    \includegraphics[width=0.59\linewidth]{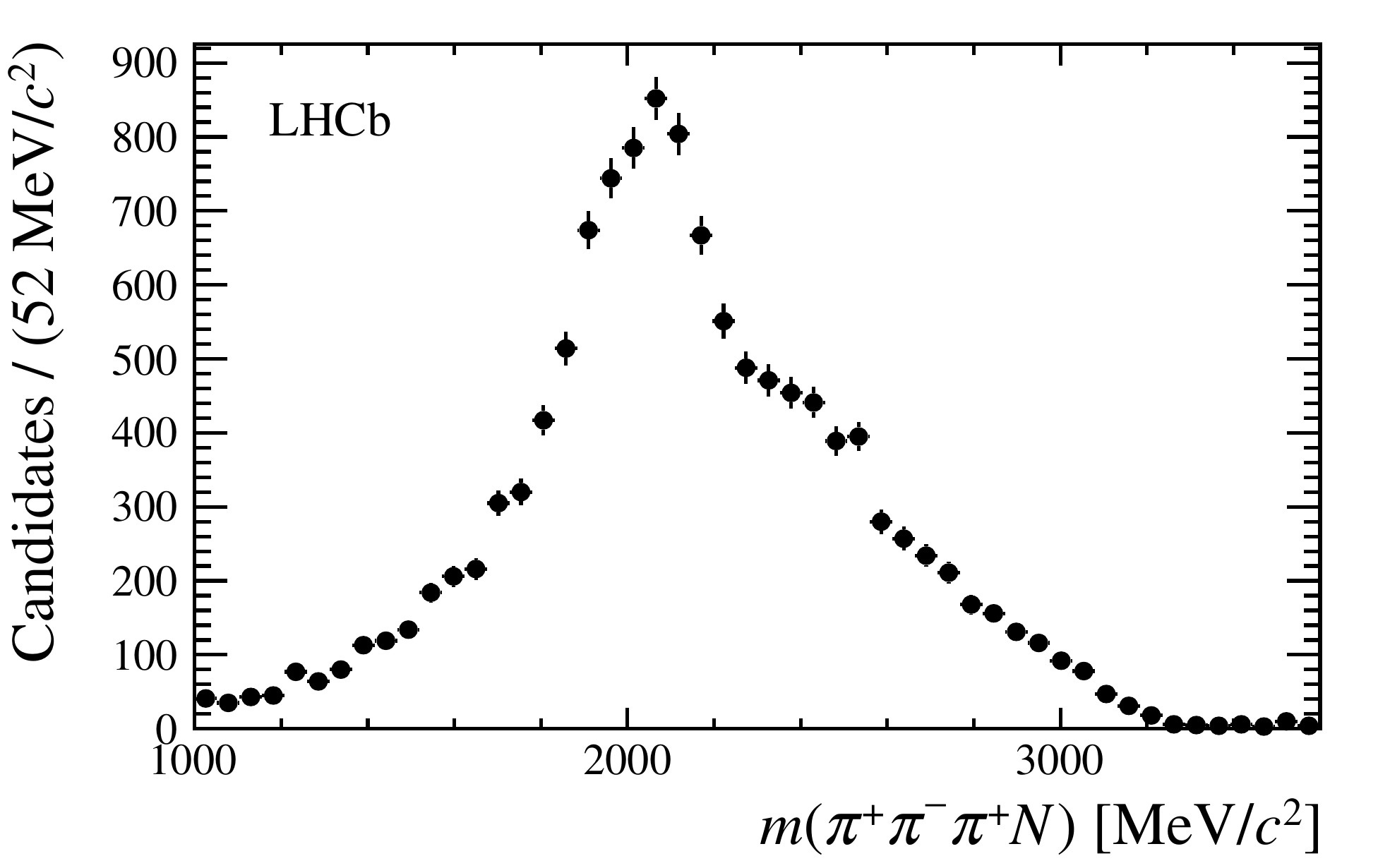}
  \end{center}
  \caption{
    \small 
    Distribution of the reconstructed $3\pi N$ mass observed in a data sample
enriched by $\B\to\Dstarm\Ds (X)$ candidates.}
  \label{fig:dsresol}
\end{figure}

 Since this partial
reconstruction works without imposing a mass to the $3\pion N$ system,
the reconstructed $3\pion N$ mass can be used as a discriminating variable.
Figure~\ref{fig:dsresol} shows the $3\pion N$ mass
distribution obtained on a sample
enriched in \decay{\PB}{\Dstarm\Dsp (X)} decays, with \decay{\Dsp}{3\pi N},
by means of the output of the MVA (see Sec.~\ref{ssec:BDT}).
A peaking structure originating from \Dsp and \Dssp decays is also present
around $2000$~MeV/$c^2$.
Due to the presence of two neutrinos at different
vertices, signal decays are not handled as well by this partial reconstruction
method, which therefore provides a useful discrimination between signal and
background due to \decay{\PB}{\Dstarm\Dsp (X)} decays. However, this method cannot
discriminate the signal from double-charm backgrounds due to \decay{\PB}{\Dstarm\Dz (X)} and
\decay{\PB}{\Dstarm\Dp (X)} decays, where  two kaons are missing at the \Bz
and 3\pion vertices.

\subsection{Multivariate analysis}
\label{ssec:BDT}
Three features are used to reject the double-charm background: the different
resonant structures of $\taup\to3\pion\neutb$ and
$\Dsp\to3\pion X$ decays, the neutral isolation and the different kinematic properties
of signal and background candidates. The latter feature is exploited by using
the reconstruction techniques described in Sec.~\ref{ssec:reconstr}.

\begin{figure}[tb]
\begin{center}
    \includegraphics[width=0.48\linewidth]{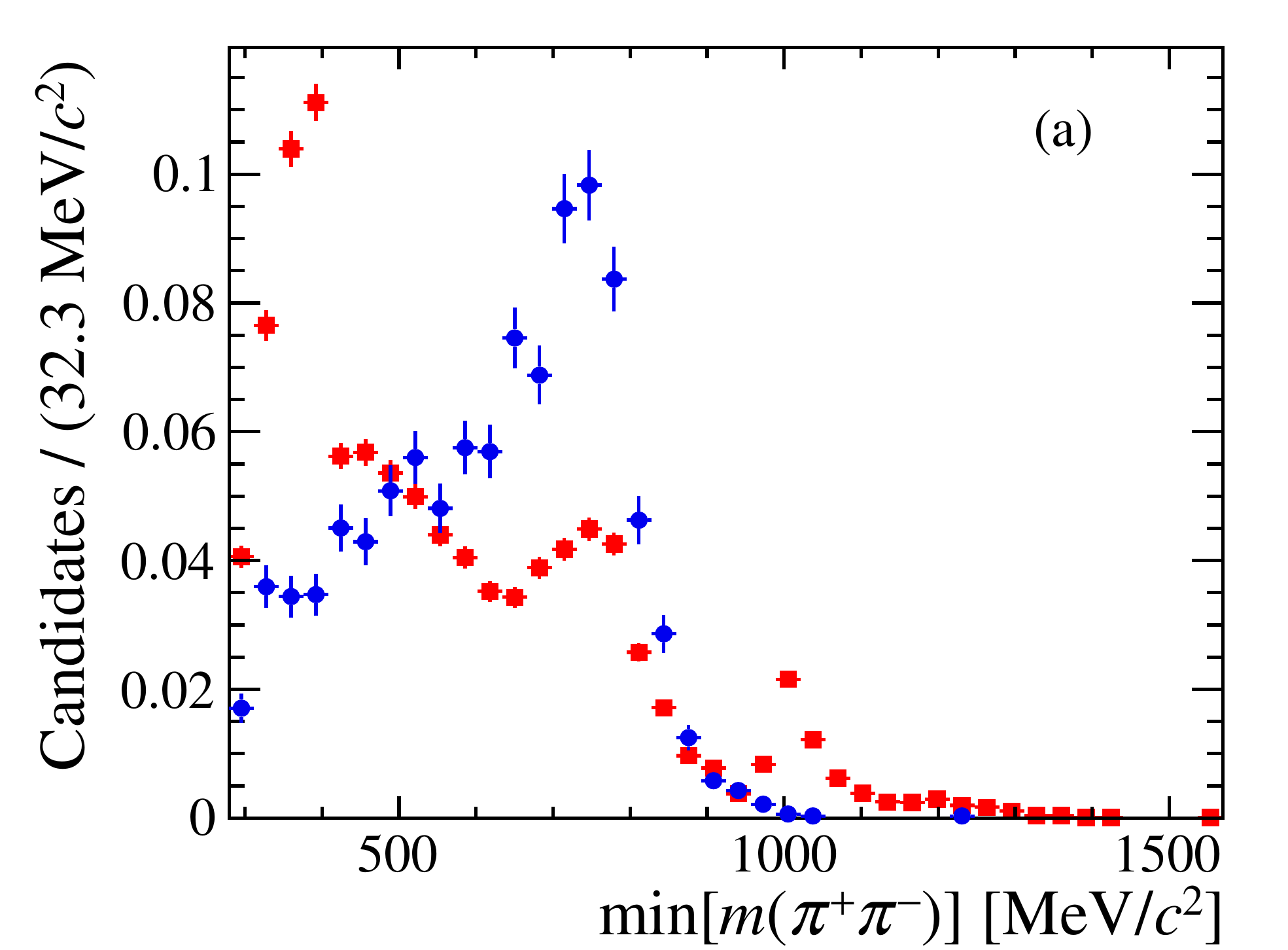}
    \includegraphics[width=0.48\linewidth]{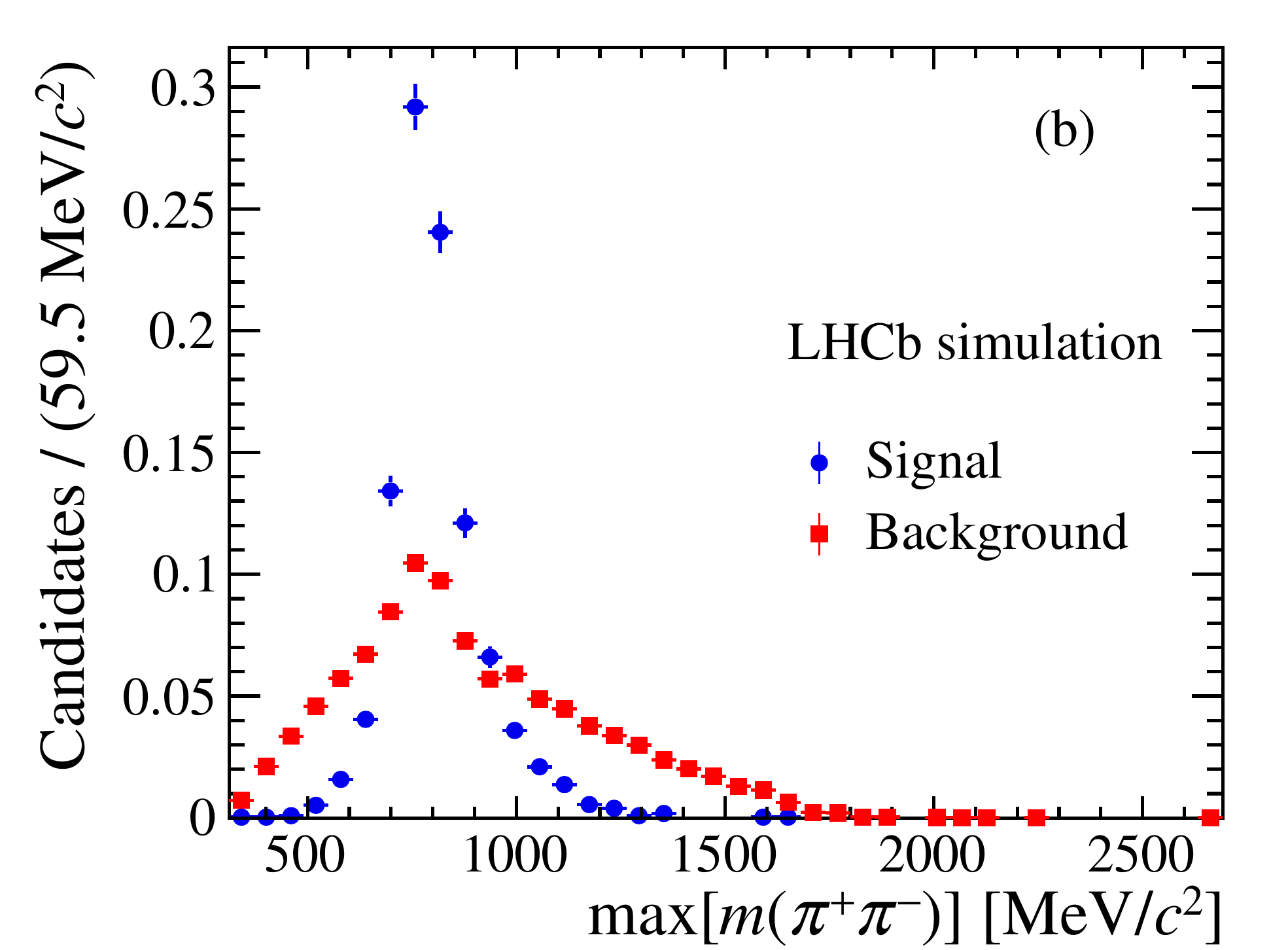}
    \includegraphics[width=0.48\linewidth]{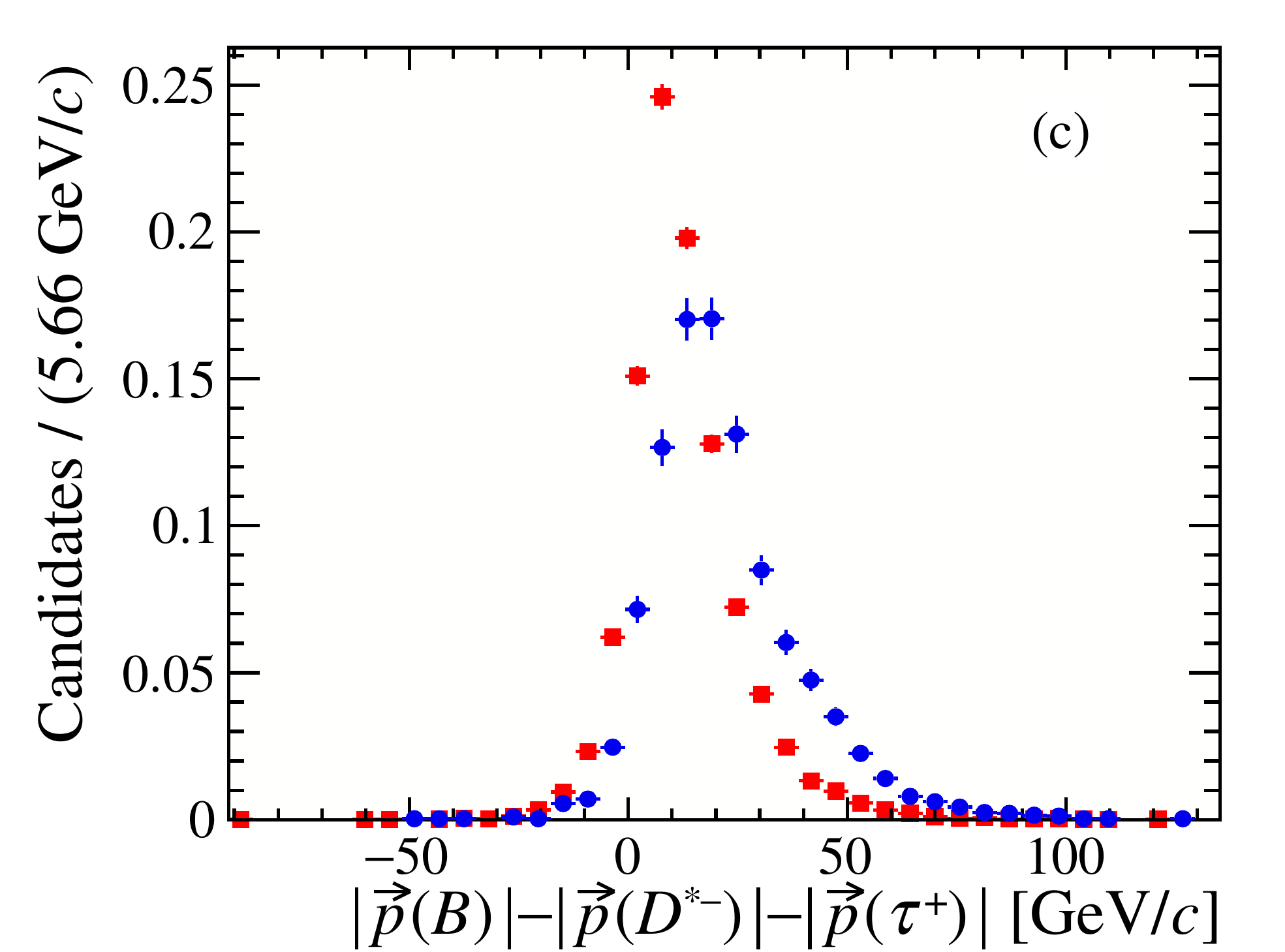}
    \includegraphics[width=0.48\linewidth]{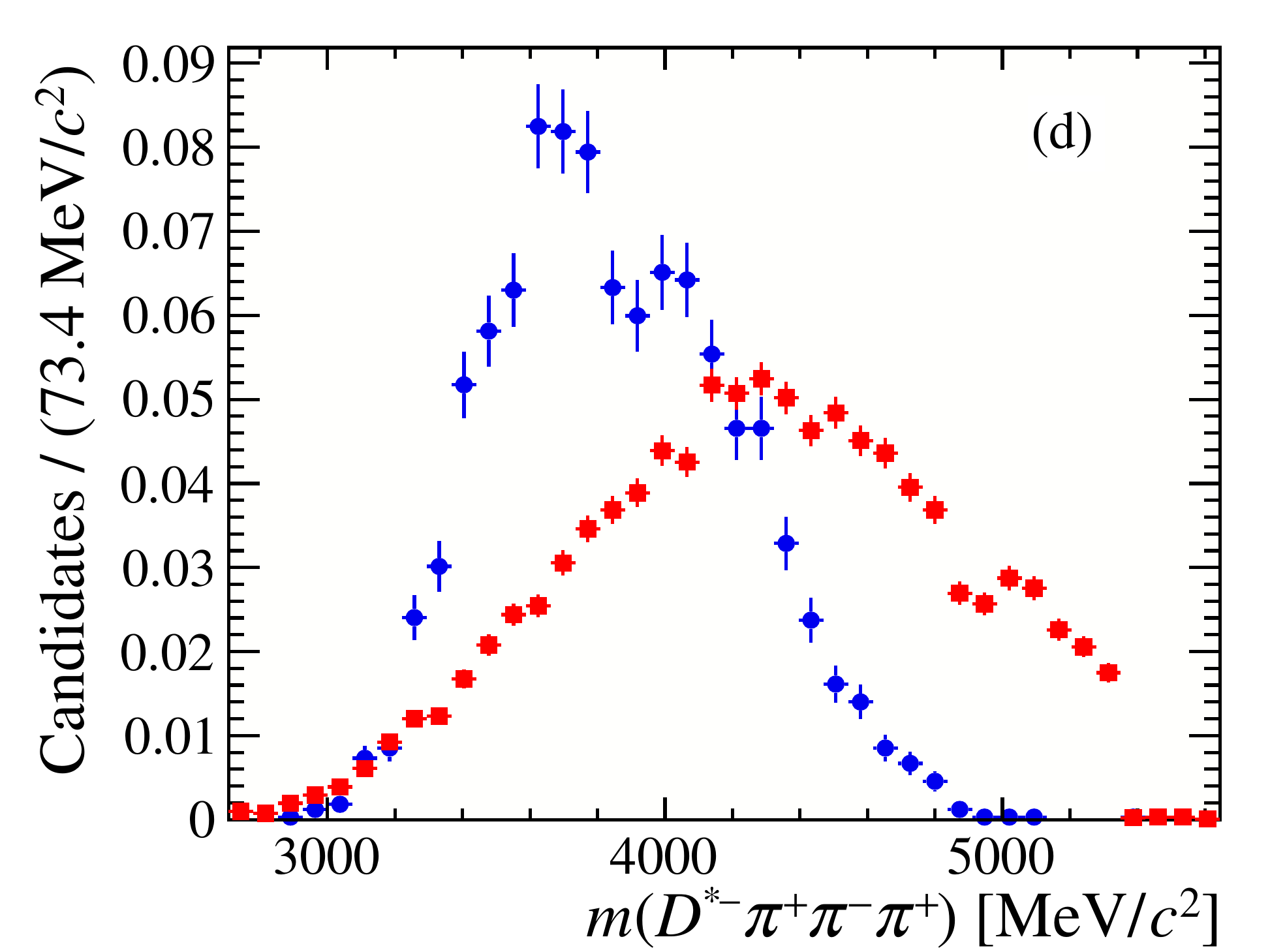}
  \end{center}
  \caption{
    \small 
    Normalized distributions of (a) $\min[m(\pi^+ \pi^-)]$, (b)
$\max[m(\pi^+\pi^-)]$, (c) approximated neutrino momentum reconstructed
in the signal hypothesis, and (d) the $D^{*-} 3\pi$ mass  in
simulated samples.}
  \label{fig:BDT_input}
\end{figure}

\begin{figure}[bt]
\begin{center}
    \includegraphics[width=0.56\linewidth]{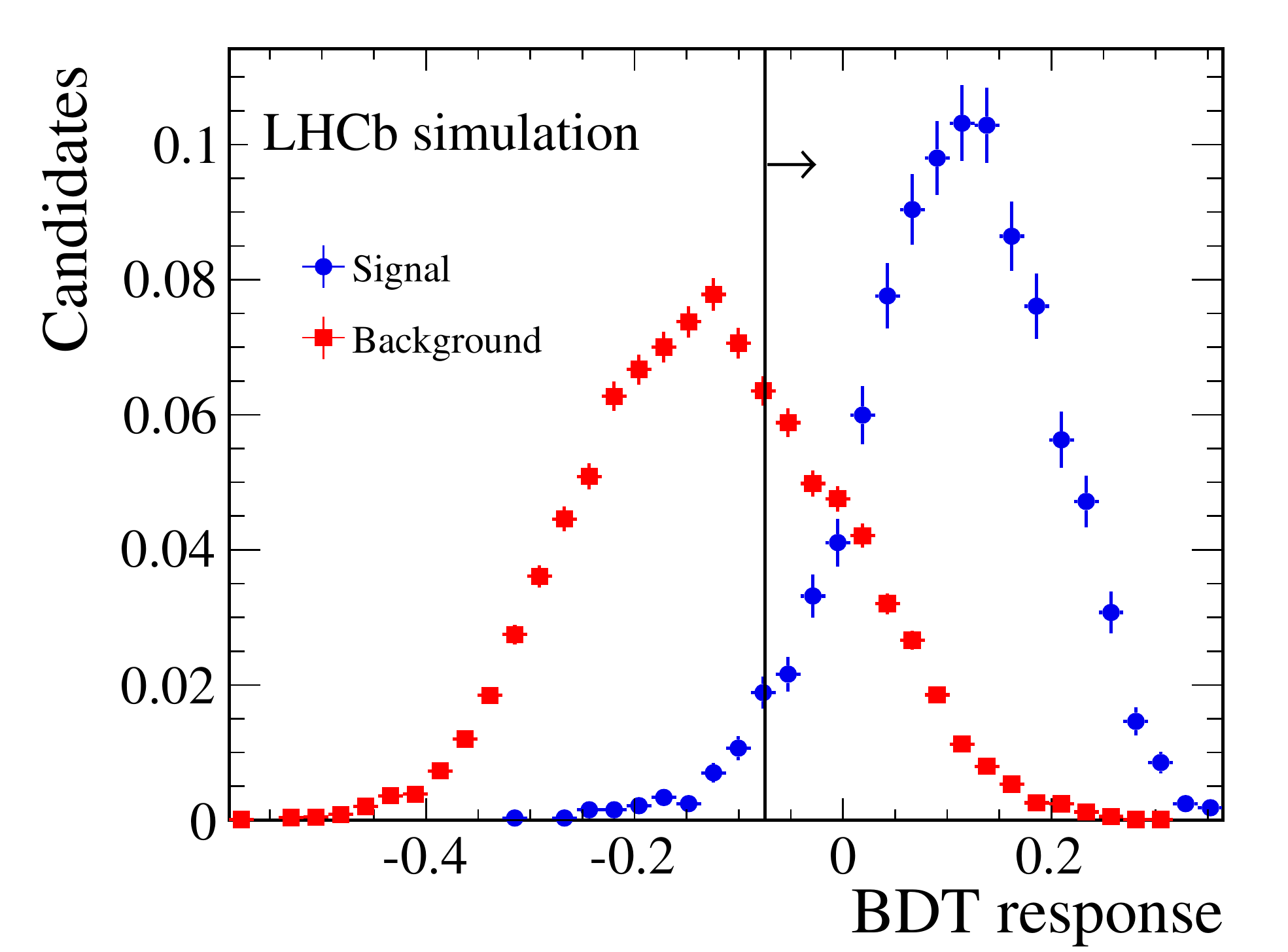}
  \end{center}
  \caption{
    \small 
    Distribution of the BDT response on the signal and background simulated samples.}
  \label{fig:BDT_output}
\end{figure}

To suppress double-charm background, a set of 18 variables is used as input to a
MVA based upon a boosted decision tree (BDT)~\cite{Breiman,AdaBoost}. This set
is as follows: the output variables of the neutral isolation algorithm; momenta,
masses and quality of the reconstruction of the decay chain under the
signal
and background
hypotheses; the masses of oppositely charged pion pairs,
the energy and the flight distance in the transverse plane of the 3\pion system;
the mass of the six-charged-tracks system.
The BDT is trained using simulated samples of signal and double-charm background decays.
Figure~\ref{fig:BDT_input} shows the normalized distributions of the four input
variables having the largest discriminating power for signal and background:
the minimum and maximum of the masses of oppositely charged pions, $\min[m(\pi^+
\pi^-)]$ and $\max[m(\pi^+ \pi^-)]$; the neutrino momentum, approximated as
the difference of the modulus of the momentum of the \Bz and the sum of the
moduli of the momenta of $D^{*-}$ and $\tau$
reconstructed in the signal hypothesis; and the $D^{*-} 3\pi$ mass.
The BDT response for signal and background is illustrated in Fig.~\ref{fig:BDT_output}.


The $\PB\to\Dstarm\Dsp (X)$, $\PB\to\Dstarm\Dz (X)$ and $\PB\to\Dstarm\Dp (X)$ control samples, described in Sec.~\ref{sec:double_charm}, are used to validate the BDT.
Good agreement between simulation and control samples is observed both for the
BDT response and the distributions of the input variables.

The signal yield is determined from candidates in the region where the BDT
output is greater than $-0.075$. According to simulation, this value gives the
best statistical power in the determination of the signal yield. Candidates with
the BDT output less than $-0.075$ are highly enriched in \Dsp decays and contain
very little signal, as shown in Fig.~\ref{fig:BDT_output}, and represent about
half of the total data sample. They are used to validate the simulation of
the various components in $\Dsp \to 3\pi X$ decays
used in the parametrization of the templates entering in the fit that
determines the signal yield, as explained in Sec.~\ref{sec:Ds_composition}.
No BDT cut is applied in the selection for the normalization channel.

\subsection{Composition of the selected sample and selection efficiencies}
\label{ssec:comp}

\begin{figure}[htb]
	\centering
	\includegraphics[width=0.85\textwidth]{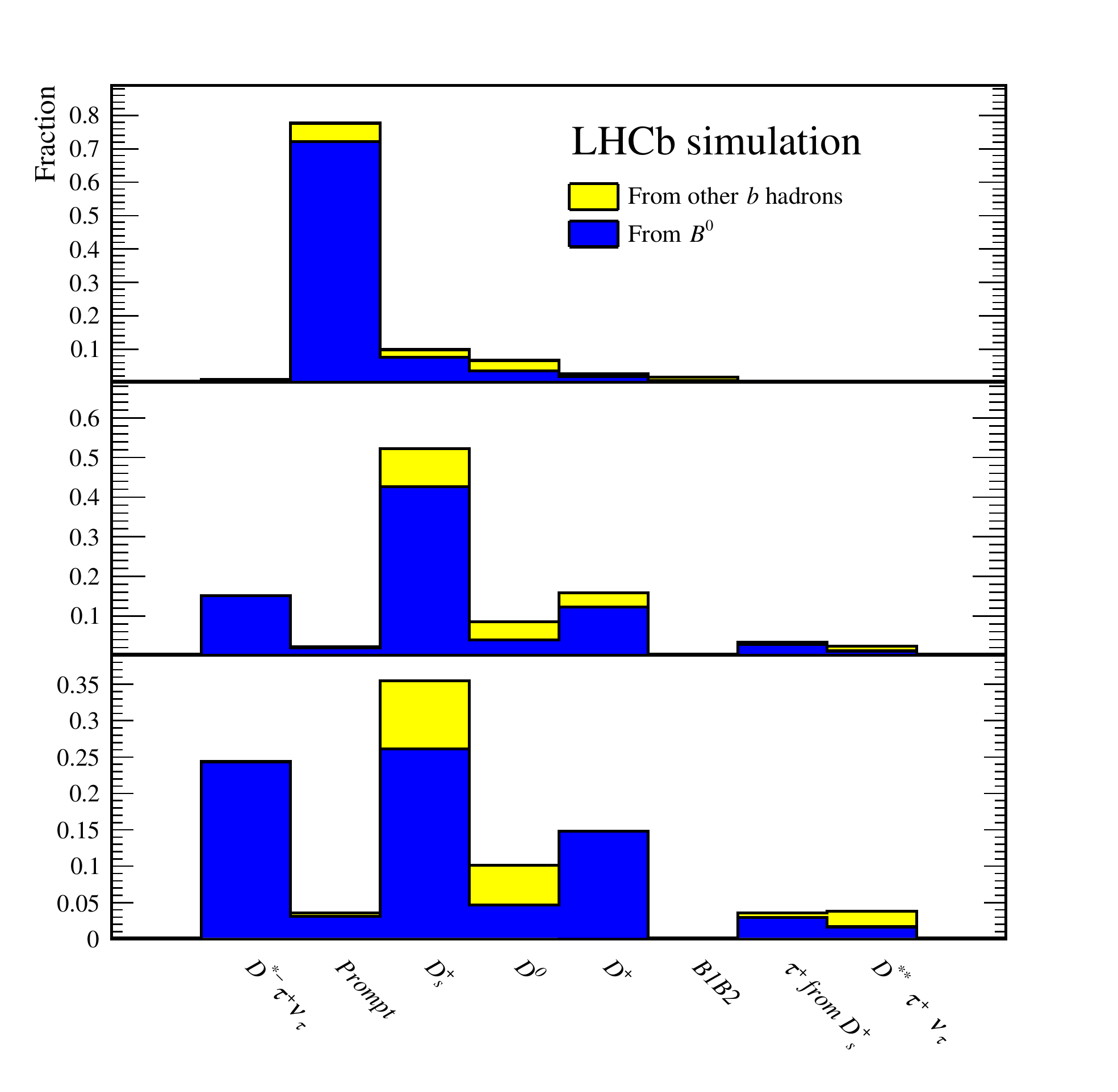}
	\caption{Composition of an inclusive simulated sample where a \Dstarm
and a 3\pion system have been produced in the decay chain of a  \bquark\bquarkbar pair
from a $pp$ collision.
Each bin shows the fractional contribution of the different possible parents of
the 3\pion system (blue from a \Bz, yellow for other \bquark hadrons):
 from signal; directly from the \bquark hadron (prompt); from a charm parent
\Ds, \Dz, or \Dp meson; 3\pion from a \PB and the \Dz from the other \PB ($B1B2$);
from $\tau$ lepton following a \Dsp decay; from a $\tau$ lepton following a
$D^{**}\taup\neut$ decay ($D^{**}$ denotes here any higher excitation of \PD
mesons). (Top) After the initial selection and the removal of spurious $3\pi$
candidates. (Middle) For candidates  entering the signal  fit. (Bottom) For
candidates populating the last 3 bins of the BDT distribution (cf.
Fig.~\ref{fig:fit_results}).}
\label{fig:compo}
\end{figure}

Figure~\ref{fig:compo} shows the composition of an inclusive sample of simulated
events, generated by requiring that a \Dstarm meson and a 3\pion system are both part of
the decay chain of a \bquark\bquarkbar pair produced in a proton-proton
collision before the detached-vertex requirement, at the level of the signal fit,  and with a tighter cut corresponding to the last three BDT bins of Fig.~\ref{fig:fit_results}. In the histograms, the first bin corresponds to the signal, representing only 1\% of the candidates at the initial stage, and the second bin to  prompt candidates, where the 3\pion
system originates from the \bquark-hadron decay. It constitutes by far the largest initial  background
source. The following three bins correspond to cases where the 3\pion system
originates from the decay of
a \Dsp, \Dz or \Dp meson, respectively. The plot in the middle corresponds to the BDT output greater
than $-0.075$ used in the analysis to define the sample in which the signal
determination takes place. One can see the suppression of the prompt background due to the detached-vertex requirement, and the dominance of the \Ds background. The bottom plot shows for illustration the sample
composition with the harder BDT output cut. The \Ds contribution is now
suppressed as well. The signal fraction represents about $25\%$ at this stage.
Figure~\ref{fig:compo} also allows contributions due
to decays of other \bquark hadrons to be compared with those of \Bz mesons.
Table~\ref{tab:efficiency-summary} presents the efficiency of the various
selection steps, both for signal and normalization channels. The signal efficiency is computed from the  efficiencies and abundances of the 3\pion
 and 3\pion\piz channels.
\begin{table}[!tbph]
\small
\caption{Summary of the efficiencies (in \%) measured  at the various steps of
the analysis for simulated samples of the $\Bz\to \Dstarm 3\pi$ channel and the
\Bz\to\Dstarm\taup\neut signal channel for both \tauon decays to 3\pion\neutb and
3\pion\piz\neutb modes. No requirement on the BDT output is
applied for $\Dstarm
3\pi$ candidates. The relative efficiency designates the individual efficiency
of each requirement.}
\begin{center}
\begin{tabular}{l c c c c c c}
\hline
Requirement& \multicolumn{3}{c}{Absolute efficiencies ($\%$)}&\multicolumn{3}{c}{Relative efficiencies ($\%$)}\\
        & $\Dstarm 3\pi$&\multicolumn{2}{c}{\Dstarm\taup\neut}&$\Dstarm 3\pi$&\multicolumn{2}{c}{\Dstarm\taup\neut}\\
 & &3\pion\neutb&3\pion\piz\neutb& &3\pion\neutb&3\pion\piz\neutb \\
\hline
\text{Geometrical acceptance} &14.65 &15.47 &14.64\\\hline
\text{After:} & & & \\
\text{\,\, initial selection} &1.382 &0.826 &0.729\\
\text{\,\, spurious 3\pion removal} &0.561 &0.308 &0.238 &40.6 &37.3 &32.6\\
\text{\,\, trigger requirements} &0.484 &0.200 &0.143 &86.3 &65.1 &59.9\\
\text{\,\, vertex selection} &0.270 &0.0796 &0.0539 &55.8 &39.8 &37.8\\
\text{\,\, charged isolation} &0.219 &0.0613 &0.0412 &81.2 &77.0 &76.3\\
\text{\,\, BDT requirement} &- &0.0541 &0.0292 &- &94.1 &74.8\\
\text{\,\, PID requirements} &0.136 &0.0392 &0.0216 &65.8 &72.4 &74.1\\
\hline

\end{tabular}
\end{center}
\label{tab:efficiency-summary}
\end{table}

\section{Study of double-charm candidates}
\label{sec:double_charm}

The fit that determines the signal yield uses templates that are taken
from simulation. It is therefore of paramount importance to verify the
agreement between data and simulation for the remaining background
processes. Control samples from data are used wherever
possible for this purpose. The relative contributions of
double-charm backgrounds and their \qsq
distributions from simulation are validated, and corrected where appropriate, by using
data control samples enriched in such processes. Inclusive decays of
\Dz, \Dp and \Dsp mesons to 3\pion are also studied in this way.

\subsection{The {\boldmath$\Dsp$} decay model}
\label{sec:Ds_composition}

The branching fraction of \Dsp meson decays with a 3\pion system in the
final state, denoted as \decay{\Dsp}{3\pion X} is about 15 times larger than that of
the exclusive
\Dsp\to3\pion decay. This is due to the large contributions from
decays involving intermediate states such as \KS, $\eta$, \etapr,
$\phi$, and $\omega$, which are generically denoted with the symbol $R$ in the
following. The branching fractions of processes of the type \Dsp\to$R$\pip are
 well known, but large uncertainties exist for several decays, such as
$\Dsp\to R(\to\pip\pim X)\pi^+\piz$  and \Dsp\to$R$3\pion.

\begin{figure}[t]
\begin{center}
   \includegraphics*[width=\textwidth]{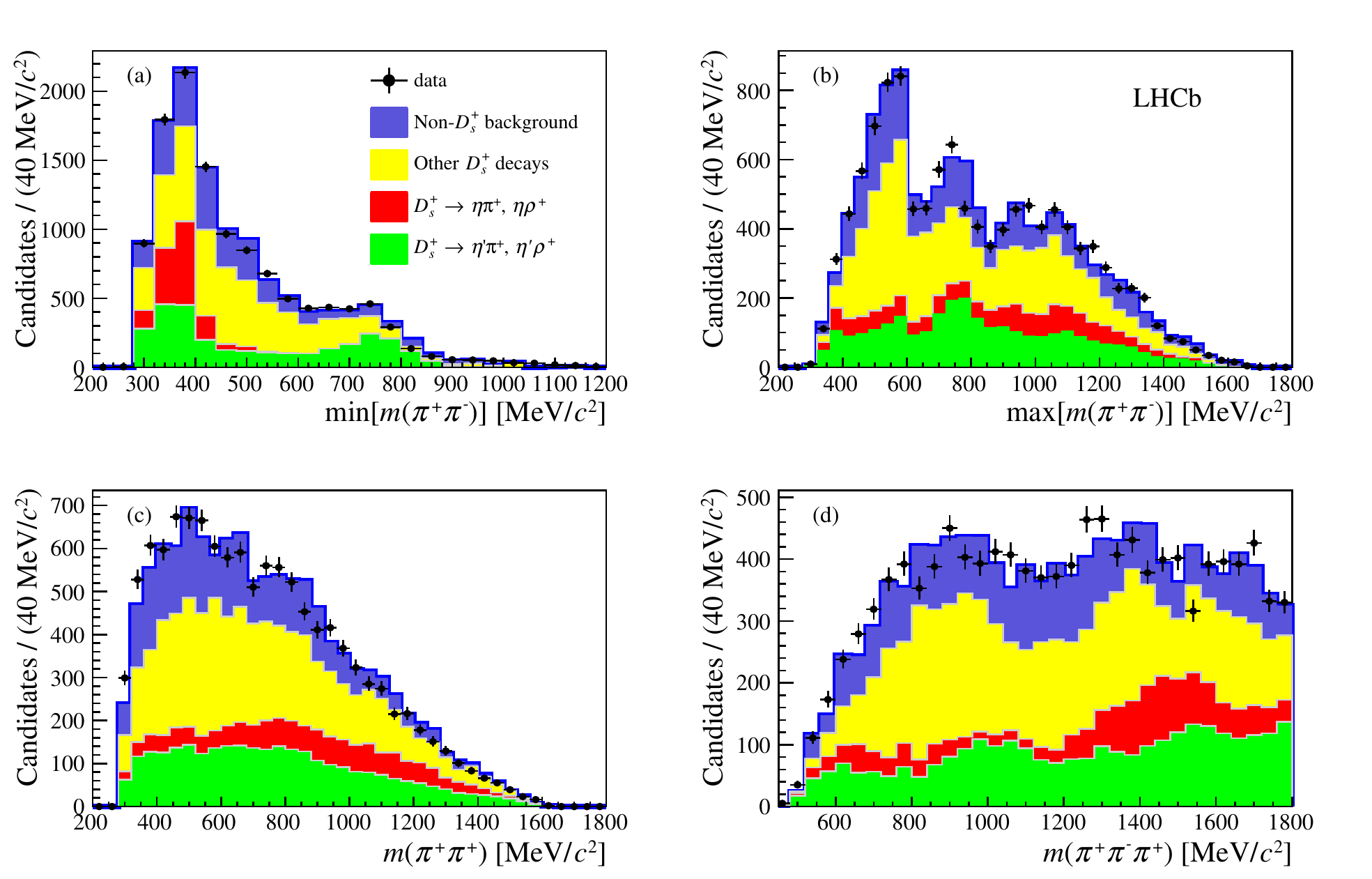}
 \end{center}
\caption{Distributions of (a) ${\mathrm{min}}[m(\pi^+\pi^-)]$,
  (b) ${\mathrm{max}}[m(\pi^+\pi^-)]$,  (c)
  $m(\pi^+\pi^+)$,  (d) $m(\pi^+\pi^-\pi^+)$ for a sample enriched in
\decay{\PB}{\Dstarm\Dsp(X)} decays, obtained by requiring the BDT output below a
certain threshold. The different fit components correspond to
\Dsp decays with (red) $\eta$ or (green) $\etapr$ in the final state,
(yellow) all the other considered \Dsp decays, and (blue) backgrounds originating from decays
not involving the \Dsp meson.}
 \label{fig:dsmodelfit}
\end{figure}

The $\tau$ lepton decays through the $a_1(1260)^+$
resonance, which leads to the $\rho^0\pion^+$ final state~\cite{Nugent:2013hxa}. The
dominant source of $\rho^0$ resonances in \Dsp  decays
is due to  \decay{\etapr}{\rhoz\gamma}
decays. It is therefore crucial to control the \etapr
contribution in \Dsp decays very accurately. The  \etapr contribution in
the $\min[m(\pi^+\pi^-)]$ distribution, obtained from simulation, is
shown in Fig.~\ref{fig:dsmodelfit}. It
exhibits a double peaking structure: at low mass, due to the endpoint of phase space
for the charged pion pair in the $\eta\to\pip\pim\piz$ and $\etapr\to\eta\pip\pim$
decays and, at higher mass, a $\rhoz$ peak. The shape of
this contribution is precisely known since the \etapr branching fractions are
known to better than 2\%. The precise measurement on data of the
low-mass excess, which consists only of \etapr and $\eta$ candidates, therefore enables the
control of the \etapr contribution in the sensitive $\rho$ region.
The \mbox{$\decay{\Dsp}{3\pion X}$} decay model is determined from a data sample
enriched in \mbox{\decay{\PB}{\Dstarm\Dsp (X)}} decays by requiring a low
value of the BDT output.
The distributions of  $\min[m(\pi^+\pi^-)]$ and
$\max[m(\pi^+\pi^-)]$, of the mass of the same-charge
pions, $m(\pi^+\pi^+)$, and of the mass of the 3\pion system,
$m(3\pion)$, are simultaneously fit with a model obtained from
simulation.
The fit model is constructed from the following components:
\begin{itemize}
\item \Dsp decays where at least one pion originates from the decay of an $\eta$ meson; the
$\Dsp\to\eta\pip$ and $\Dsp\to\eta\rhop$ components are in this category.
\item \Dsp decays where, in analogy with the previous category, an \etapr meson is involved.
\item \Dsp decays where at least one pion originates from an intermediate
resonance other than $\eta$ or $\etapr$; these are then
subdivided into $R\pip$ and $R\rhop$ final states; these
decays are dominated by $R=\omega$, $\phi$ resonances.
\item Other \Dsp decays, where none of the three pions
originates from an intermediate state; these are then
subdivided into $K^03\pi$, $\eta3\pi$, $\etapr3\pi$,
$\omega3\pi$, $\phi3\pi$, $\tau^+(\to 3\pi(N)\overline{\nu}_{\tau})\nu_{\tau}$, and $3\pi$ nonresonant final
states, $X_{\rm nr}$. Regarding the tauonic $\Ds\to\tau^+\nu_{\tau}$ decay, the
label $N$ stands for any potential extra neutral particle.
\end{itemize}
 Templates for each category and for the non-\Ds candidates are determined from
\mbox{$\PB\to\Dstarm\Dsp (X)$} and $\PB\to\Dstarm 3\pion X$ simulation samples, respectively.
 Figure~\ref{fig:dsmodelfit} shows the fit results
 for the four variables. The fit measures the
$\eta$ and $\etapr$ inclusive fractions very precisely because, in the
$\min[m(\pip\pim)]$ histogram, the
low-mass peak is the sum of the $\eta$ and $\etapr$ contributions, while
only the \etapr meson contributes to the $\rho^0$ region.
The ratio between decays with a \pip and a \rhop meson in the final state
is not precisely determined because of the limited sensitivity of the fit
variables to the presence of the extra \piz. The sensitivity only comes from
the low-yield high-mass tail of the 3\pion mass distribution which exhibits different endpoints
for these two types of decays. Finally, the kinematical endpoints of
the 3\pion mass for each $R3\pion$ final state enable the fit to
determine their individual contributions, which are  presently either
poorly measured or not measured at all. The \mbox{\Ds\to$\phi$3\pion} and \Ds\to$\tau^+(\to 3\pi(N)\overline{\nu}_{\tau})\nu_{\tau}$ branching fractions, known with a 10\% precision, are fixed to their measured values~\cite{PDG2017}.

The fit is in good agreement with the data, especially in the critical
$\min[m(\pip\pim)]$ distribution. {The \chisq per degree of freedom of each fit is  0.91, 1.25, 1.1 and 1.45 for each histogram, respectively, when taking into account the simulation sample size.} The fit parameters and
their ratios, with values from simulation, are
reported in Table~\ref{tab:dsmodelfit}.
These are used to correct the corresponding contributions
from simulation.
In the final fit performed in the high BDT output region, the shape of each contribution
is scaled according to the ratio of candidates in the two BDT regions, which is taken from simulation.

The fit determines that ($47.3\pm 2.5$)\% of the \Dsp decays in this sample
contain $\eta$ and $\etapr$ mesons with an additional charged pion, ($20.6\pm
4.0$)\% contain $\phi$ and $\omega$ mesons with an additional
charged pion and  ($32.1\pm 4.0$)\% are due to $R$3\pion modes.
This last contribution is dominated by the $\eta 3\pi$ and $\etapr 3\pi$ modes.{~The large weighting factors observed in this \Ds decay-model fit correspond to channels whose  branching fractions are not precisely known.}
\begin{table}[t]
	\centering
	\caption{Results of the fit to the \Dsp decay model. The relative contribution of each decay and the correction to be applied
        to the simulation are reported in the second and third columns, respectively.}
	\label{tab:dsmodelfit}
	\begin{tabular}{c c c}
          \hline
          \Dsp decay                  & Relative              & Correction \\
                                      & contribution          & to simulation \\ \hline
          $\eta\pip(X)$             & 0.156 $\pm$ 0.010     & \\
          \multicolumn{1}{r}{$\eta\rhop$}                & \multicolumn{1}{r}{~~~~~~~~~~0.109 $\pm$ 0.016}     & 0.88 $\pm$ 0.13 \\
          \multicolumn{1}{r}{$\eta\pip$}                  & \multicolumn{1}{r}{0.047 $\pm$ 0.014}     & 0.75 $\pm$ 0.23 \\ \hline
          $\etapr\pip(X)$           & 0.317 $\pm$ 0.015     & \\
          \multicolumn{1}{r}{$\etapr\rhop$}               & \multicolumn{1}{r}{0.179 $\pm$ 0.016}     & 0.710 $\pm$ 0.063 \\
          \multicolumn{1}{r}{$\etapr\pip$}                & \multicolumn{1}{r}{0.138 $\pm$ 0.015}     & 0.808 $\pm$ 0.088 \\ \hline
          $\phi\pip(X),\,\omega\pip(X)$    &0.206 $\pm$ 0.02&\\
          \multicolumn{1}{r}{$\phi\rhop,\,\omega\rhop$}   &\multicolumn{1}{r}{0.043 $\pm$ 0.022}     & 0.28 $\pm$ 0.14\\
          \multicolumn{1}{r}{$\phi\pip,\,\omega\pip$}     &\multicolumn{1}{r}{0.163 $\pm$ 0.021}     & 1.588 $\pm$ 0.208 \\ \hline
          $\eta 3\pi$                 & 0.104 $\pm$ 0.021     & 1.81 $\pm$ 0.36 \\
          $\etapr 3\pi$               & 0.0835 $\pm$ 0.0102   & 5.39 $\pm$ 0.66 \\
          $\omega 3\pi$               & 0.0415 $\pm$ 0.0122   & 5.19 $\pm$ 1.53 \\
          $K^0 3\pi$                  & 0.0204 $\pm$ 0.0139   & 1.0 $\pm$ 0.7 \\
          $\phi 3\pi$                 & 0.0141                & 0.97 \\
          $\taup(\to 3\pi(N)\neutb)\neut$   & 0.0135                & 0.97 \\
          $X_{\rm nr}3\pi$                & 0.038 $\pm$ 0.005     & 6.69 $\pm$ 0.94\\\hline
 	\end{tabular}
\end{table}

\subsection{The {\boldmath$\PB\to\Dstarm\Dsp (X)$} control sample}
\label{sec:Ds_sample}

Candidates where the \Dsp meson decays exclusively to the
$\pi^+\pi^-\pi^+$ final state give a pure sample of $\PB \to \Dstarm\Dsp (X)$ decays.
This sample includes three types of processes:\footnote{In this
  Section, $D^{**}$ and  $D_s^{**}$ are used to refer to any higher-mass
excitations of \Dstarm or \Dsp mesons decaying to $D^{*-}$ and $D_s^+$ ground states.}
\begin{itemize}
\item $\Bz\to \Dstarm D_s^{(*,**)+}$ decays, where a neutral particle is emitted
in the decay of the excited states of the $D_s^{+}$ meson.
The corresponding $q^2$ distribution
peaks at the squared mass, $(p_{B^0}-p_{D^{*-}})^2$, of the given
states.
\item $\Bs\to \Dstarm\Dsp X$ decays, where at least one additional particle
is missing. This category contains feed-down from excited states, both for
\Dstarm or \Dsp mesons. The $q^2$ distribution is shifted to higher values.
\item $B^{0,-}\to \Dstarm\Dsp X^{0,-}$ decays, where at least one additional particle
originates from either the $B^{0,-}$ decay, or the deexcitation of
charm-meson resonances of higher mass, that results in a \Dstarm meson in the final state. These additional missing particles shift the
$q^2$ distribution to even higher values.
\end{itemize}

The \decay{\PB}{\Dstarm\Dsp (X)} control sample is used to evaluate the
agreement between data and simulation, by performing a fit to the
distribution of the mass of the \mbox{$\Dstarm 3\pion$} system,
$m(\Dstarm3\pion)$. The fitting probability density function $\cal{P}$ is parametrized as

\begin{eqnarray}
{\cal{P}} = f_{\rm c. b.}~{\cal{P}}_{\rm c. b.} + \frac{(1-f_{\rm
c. b.})}{k}\sum_j f_j {\cal{P}}_j
\end{eqnarray}

\noindent where
$i,j=\{D_s^{*+};~D_s^{+};~D_{s0}^{*+};~D_{s1}^{+};~D_s^{+}X;~(D_s^{+}X)_s\}$
and $k=\sum_i f_i$.
The fraction of combinatorial background, $f_{\rm c. b.}$, is fixed in the fit.
Its shape is taken from a sample where the $D^{*-}$ meson and the $3\pi$ system have the
same charge.
Each component $i$ is described by the probability density function
${\cal{P}}_i$, whose shapes are taken from simulation.
The parameters $f_i$ are
the relative yields of
$B^0\to D^{*-}D_s^{+}$, $B^0\to D^{*-}D^*_{s0}(2317)^{+}$, $B^0\to D^{*-}D_{s1}(2460)^{+}$,
\mbox{$B^{0,+}\to D^{*-}D_s^{+}X$} and $B_s^0\to D^{*-}D_s^{+}X$ decays with
respect to the number of \mbox{$B^0\to D^{*-}D_s^{*+}$} candidates. They
are floating in the fit, and $f_{D_s^{*+}}=1$ by definition.

\begin{figure}[!htbp]
  \begin{center}
    \includegraphics[width=0.495\textwidth]{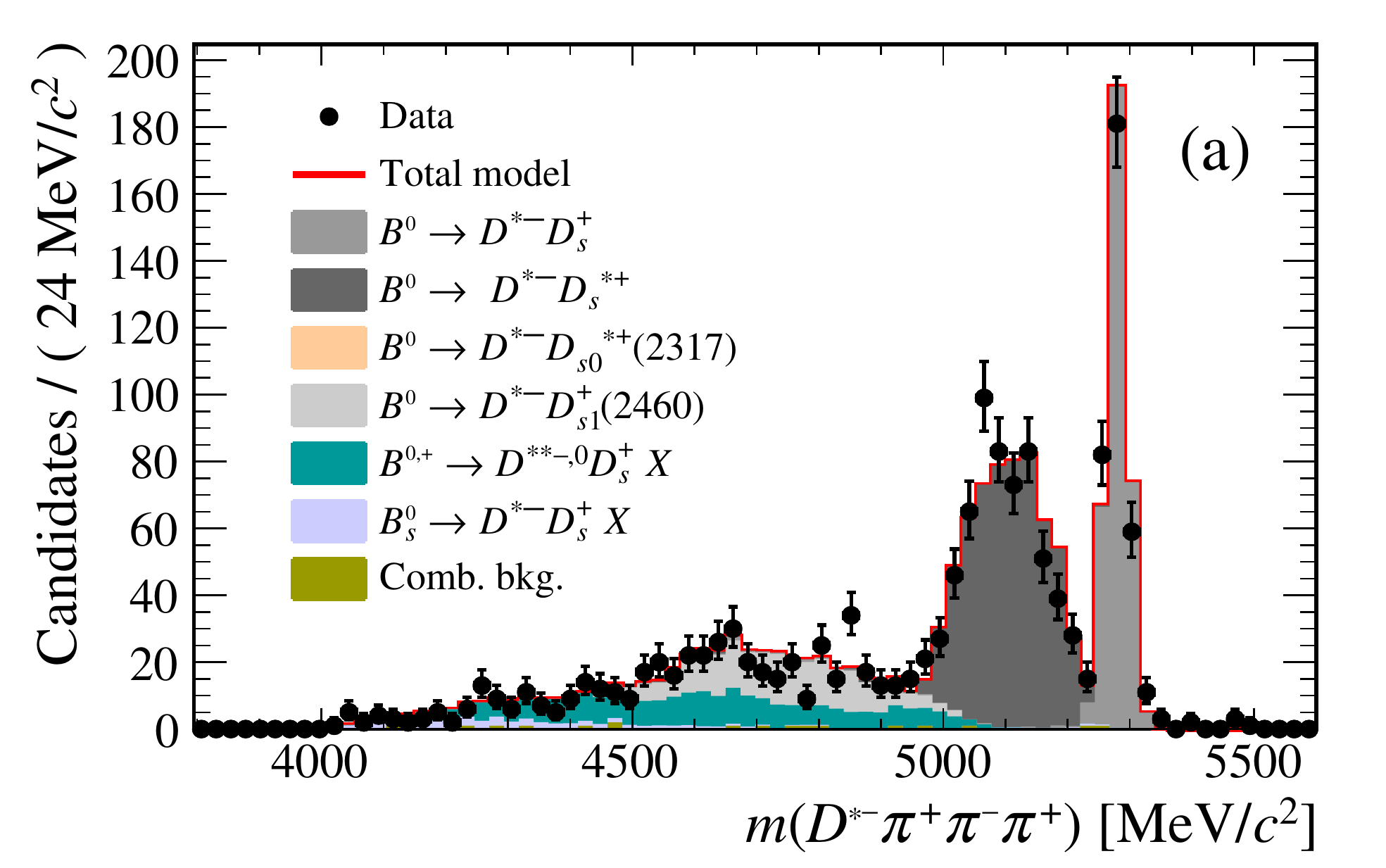}
    \includegraphics[width=0.495\textwidth]{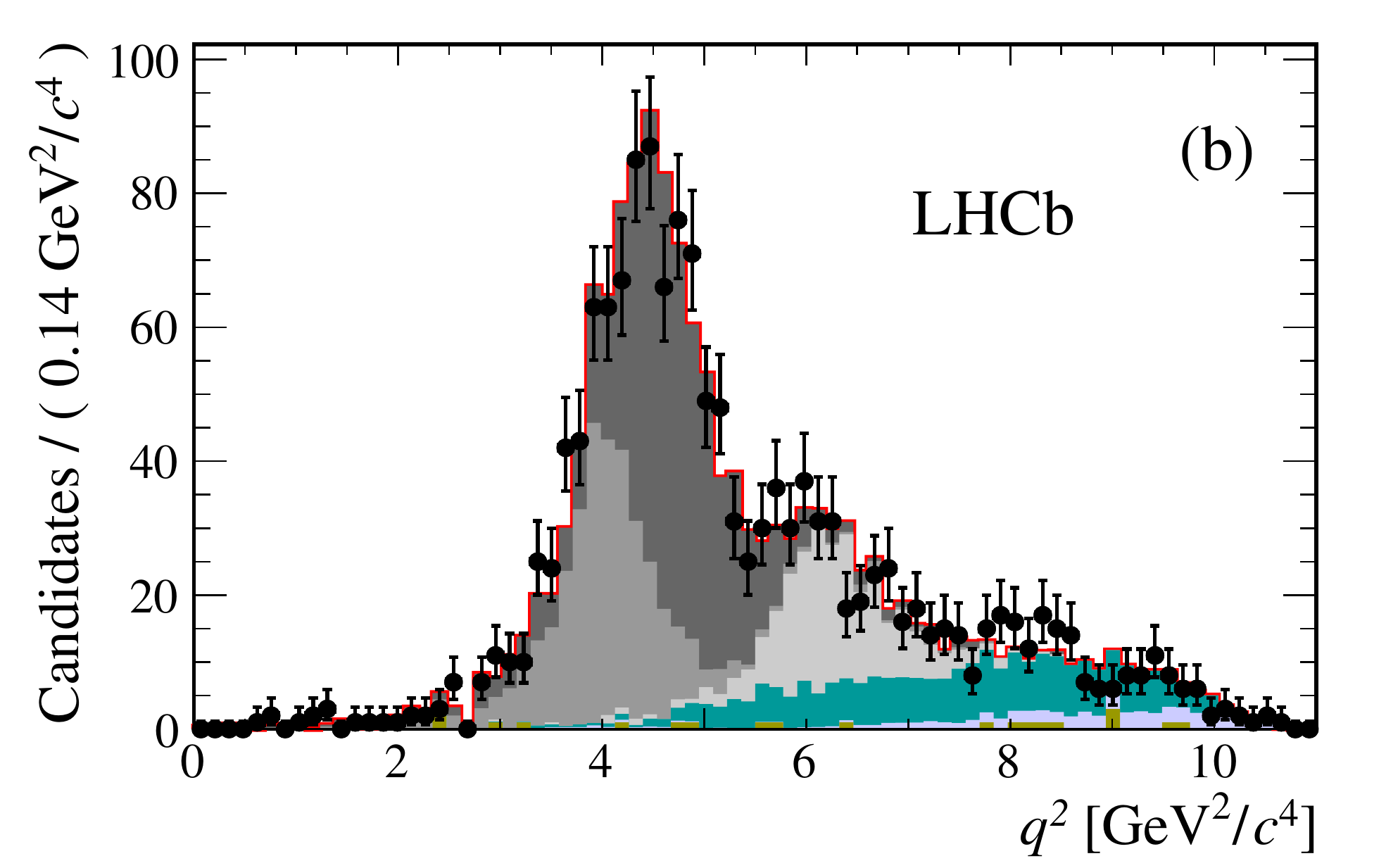}
    \includegraphics[width=0.495\textwidth]{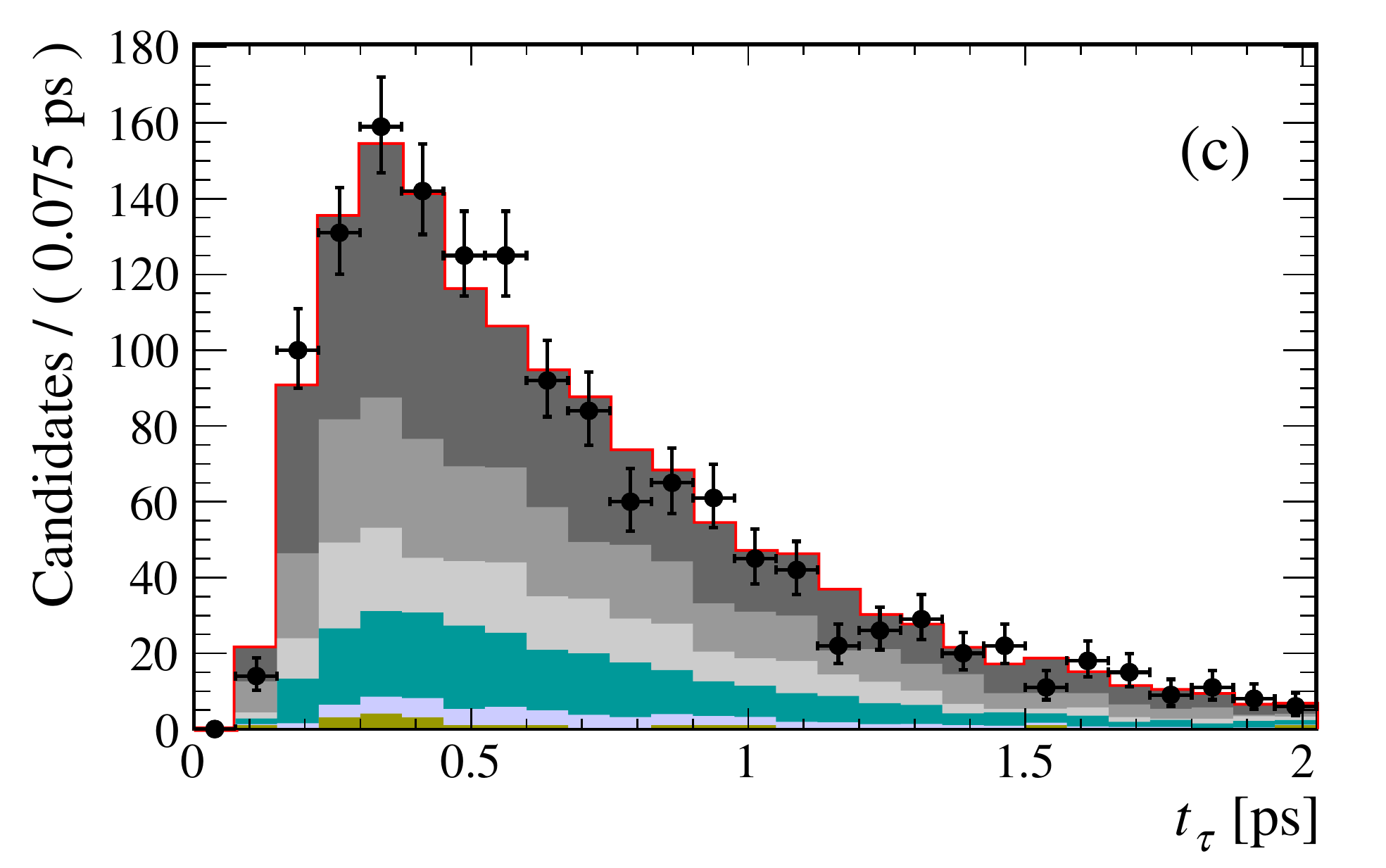}
    \includegraphics[width=0.495\textwidth]{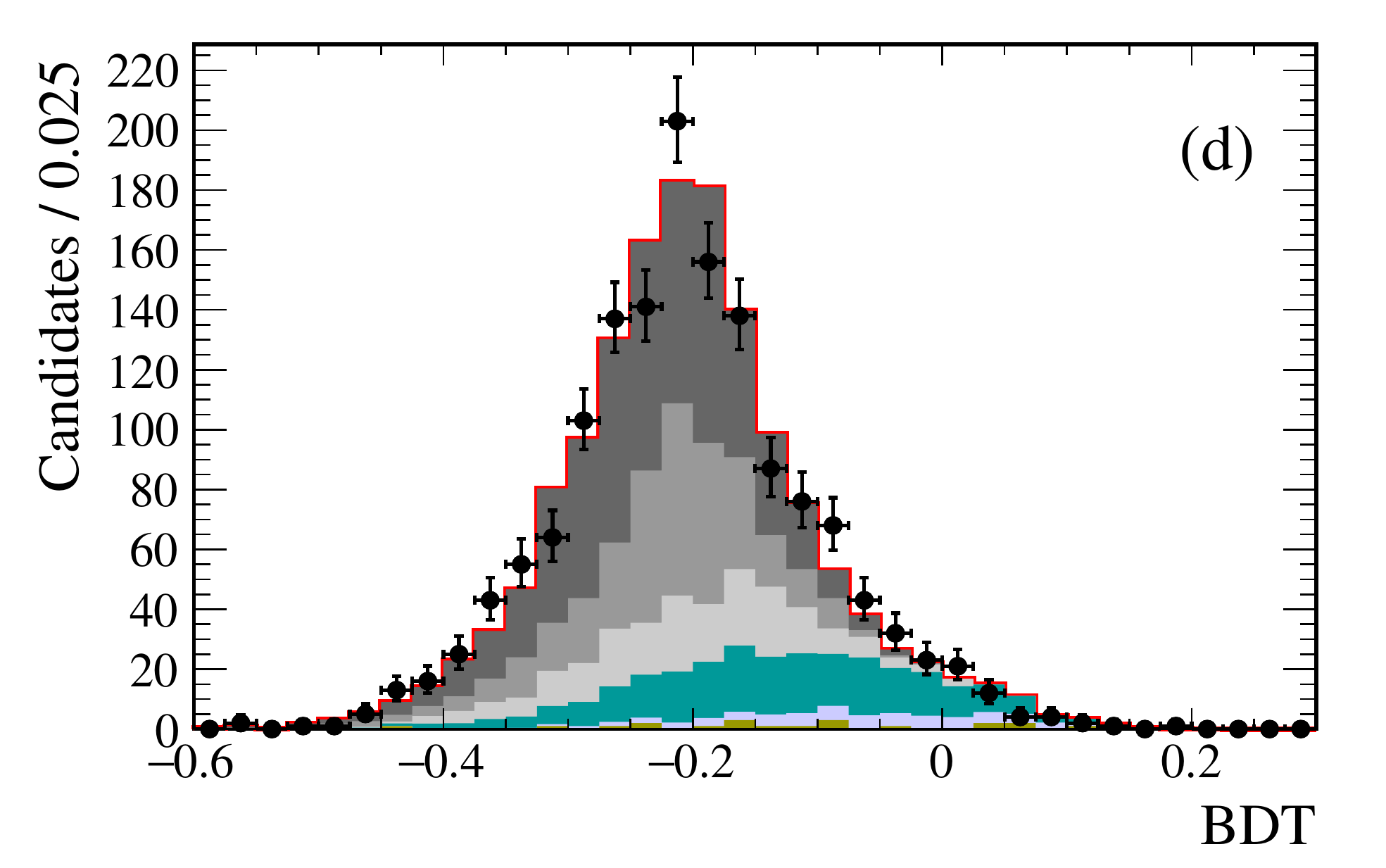}
  \end{center}
  \caption{
    \small 
Results from the fit to data for candidates containing a \Dstarm\Dsp pair, where $\Dsp\to 3\pi$.
The fit components are described in the legend.
 The figures correspond to the fit projection on (a)
$m(D^{*-}3\pi)$, (b) $q^2$, (c) 3\pion decay
time $t_{\tau}$ and
(d) BDT output distributions.
    }
  \label{fig:Ds_control_Fit}
\end{figure}

The fit results are shown in Fig.~\ref{fig:Ds_control_Fit} and
reported in Table~\ref{tab:Ds_control_Fit}, where a comparison with the
corresponding values in the simulation is also given, along with their ratios.
 The measured ratios, including the uncertainties and correlations, are used to constrain these
contributions in the final fit.{~The large weighting factors observed in this  fit correspond to channels whose branching fractions are not precisely known.}

\begin{table}[t]
  \centering
  \caption{Relative fractions of the various components obtained from the fit to the $B\to D^{*-}D_{s}^{+}(X)$ control sample. The values used in the simulation and the ratio of the two are also shown.}
  \label{tab:Ds_control_Fit}
    \begin{tabular}{lccc}
      \hline
      Parameter          & Simulation & Fit                      & Ratio  \\
      \hline
      $f_{\rm c. b.}$    & ---      & $0.014$                    & ---             \\
      $f_{D_s^{+}}$      & $0.54$   & $0.594 \pm 0.041$          & $1.10 \pm 0.08$ \\
      $f_{D_{s0}^{*+}}$  & $0.08$   & $0.000^{+0.040}_{-0.000}~~$  & $0.00^{+0.50}_{-0.00}~$ \\
      $f_{D_{s1}^{+}}$   & $0.39$   & $0.365 \pm 0.053$          & $0.94 \pm 0.14$ \\
      $f_{D_{s}^+X}$     & $0.22$   & $0.416 \pm 0.069$          & $1.89 \pm 0.31$ \\
      $f_{(D_{s}^+X)_s}$ & $0.23$ & $0.093 \pm 0.027$ & $0.40 \pm 0.12$ \\
      \hline
    \end{tabular}
\end{table}

\begin{figure}[b]
  \begin{center}
    \includegraphics[width=0.67\textwidth]{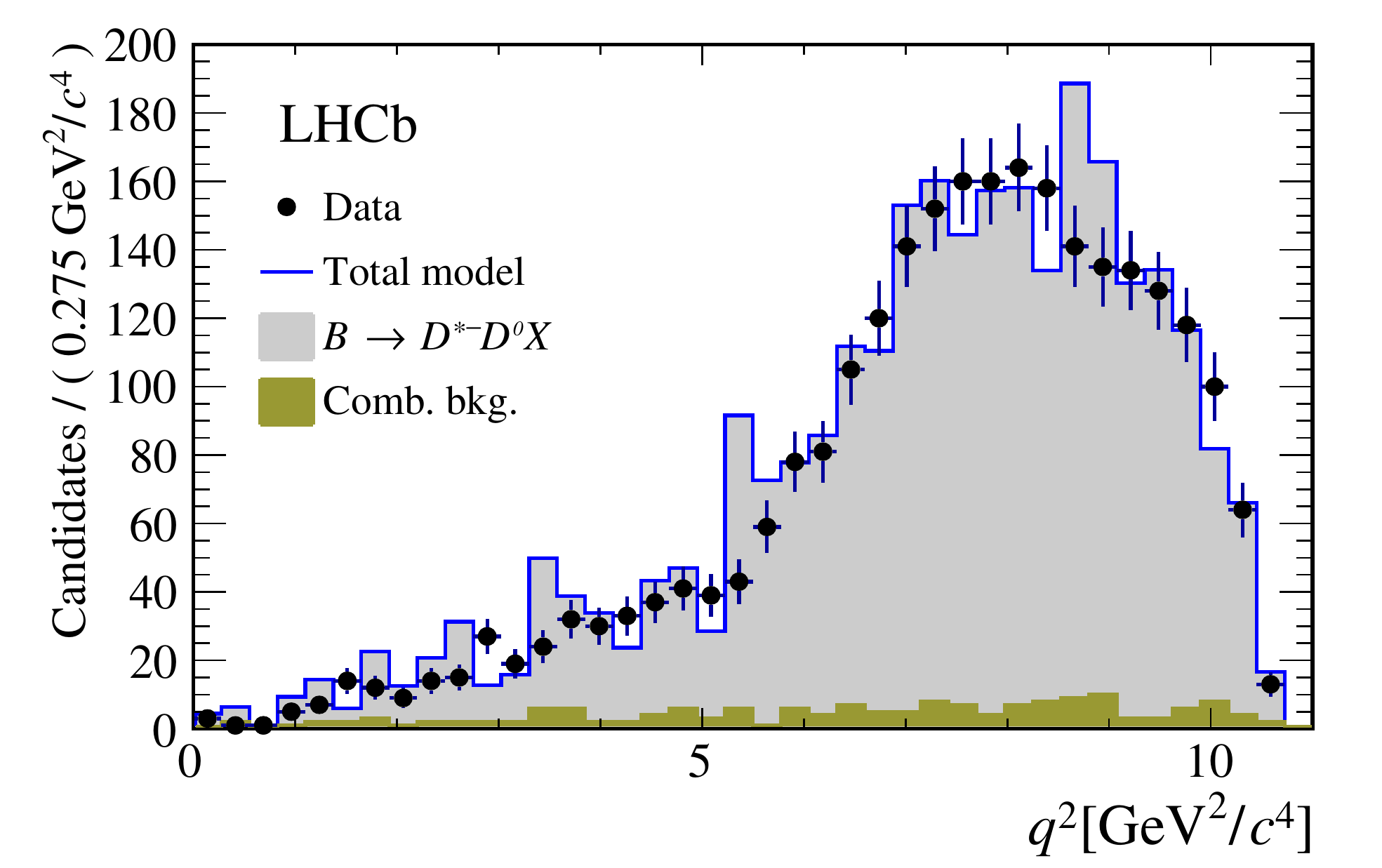}
  \end{center}
  \caption{
    \small 
Distribution of $q^2$ for candidates in the
$B\to D^{*-}D^0(X)$
control sample, after correcting for the disagreement between data and simulation.}
  \label{fig:D0_control_MC}
\end{figure}

\subsection{The {\boldmath$\PB\to\Dstarm\Dz (X)$} and {\boldmath$\PB\to\Dstarm\Dp (X)$} control samples}

The decays of \Dz and \Dp mesons into final states with three pions are
dominated by the \decay{D^{0,+}}{K^{-,0}3\pi(\piz)} modes, whose
subresonant structure is known.
The agreement between data and simulation is validated in the \Dz case
by using a control sample. The isolation algorithm identifies a
kaon with charge opposite to the total charge of the 3\pion system,
and compatible with originating from the 3\pion vertex. The
mass of the $\Km3\pion$ system must be compatible with the known \Dz mass.
Disagreement between data and simulation is found in the $q^2$ and
\Dstarm\Dz mass distributions, and corrected for.
Figure~\ref{fig:D0_control_MC} shows the $q^2$ distribution after this correction.

A pure sample of \decay{\PB}{\Dstarm\Dp (X)} decays is obtained by inverting the
PID requirements on
the negative pion of the 3\pion system, assigning to this particle the kaon mass
and selecting 3\pion candidates
with mass compatible with the known \Dp mass.
As in the \decay{\PB}{\Dstarm\Dz (X)} control sample, disagreement
between data and simulation is found. The limited size of this sample does not
allow the determination of a specific correction. The same correction found
in the \decay{\PB}{\Dstarm\Dz (X)} case is therefore applied, since the dominant
decay $\B\to\Dstarm\D K$ is identical for both cases.

\section{Determination of the signal yield}
\label{sec:signal_yield}

  The yield of $B^0\to D^{*-}\tau^+\nu_{\tau}$ decays is determined from a
three-dimensional binned maximum likelihood fit
to the distributions of $q^2$, $3\pi$ decay time, and
BDT output. Signal
and background templates are produced with eight bins in $q^2$, eight bins in
$t_{\tau}$, and four bins in the BDT output,
 from the corresponding simulation samples. The model used to
fit the data is summarized in Table \ref{tab:fit_model}. In the table,
\begin{table}[b]
  \centering
  \caption{Summary of fit components and their corresponding normalization
parameters. The first three components correspond to parameters related to the signal.}
    \begin{tabular}{ll}
      \hline
      Fit component & Normalization \\
      \hline
      $B^0\to D^{*-}\tau^+(\to 3\pi\overline{\nu}_{\tau})\nu_{\tau}$        & ${N_{\rm{sig}}}\times {f_{\tau \to 3\pi\nu}}$     \\
      $B^0\to D^{*-}\tau^+(\to 3\pi\pi^0\overline{\nu}_{\tau})\nu_{\tau}$   &
${N_{\rm{sig}}}\times (1-{f_{\tau \to 3\pi\nu}})$     \\
      $B\to D^{**}\tau^+\nu_{\tau}$                                         & ${N_{\rm{sig}}}\times {f_{D^{**}\tau\nu}}$     \\
      \hline
      $B\to D^{*-}D^+X$                              & ${f_{D^+}} \times {N_{D_s}}$     \\
      $B\to D^{*-}D^0X$ different vertices           & ${f_{D^0}^{v_1v_2}}\times {N_{D^0}^{\rm{sv}}}$     \\
      $B\to D^{*-}D^0X$ same vertex                  & ${N_{D^0}^{\rm{sv}}}$     \\
      $B^0 \to D^{*-}D_s^+$                          & ${N_{D_s}} \times f_{D_s^+}/k$     \\
      $B^0 \to D^{*-}D_s^{*+}$                       & ${N_{D_s}} \times 1/k$     \\
      $B^0 \to D^{*-}D_{s0}^{*}(2317)^+$             & ${N_{D_s}} \times f_{D_{s0}^{*+}}/k$     \\
      $B^0 \to D^{*-}D_{s1}(2460)^{+}$               & ${N_{D_s}} \times f_{D_{s1}^+}/k$     \\
      $B^{0,+} \to D^{**}D_{s}^{+}X$                 & ${N_{D_s}} \times f_{D_s^+X}/k$     \\
      $B_s^0 \to D^{*-}D_{s}^{+}X$                   & ${N_{D_s}} \times f_{(D_s^{+}X)_s}/k$     \\
      $B\to D^{*-}3\pi X$                            & ${N_{B\to D^\ast 3\pi X}}$ \\
      \hline
      B1B2 combinatorics                             & ${N_{B1B2}}$ \\
      Combinatoric $D^{*-}$                          & ${N_{{\rm{not}} D^*}}$ \\
      \hline
    \end{tabular}
\label{tab:fit_model}
\end{table}

\begin{itemize}
\item{ ${N_{\rm{sig}}}$} is a free parameter accounting for the yield
of signal candidates.
\item{ ${f_{\tau\to 3\pi\nu}}$ is the fraction of $\taup\to
3\pi\neutb$ signal candidates with respect to the sum
of the $\taup\to 3\pi\neutb$ and $\taup\to 3\pi\piz\neutb$ components. This
parameter is fixed to $0.78$, according to the different branching
fractions and efficiencies of the two modes.
}
\item{ ${f_{D^{**}\tau\nu}}$, fixed to $0.11$, is the ratio of the yield of
$B\to D^{**}\taup\neut$
decay candidates to the signal decays. This yield is
computed assuming that the ratio of the
decay rates lies between the ratio of available phase space ($0.18$) and the
predictions of Ref.\cite{Bernlochner:2016bci} ($0.06$), and taking into account the relative efficiencies of the different channels.
}
\item{ ${N_{D^{0}}^{\rm{sv}}}$ is the yield of $B\to D^{*-}D^0X$ decays where the
three pions originate from the same vertex (SV) as the the $D^0$ vertex. The $D^0\to
K^+\pi^-\pi^+\pi^-(\pi^0)$ decays are
reconstructed by recovering a charged kaon pointing to the $3\pi$ vertex in nonisolated
events. The exclusive $D^0\to K^+\pi^-\pi^+\pi^-$  peak is used to apply a $5\%$ Gaussian
constraint to this parameter, accounting for
the knowledge of the efficiency in finding the additional kaon.
}
\item{ ${f_{D^{0}}^{v_1v_2}}$ is the ratio of $B\to D^{*-}D^0X$ decays where
at least one pion originates from the $D^0$ vertex and the other pion(s) from a different
vertex, normalized to ${N_{D^{0}}^{\rm{sv}}}$. This is the case when the soft pion from a $D^{*-}$
decay is reconstructed as it was produced at the $3\pi$ vertex.
}
\item{ ${f_{D^+}}$ is the ratio of $B\to \Dstarm D^+X$ decays with respect to
those containing a $D_s^+$ meson.
}
\item{ $N_{D_s}$ is the yield of events involving a $D_s^+$.
The parameters $f_{D_s^+}$, $f_{D_{s0}^{*+}}$,
$f_{D_{s1}^+}$, $f_{D_s^+X}$,
$f_{(D_s^{+}X)_s}$ and $k$, defined in Sec.~\ref{sec:Ds_sample},  are used after
correcting for efficiency.
}
\item{ ${N_{B\to D^*3\pi X}}$ is the yield of $B\to D^{*-}3\pi X$ events
where the three pions come from the $B$ vertex. This value is constrained by
using the observed ratio between $B^0\to D^{*-}3\pi$ exclusive and $B\to D^{*-}3\pi X$
inclusive decays, corrected for efficiency.
}
\item{ ${N_{B1B2}}$ is the yield of combinatorial background events
where the $D^{*-}$ and the 3\pion system come from different $B$ decays. Its yield is
fixed by using the yield of wrong-sign events $D^{*-}\pim\pip\pim$ in the region
$m(D^{*-}\pim\pip\pim)>5.1$ GeV/${\it{c}}^2$.
}
\item{ ${N_{{\rm{not}} D^{*}}}$ is the combinatorial background yield
with a fake $D^{*-}$. Its value is fixed by using the number of events in the
$\Dzb$ mass sidebands of the $D^{*-}\to \Dzb \pi^-$ decay.}
\end{itemize}

\subsection{Fit results}
\begin{figure}[t]
  \begin{center}
    \includegraphics[width=0.6\textwidth]{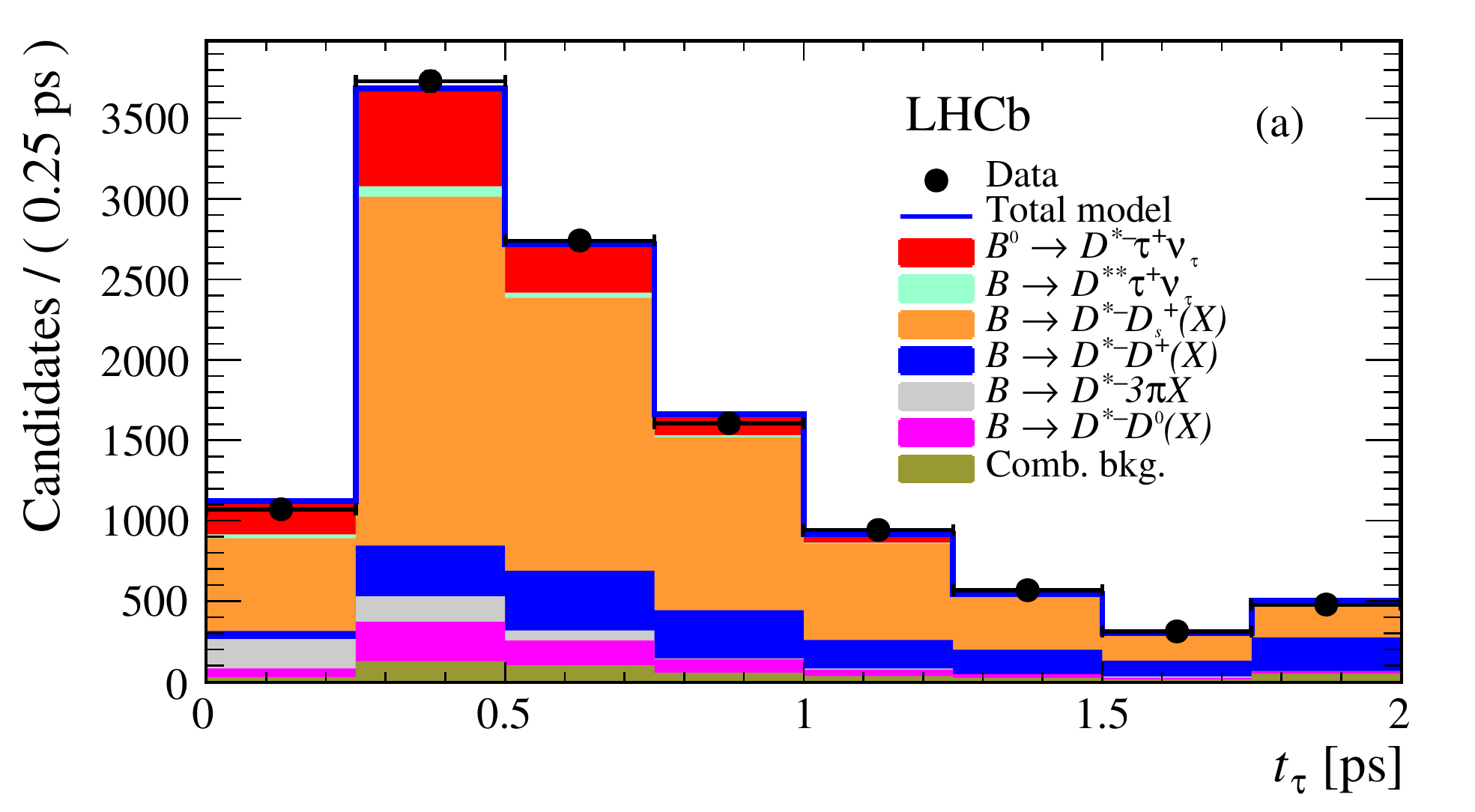}
    \includegraphics[width=0.6\textwidth]{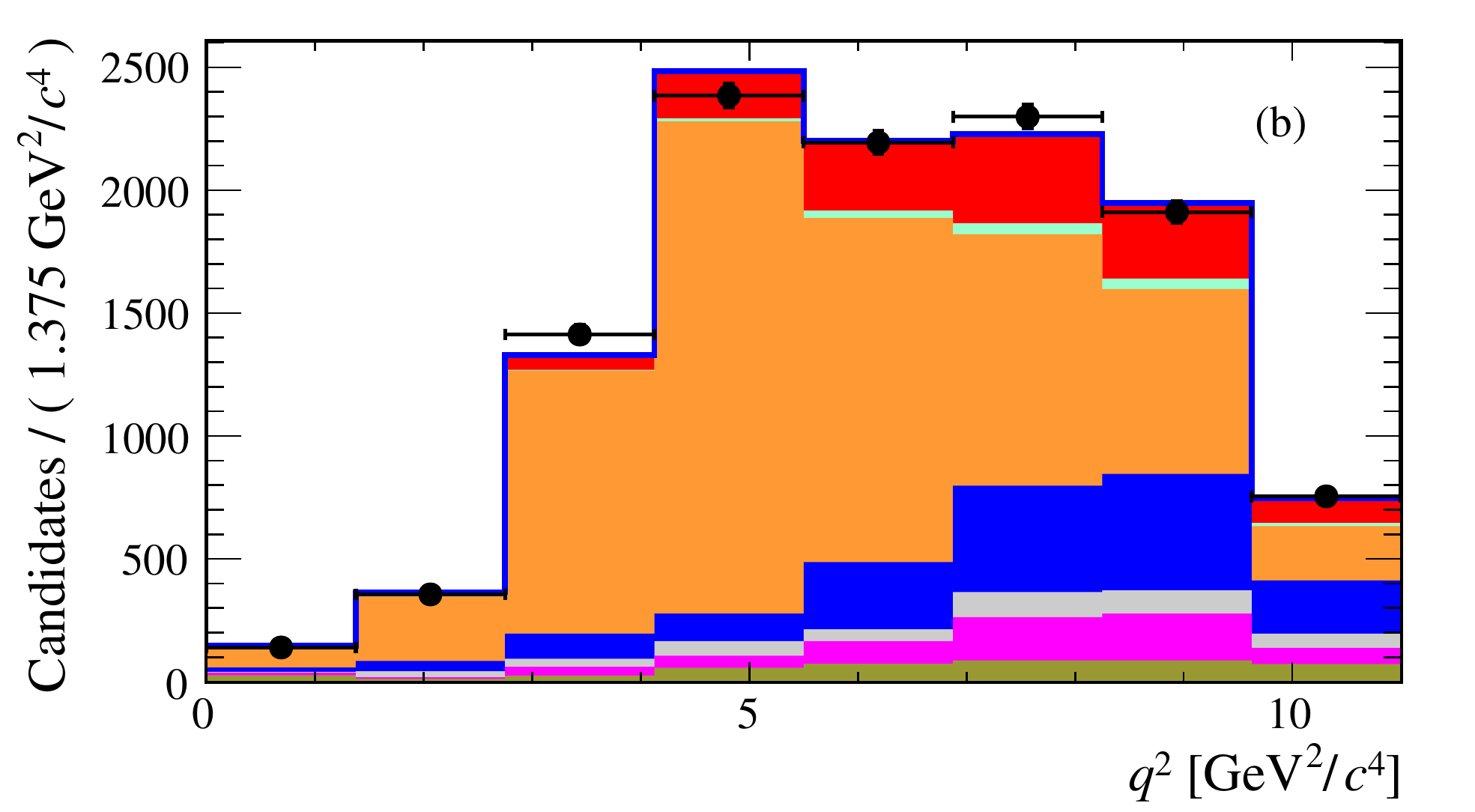}
    \includegraphics[width=0.6\textwidth]{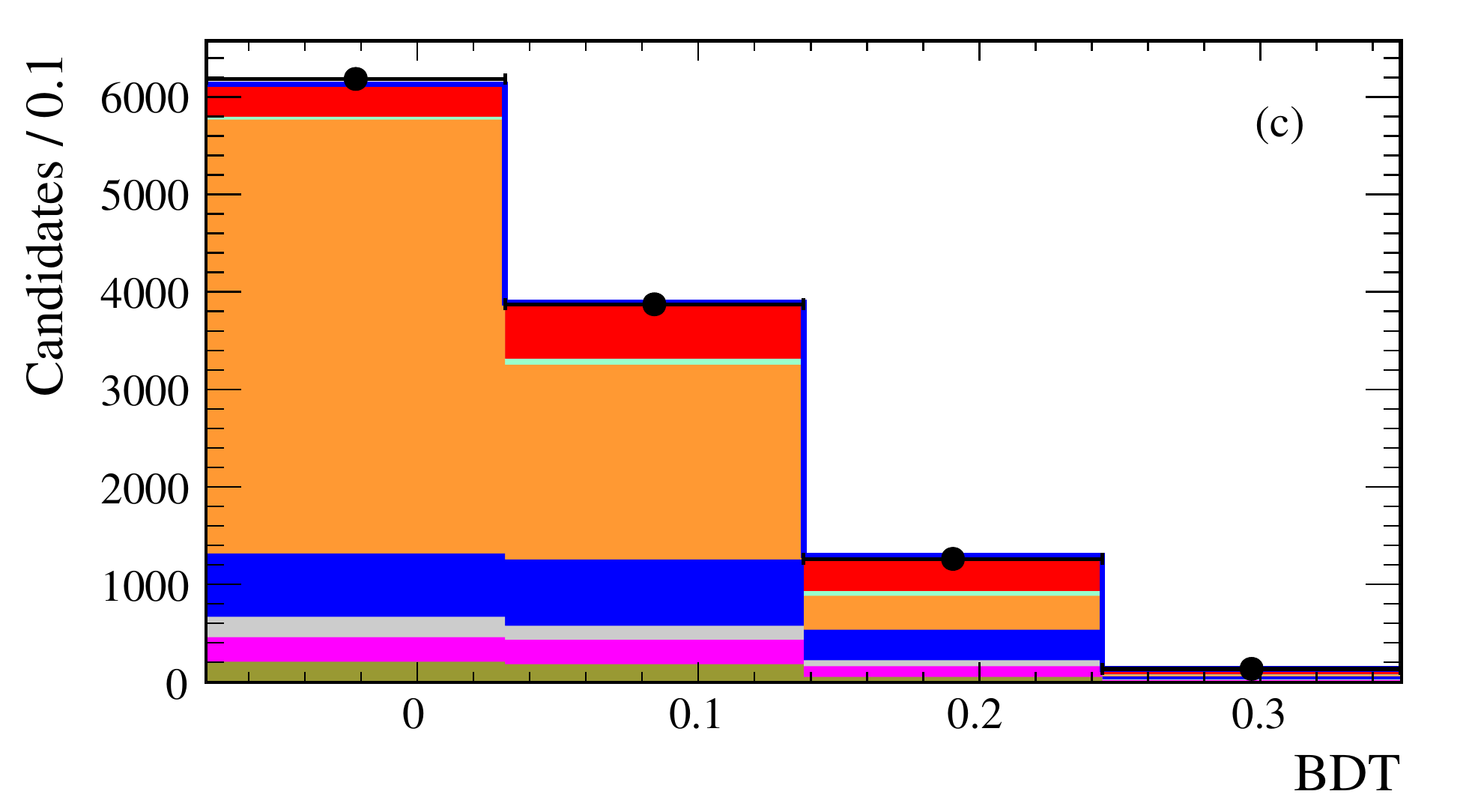}
  \end{center}
  \caption{
    \small 
  Projections of the three-dimensional fit on the (a) $3\pi$ decay time,
(b) $q^2$ and (c) BDT output  distributions. The fit components
are described in the legend.
 }
  \label{fig:fit_results}
\end{figure}
\begin{figure}[t]
  \begin{center}
    \includegraphics[width=0.9\textwidth]{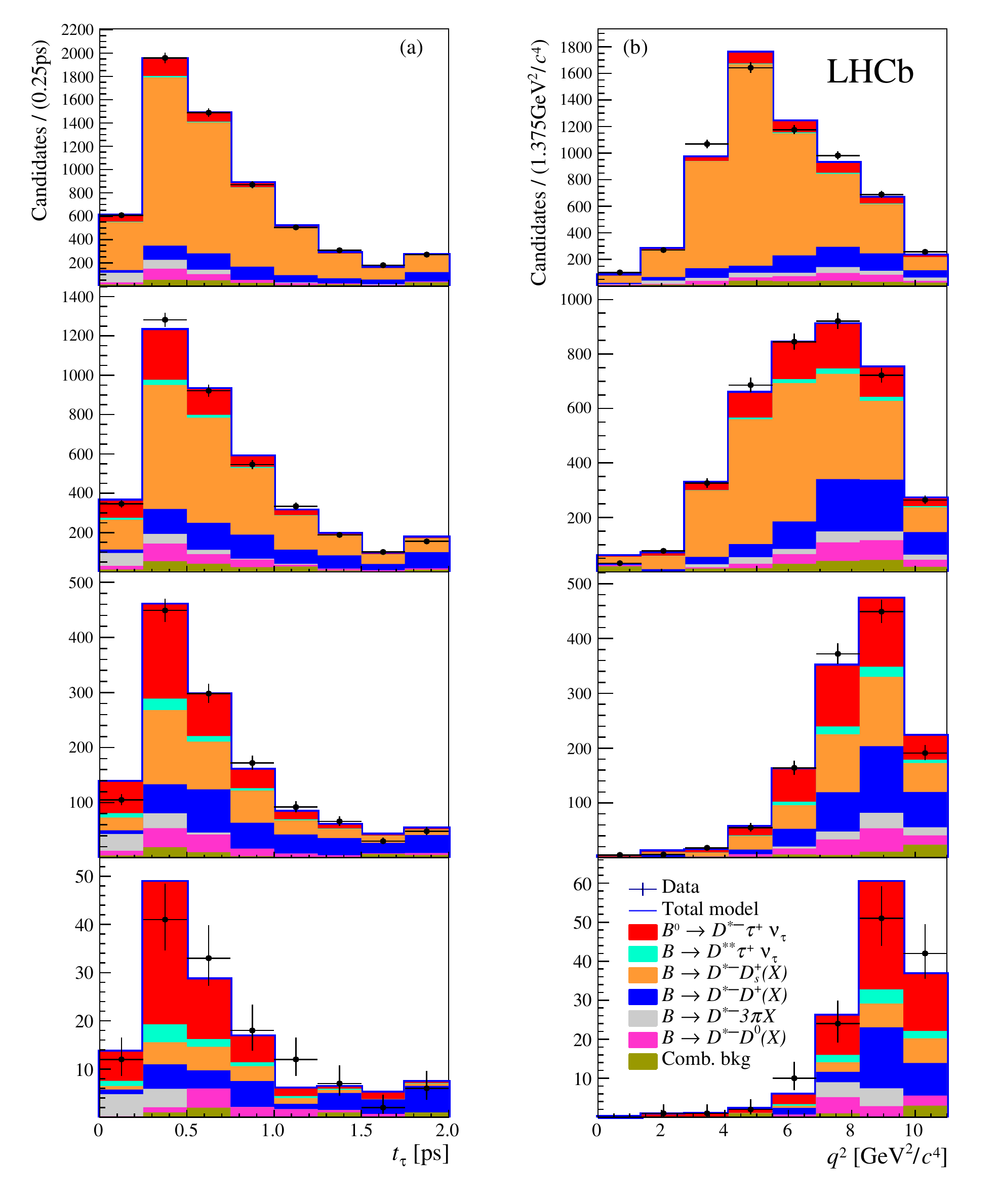}
  \end{center}
  \caption{
    \small 
Distributions of (a) $t_{\tau}$ and (b) $q^2$ in
four different BDT bins, with increasing values of the BDT response from top to
bottom. The fit components are described in the legend.
 }
  \label{fig:fit_bdtbins}
\end{figure}
\begin{figure}[b]
  \begin{center}
    \includegraphics[width=0.48\textwidth]{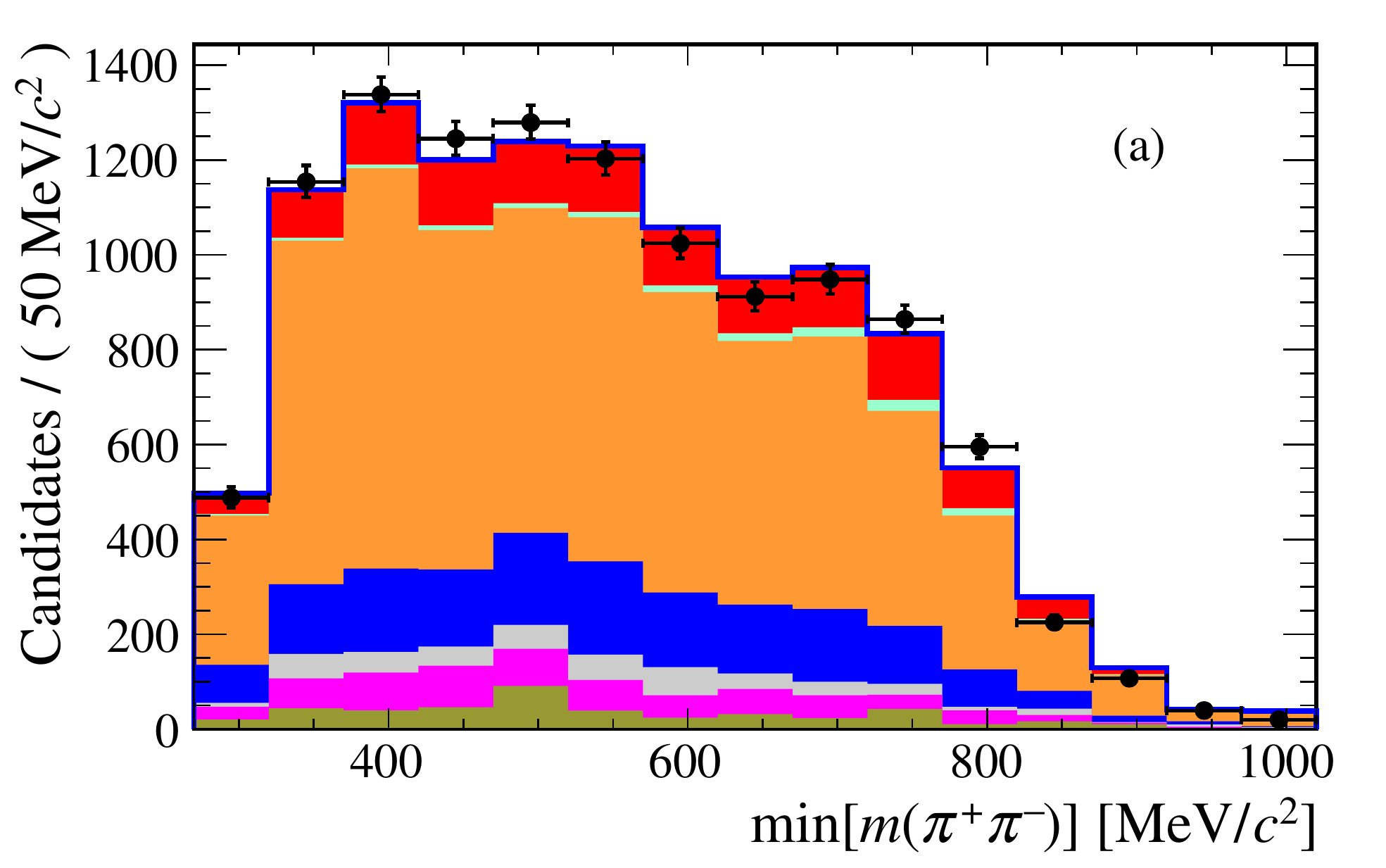}
    \includegraphics[width=0.48\textwidth]{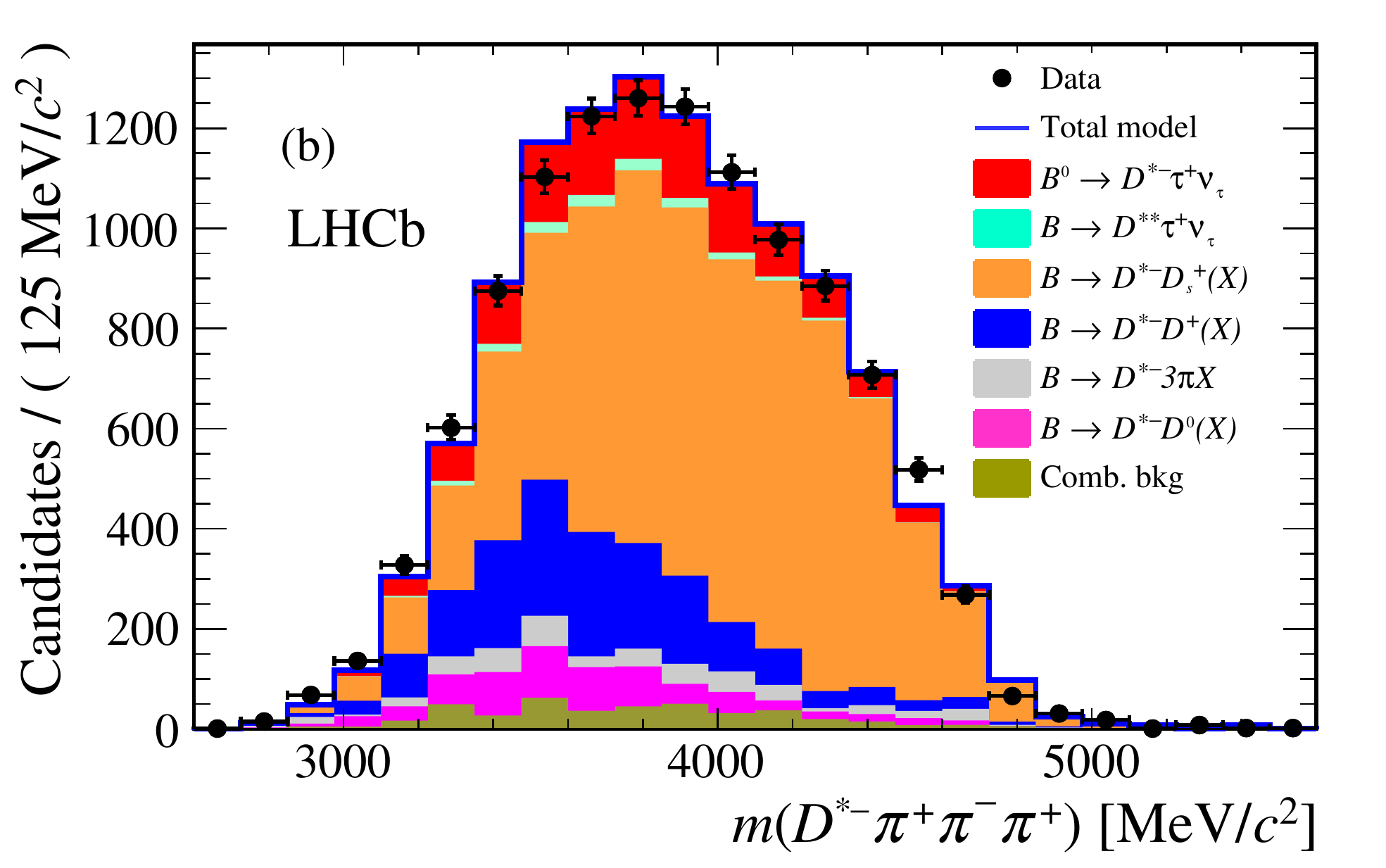}
  \end{center}
  \caption{
    \small 
Projection of the fit results on (a)
${\mathrm{min}}[m(\pi^+\pi^-)]$ and (b) $m(D^{*-}3\pi)$
distributions. The fit components are described in the legend.
 }
  \label{fig:fit_projs}
\end{figure}

 The results of the three-dimensional fit are shown in
Table~\ref{tab:fit_results} and Fig.~\ref{fig:fit_results}. A raw number of
$1336$ decays translates into a yield of $N_{\rm{sig}} = 1296 \pm 86$ 
$B^0\to D^{*-}\tau^+\nu_{\tau}$ decays, after a correction of $-3\%$
due to a fit bias is applied, as detailed below.
Figure~\ref{fig:fit_bdtbins} shows the results of the fit in bins of the
BDT output. The two most discriminant variables of the BDT response are the variables
$\min[m(\pi^+\pi^-)]$ and $m(D^{*-}3\pi)$. Figure~\ref{fig:fit_projs} shows the
fit results projected onto these variables. A good agreement with data and
the post-fit model is found. The fit  \chisq is 1.15 per degree of
freedom, after taking into account the statistical fluctuation in the
simulation templates, and 1.8 without.
  Due to the limited size of the simulation samples used to build the templates
(the need to use templates from inclusive \bquark-hadron decays requires extremely large simulation samples),
the existence of
empty bins in the templates introduces potential biases in the determination of the signal yield that
must be taken into account. To study this effect, a method based on the use of
kernel density estimators (KDE)~\cite{Cranmer:2000du} is used.
For each simulated sample, a three-dimensional density function is
produced.  Each KDE is then transformed in a three-dimensional
template, where bins that were previously empty may now be filled.
These new templates are used to build a smoothed fit model.
The fit is repeated with different signal yield hypotheses.
The results show that a bias is observed for low values of the
generated signal yield that decreases when the generated signal yield increases.
For the value found
by the nominal fit, a bias of $+40$ decays is found, and is used to correct the
fit result.

The statistical contribution to the total uncertainty is determined by performing a
second fit where the parameters governing the templates shapes of the double-charmed decays,
$f_{D_s^+}$, $f_{D_{s0}^{*+}}$, $f_{D_{s1}^+}$, $f_{D_s^+X}$, $f_{(D_s^{+}X)_s}$ and
$f_{D^0}^{v_1v_2}$, are fixed to the values obtained in the first fit.
The quadratic difference between the uncertainties provided by the two fits is taken as systematic
uncertainty due to the knowledge of the $B\to D^{*-}D_s^+X$ and
$B\to D^{*-}D^0X$ decay models, and reported in
Table~\ref{tab:systematics}.

\begin{table}[t]
  \centering
  \caption{Fit results for the three-dimensional fit. The constraints on the
parameters $f_{D_s^+}$, $f_{D_{s0}^{*+}}$, $f_{D_{s1}^+}$,
$f_{D_s^+X}$ and $f_{(D_s^{+}X)_s}$ are applied taking into account their correlations.}
  \label{tab:fit_results_nominal}
    \begin{tabular}{lcc}
      \hline
      Parameter & Fit result & Constraint \\
      \hline
      $N_{\rm sig}$                      & $1296 \pm 86$                         &  \\
      $f_{\tau \to 3\pi\nu}$             & \phantom{1}$0.78$                     & \phantom{11}$0.78$ (fixed) \\
      $f_{D^{**}\tau\nu}$                & \phantom{1}$0.11$                     & \phantom{11}$0.11$ (fixed) \\
      $N_{D^0}^{\rm{sv}}$                & \phantom{1}$445 \pm 22$               & $445 \pm 22$ \\
      $f_{D^0}^{v_1v_2}$                 & \phantom{11}$0.41 \pm 0.22$           & \\
      $N_{D_s}$                          & \phantom{1}$6835 \pm 166$             & \\
      $f_{D^+}$                          & \phantom{11}$0.245 \pm 0.020$         & \\
      $N_{B\to D^*3\pi X}$               & \phantom{1}$424 \pm 21$               & $443 \pm 22$ \\
      $f_{D_s^+}$                        & \phantom{11}$0.494 \pm 0.028$         & \phantom{1}$0.467 \pm 0.032$ \\
      $f_{D_{s0}^{*+}}$                  & \phantom{.4711}$0^{+0.010}_{-0.000}$ & \phantom{.461}$0^{+0.042}_{-0.000}$  \\
      $f_{D_{s1}^+}$                     & \phantom{11}$0.384 \pm 0.044$         & \phantom{1}$0.444 \pm 0.064$ \\
      $f_{D_s^+X}$                       & \phantom{11}$0.836 \pm 0.077$         & \phantom{1}$0.647 \pm 0.107$ \\
      $f_{(D_s^{+}X)_s}$                     & \phantom{11}$0.159 \pm 0.034$         & \phantom{1}$0.138 \pm 0.040$ \\
      $N_{B1B2}$                         & \phantom{1}$197$                      & \phantom{11}$197$ (fixed) \\
      $N_{{\rm{not}}D^*}$                & \phantom{1}$243$                      & \phantom{11}$243$ (fixed) \\
      \hline
    \end{tabular}
\label{tab:fit_results}
\end{table}

\section{Determination of normalization yield}
\label{sec:normfit}

\begin{figure}[tb]
  \begin{center}
    \includegraphics[width=0.49\linewidth]{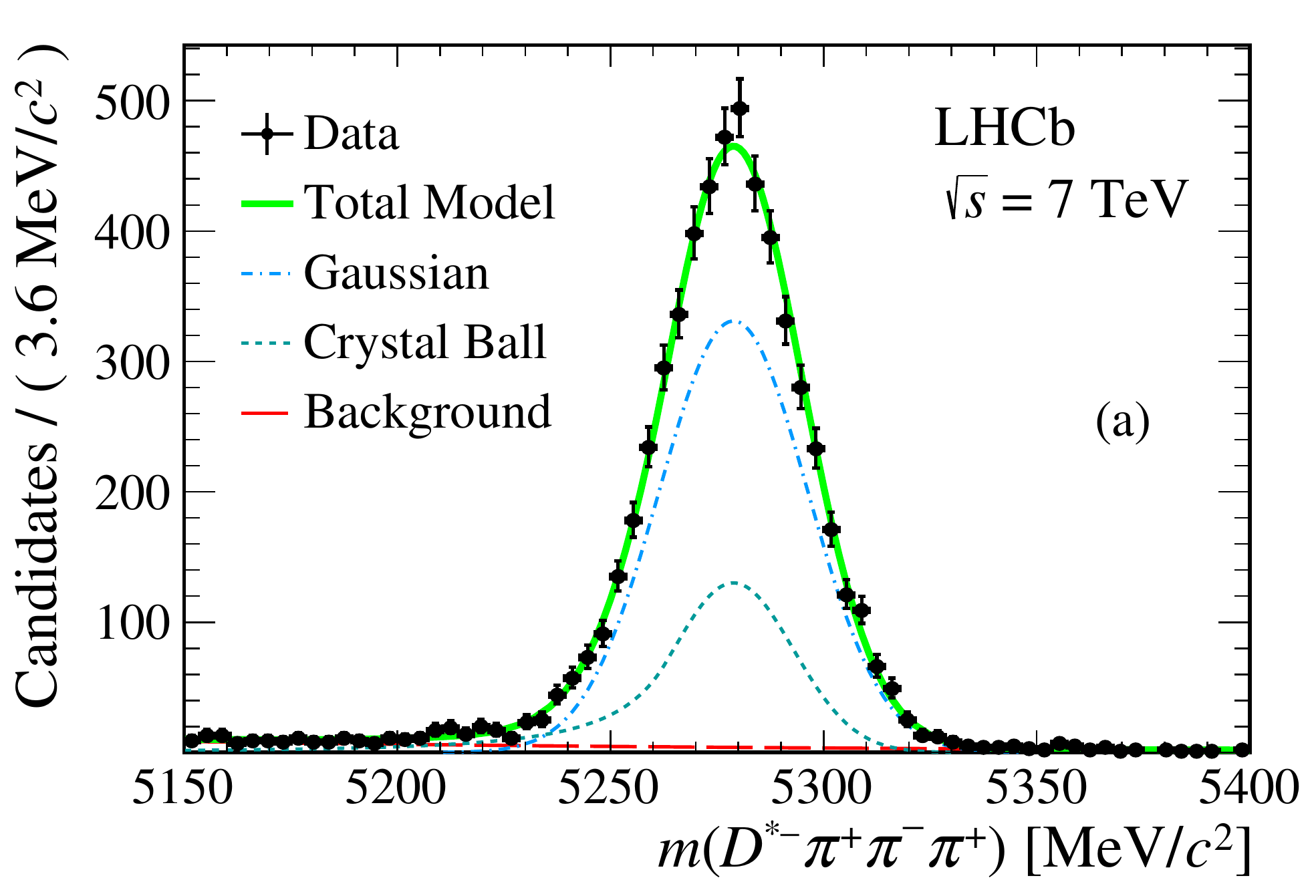}
    \includegraphics[width=0.49\linewidth]{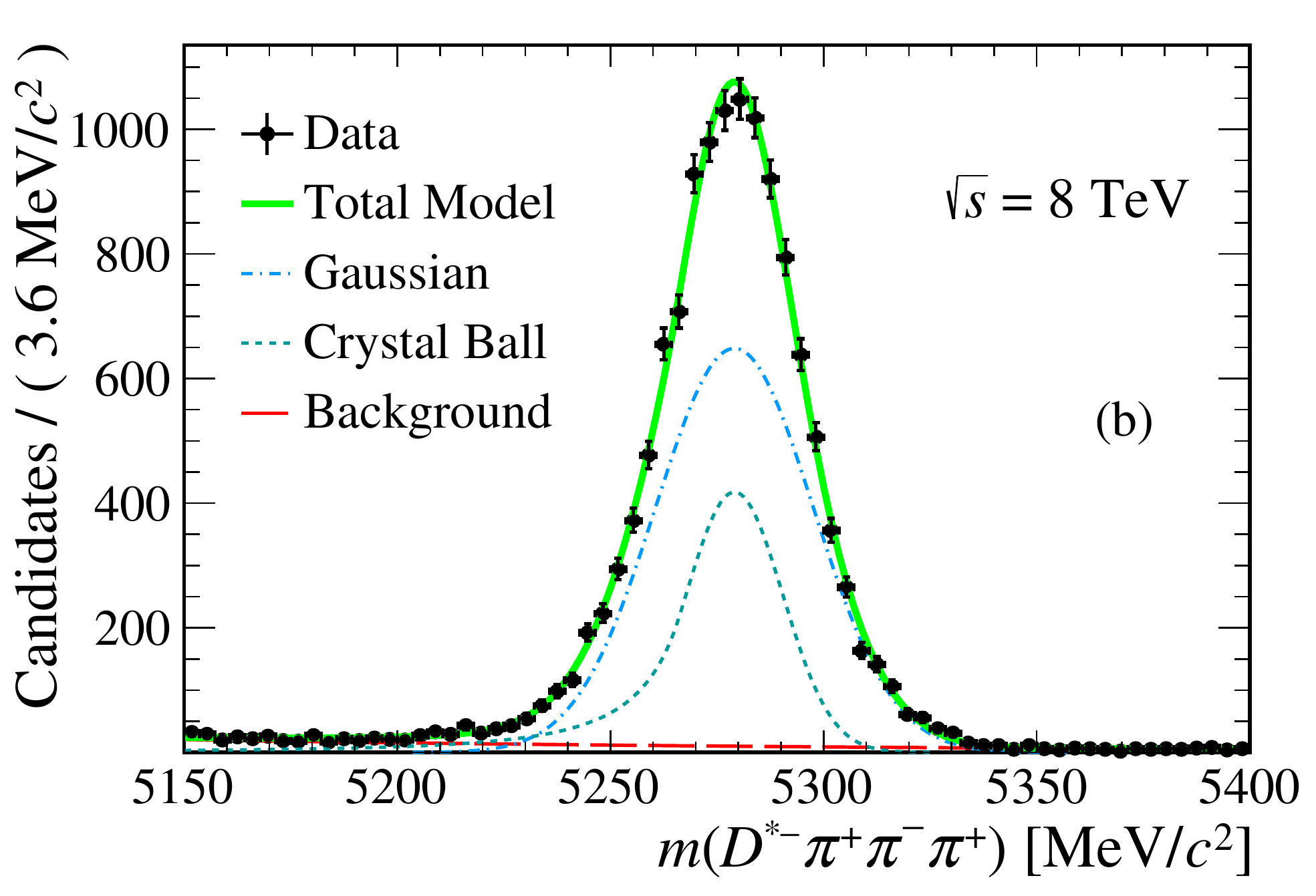}
  \end{center}
  \caption{
    \small 
    Fit to the $m(D^{*-}3\pi)$ distribution after the
    full selection in the (a) $\sqrt{s}=7$ TeV and (b) $8$ TeV data
    samples.}
  \label{fig:fit_norm}
\end{figure}

Figure~\ref{fig:largebmass} shows the  $D^{*-}3\pi$
mass after the selection of the normalization sample. A
clear \Bz signal peak is seen. In order to determine the normalization
yield, a fit is performed in the region between 5150 and 5400\mevcc.
The signal component is described by the sum of
a Gaussian function and a Crystal Ball function~\cite{Skwarnicki:1986xj}.
An exponential function is used to describe the background.
The result of the fit is shown in Fig.~\ref{fig:fit_norm}.
The yield obtained is $17\ 808 \pm 143$.

The fit is also performed with alternative configurations, namely with a
different fit range or requiring the common mean value of the signal
functions to be the same in the $7$ and $8$\tev data samples.
The maximum differences between signal yields in alternative and
nominal configurations are $14$ and $62$ for the $7$ and $8$\tev data samples,
respectively, and are used to assign systematic uncertainties to the
normalization yields.

\begin{figure}[htb]
\begin{center}
    \includegraphics[width=0.49\linewidth]{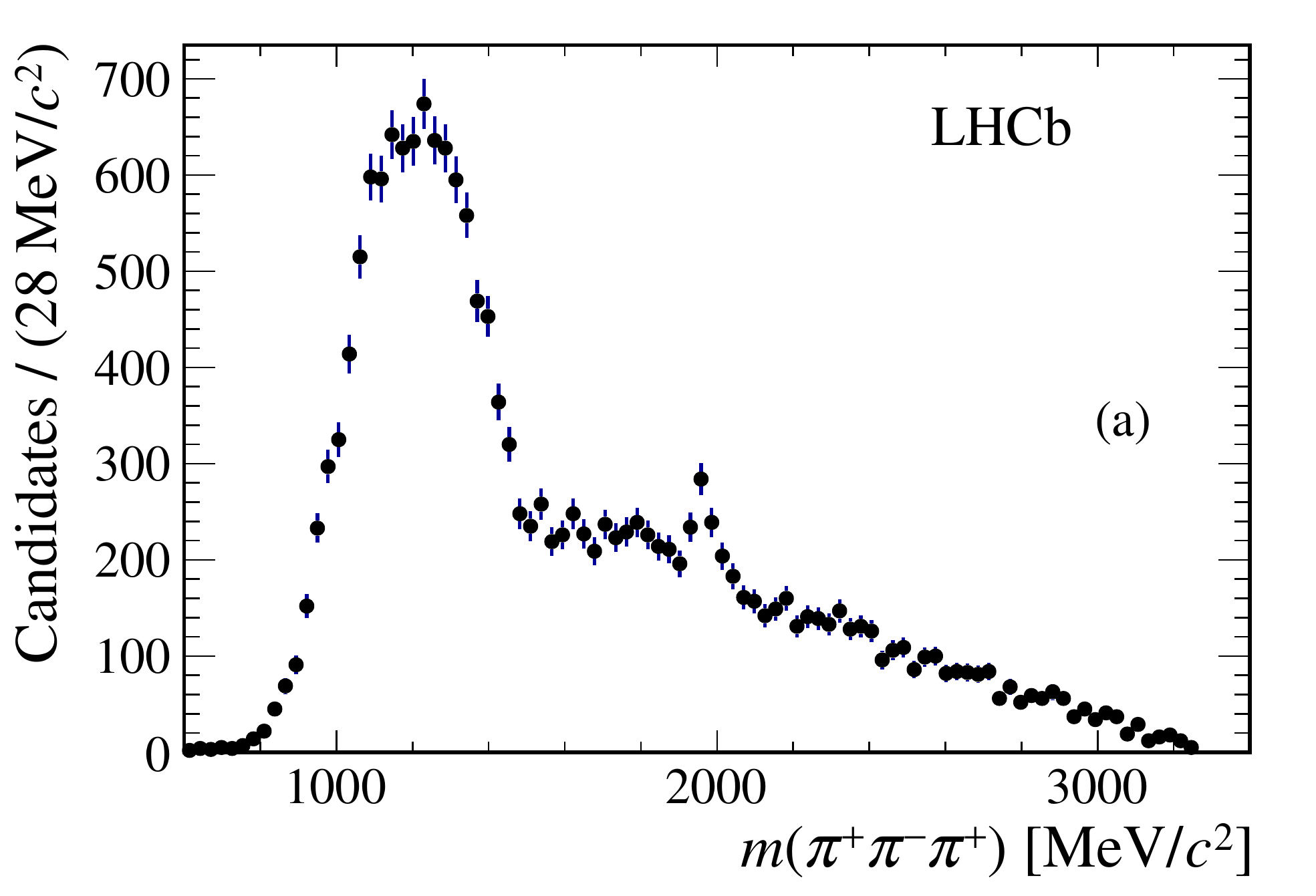}
    \includegraphics[width=0.49\linewidth]{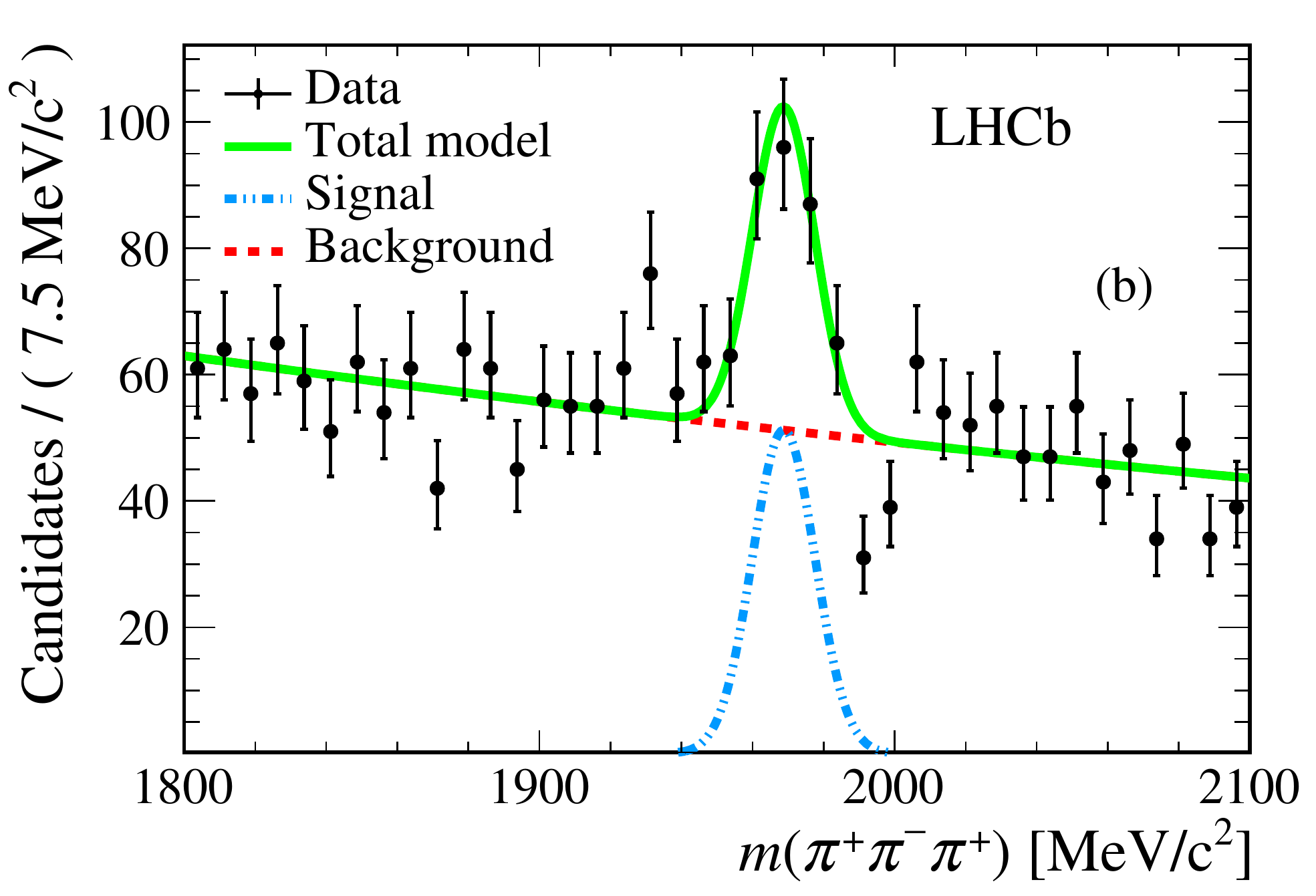}
  \end{center}
  \caption{
    \small 
   (a) Distribution of $m(3\pi)$ after selection, requiring $m(D^{*-} 3\pi)$
to be between 5200 and 5350\mevcc ; (b) fit in the mass region
around the \Ds.}
  \label{fig:data-3pimassB0}
\end{figure}

Figure~\ref{fig:data-3pimassB0} shows the
$m(3\pi)$  distribution for candidates with  $D^{*-}
3\pi$ mass between $5200$ and $5350$\mevcc for the full data
sample.
The spectrum is dominated by the $a_1(1260)^+$ resonance but also a smaller peak
due to the $\Ds\to 3\pi$ decay is visible and is subtracted.
A fit with the sum of a Gaussian function modeling the \Ds mass peak, and an exponential describing the combinatorial background, is
performed  to estimate this \Ds contribution,
giving $151 \pm 22$ candidates.
As a result, the number of normalization decays in the full data
sample is $N_\mathrm{norm} = 17\ 660 \pm 143 \stat \pm 64 \syst \pm 22 \mathrm{\,(sub)}\xspace$,
where the third uncertainty is due to the subtraction of the
$\Bz\to\Dstarm\Ds$ component.

\section{Determination of {\boldmath${\cal{K}}(\Dstarm)$}}
\label{sec:kDstar}

 The result
\begin{equation*}
{\cal{K}}(\Dstarm) =
1.97 \pm 0.13\stat \pm 0.18\syst,
\end{equation*}
\noindent is obtained
using Eq.~\ref{eqn:kappa}.
The ratio of efficiencies between
the signal and normalization modes, shown in Table~\ref{tab:efficiency-summary},
differs from unity due to the softer momentum spectrum of the signal
particles and the correspondingly lower trigger efficiency.
The effective sum of the branching fractions for the $\taup\to 3\pi\neutb$ and $\taup\to
3\pi\piz\neutb$ decays is $(13.81\pm 0.07)\%$~\cite{PDG2017}. This includes the
3\pion mode (without \Kz), a very small feed-down from \tauon five-prong decays, the 3\pion\piz mode
(without \Kz), and only 50\% of the 3\pion\piz\piz mode due to the smaller efficiency of this decay mode. This latter
contribution results in a $1\%$ correction (see Sec.~\ref{ssec:mod_uncertainties}).
Finally, a correction factor
$1.056 \pm 0.025$ is applied when computing ${\cal{K}}(\Dstarm)$ in order
to account for residual efficiency discrepancies between data and simulation regarding PID and trigger. The event multiplicity, measured by the
scintillating-pad detector, affects the efficiency for the
fraction of the data sample  which is triggered at the hardware
trigger level by particles in the event other than those from the
$\Dstarm\tau^+\neut$
candidate. An
imperfect description of this multiplicity in the simulation does not cancel
completely in ${{\cal{K}}(\Dstarm)}$. The correction factor also includes a
small feed-down contribution from
{\decay{\Bs}{D_s^{**-}\taup\neut}} decays, where
{\decay{D_s^{**-}}{\Dstarm K^0}}, that is taken into account according to simulation.

As a further check of the analysis, measurements of ${\cal{K}}(\Dstarm)$ are performed in mutually exclusive subsamples, obtained by requiring different trigger conditions and center-of-mass energies.
All of these results  are found to be compatible with the result obtained with the full sample.  {Changing the requirement on the minimal BDT output value, as well as the bounds of the nuisance parameters, does not change  the final result.}

\section{Systematic uncertainties}
\label{sec:systematic}

The systematic uncertainties on ${\cal{K}}(\Dstarm)$ are subdivided into four
categories: the knowledge of the signal model,
including $\tau$ decay models; the modeling of
the various background sources; possible biases in the fit procedure
due to the limited size of the simulated samples;
and trigger selection efficiencies, external inputs and particle identification efficiency. Table~\ref{tab:systematics} summarizes the results.

\subsection{Signal model uncertainties}
\label{ssec:mod_uncertainties}

The uncertainty in the relative proportion of signal events in the mode $\taup\to
3\pi\neutb$ and $\taup\to 3\pi\piz\neutb$ affects the fit results. Taking into account the relative efficiencies, an  uncertainty of $0.01$  is assigned to
$f_{\tau\to 3\pi\nu}$.
A fit is performed with this fraction
constrained to $0.78 \pm 0.01$ using a Gaussian function. A  second fit is done
 fixing the fraction to the value found by the first fit. The squared
difference between the uncertainties of the two fits is taken as the systematic uncertainty
due to the signal composition, resulting in a $0.7\%$ systematic uncertainty.

To estimate the systematic uncertainty due to the knowledge of the $B^0\to
D^{*-} \tau^+ \nu_{\tau}$ form factors, a study based on  pseudoexperiments is performed. A total of 100 fits
to generated samples is done by varying the values of the parameters $R_1(1)$,
$R_2(1)$ and $\rho^2$ of Ref. \cite{Fajfer:2012vx} which govern the fraction
of each spin configuration in the form-factor templates. The
parameters are varied according to a multivariate Gaussian
distribution using their uncertainties and correlations. The parameter $R_0(1)=1.14\pm 0.11$
is varied under the conservative assumption that it is not correlated with the other parameters. A systematic uncertainty of $0.7\%$
on the signal yield is obtained by taking the standard deviation of the distribution
of the fitted signal yields.

A value of 1\% systematic uncertainty of the efficiency due to
  the form factor reweighting is computed by repeating the fit
  without it.

The effect of the $\tau$ polarization is studied separately for $\taup\to
3\pi\neutb$ and $\taup\to 3\pi\piz\neutb$ decays. Due to the
$a_1(1260)^+$ dominance observed in the $\taup\to
3\pi\neutb$ decay, the sensitivity of the 3\pion momenta to the polarization is
negligible and therefore
no systematic uncertainty is assigned due to this effect.
 For the $\taup\to 3\pi\piz\neutb$ decay mode,
the signal is simulated in two configurations: using either the
TAUOLA \cite{Was:2011tv}
model or a pure phase-space
model.
The effect of the $\tau$ polarization is
evaluated by multiplying the efficiency by the ratio of the
distributions of the cosine of $\alpha$
(the angle between the 3$\pi$ momentum in the $\tau$ rest frame and the $\tau$
direction in the laboratory frame)
generated with the two configurations. This produces a
relative change in the efficiency of $1.5\%$. This value, scaled by the
relative fraction of the $\taup\to 3\pi\piz\neutb$ component with respect
to the total, gives a systematic uncertainty of 0.4\%.

Other $\tau$ decays could contribute to the signal yield. They are either decays
with three charged tracks in the final states ($K^+$\pim\pip, $K^+K^-$\pip,
\pip\pim\pip\piz\piz) or five charged tracks, all of them having very small branching
fractions compared to the $\taup\to 3\pi(\piz)\neutb$ decay mode. The study of a dedicated simulation sample with inclusive $\tau$
decays indicates an effect of $1\%$ that is taken as a systematic
uncertainty.

The $B\to D^{**}\tau\nu_{\tau}$ fraction used for the nominal fit,
$0.11$,
is assigned a  40\% uncertainty, based on the results of an auxiliary
study of \decay{\Bm}{D_1(2420)^0\taup\neut} decays, where
$D_1(2420)^0\to\Dstarm\pip$. These results give a systematic
uncertainty on the signal yield of 2.3\%.

An additional systematic uncertainty of 1.5\% due to the feed-down from
$\Bs\to D_s^{**}\tau^+\neut$ decays is assigned, under the
assumption that the yield of these decays in the simulation has an uncertainty of
50\%, determined to be the upper limit from a study performed on simulated data.

\subsection{Background-related systematic uncertainties}
This section lists the systematic uncertainties due to the modeling of
different background sources, such as the \Dsp decay model, double-charm and combinatorial contributions.

Candidates in the low BDT output region are used to correct the
composition of \Dsp decays in simulation. From the fit to this data
sample corrections are obtained, which are used to generate $1000$ alternative $D_s^+$
templates for each $D_s^+$ component in the nominal three-dimensional fit. Each alternative template is
produced by varying the nominal template accounting for the uncertainty
and correlations between the $D_s^+$ subcomponents according to a Gaussian distribution. These alternative
templates are employed to refit the model to the data. The difference between the signal yield of the
alternative and the nominal fits, divided by the yield of the nominal
fit, is fitted with a Gaussian function and a systematic uncertainty
of $2.5\%$ is determined.

The mass variables that are expected to be significantly
correlated with the fit variables $q^2$ and BDT, are  $m(\Dstarm 3\pi)$, $m(3\pi)$,
$\text{min}[m(\pi^+\pi^-)]$, $\text{max}[m(\pi^+\pi^-)]$ and
$m(\pi^+\pi^+)$.\footnote{Only $m(\Dstarm 3\pion)$ is considered for the
  $\Dstarm\Ds$ case, since the effect of the other three variables is
  included in the systematic uncertainty due to the \Dsp decay model.}
The corresponding effect on the fit result of
these variables is empirically studied
by varying the distributions using a quadratic interpolation
method:
for each template, two alternative templates are produced, with a variation of $\pm 1 \sigma$.
Then, the fit enables the interpolation between the nominal and the
alternative templates to be made with a linear weight. Each nuisance parameter is allowed to float in
the range $[-1,+1]$ and a loose Gaussian constraint with $\sigma=1$ is included. This method is used to compute systematic uncertainties due to the knowledge of the shape of the templates. The corresponding systematic uncertainty is $2.9\%$.
A systematic uncertainty of 2.6\% arises due to the composition of the
$B\to D^{*-} D_s^+ (X)$ and $B\to D^{*-}D^0(X)$ decays, as discussed in
Sec.~\ref{sec:double_charm}. The use of the \Dsp  exclusive reconstruction in
$3\pi$ helps to limit the size of this uncertainty.

The systematic uncertainty due to the knowledge of the shape of the residual prompt
background component is estimated by applying the same interpolation
technique
to the corresponding
template. When combined with the knowledge of the normalization
of this background, this gives an overall uncertainty of 2.8\%.

The same method
is again used
to assess the systematic uncertainty due to the shape of the combinatorial
background. The change in the signal yield provides a systematic uncertainty of $0.7\%$.

Another systematic uncertainty is due to the normalization of this background. This
uncertainty is computed by performing the fit with a $30\%$ Gaussian
constraint around the nominal value.
The resulting difference with respect to the nominal fit is $0.1\%$, which is assigned as systematic uncertainty. This uncertainty has a negligible effect on the total systematic uncertainty associated with the shape of the combinatorial background.

\subsection{Fit-related systematic uncertainties}
To assess the systematic uncertainty relative to the bias due to empty bins in the templates used in the fit the study performed using the KDE method is
repeated implementing different smoothing parameters. A difference in the
signal yield of 1.3\% is assigned as the systematic uncertainty due to the bias
observed in the fit.

In order to estimate the systematic uncertainty due to the limited size of the
simulated samples, a
bootstrap method is used.
Each template from the nominal model is used to produce new templates sampled
from the originals by using a bootstrap procedure based on random selection with
replacement, varied bin-by-bin according to a Poisson distribution. This procedure is repeated $500$ times.
A Gaussian fit to the distributions of signal yields provides a $4.1\%$ effect
taken as the systematic uncertainty due to the limited size of the simulated
samples.

\subsection{Uncertainties related to the selection}
In this section systematic uncertainties related to the selection criteria are
discussed. Such uncertainties stem from  the choice of the
trigger strategy, the online and offline selection of the candidates, the
normalization and external inputs, and the efficiency of the PID criteria.

The trigger efficiency is studied on data using the fraction of the events where
the trigger was fired by particles other than the six tracks forming the signal candidate, as a function of the two most important
variables in this analysis, $t_{\tau}$ and  $m(\Dstarm 3\pi)$, the latter being
highly correlated with
$q^2$. Corrections on the $t_{\tau}$ and $m(\Dstarm 3\pi)$ distributions
due to different
trigger efficiency between data and simulation are applied. This gives a change
in the number of signal candidates of $1.0\%$ for the $t_{\tau}$ and
$0.7\%$ for the $m(\Dstarm 3\pi)$ corrections.
The sum in quadrature of these two contributions, taken as
systematic uncertainty related to the trigger efficiency, is $1.2\%$.

An additional $1\%$ systematic uncertainty arises from a mismatch
between data and simulation in the occupancy of the event.

The relative efficiency between the signal and the normalization
channels is precisely determined from simulated samples.
Discrepancies between data and simulation, due to online and offline selection criteria, introduce a
2\% of systematic uncertainty for both.

A 1\% systematic uncertainty is
assigned on the charged isolation criterion, due to differences observed between
the $B^0\to \Dstarm\tau^+\neut$ and the
$B^0\to \Dstarm 3\pi$ simulations.

All selection criteria, except the detached-vertex topology requirement, are common to the signal
and normalization decays. The corresponding efficiencies are therefore directly
determined from data by fitting the number of events in the $\Bz\to\Dstarm
3\pion$ mass peak before and after each selection, and no systematic uncertainty
is assigned.
To compute the systematic uncertainty attributed to the knowledge of
the relative efficiencies corresponding to the different signal and
normalisation vertex topologies, the vertex position uncertainty distribution is
split into three regions: between $-4 \sigma$ and $-2 \sigma$,
between $-2 \sigma$ and $2 \sigma$ and between $2 \sigma$ and $4 \sigma$, where $\sigma$ is the reconstructed uncertainty on the distance along the beam line of the \Bz and 3\pion vertices.
Then a ratio between the number of candidates in the outer regions and the number
of candidates in the inner region is computed for the candidates which have
$m(\Dstarm 3\pi)$ in the exclusive $B^0\to \Dstarm 3\pi$ peak.
The same procedure is performed for the candidates outside the $B^0\to D^{*-}3\pi$ peak, which exhibit a signal-like behavior.
The procedure is repeated for data, and the ratio between data and
simulation gives rise to a 2\% systematic uncertainty.

The simulation is corrected in order to match the performance of PID criteria measured in data. Correction factors are applied in
bins of momentum, pseudorapidity and global event multiplicity, after having
adjusted the simulated event multiplicity to that observed using real data.
To assess the systematic uncertainty due to the choice of the binning
scheme used to correct simulation,
two new schemes are derived from the default with
half and twice the number of bins, the default configuration
consisting of fifteen bins in momentum, seven in pseudorapidity and three in the global event multiplicity.
The correction procedure is repeated with these two alternate schemes, leading to a
systematic uncertainty related to PID of $1.3\%$.

The normalization channel consists of exactly the same final state as the signal.
In this way, differences between data and simulation are minimized.
The systematic uncertainty in the normalization yield is determined to be equal to  1\%.
The statistical uncertainty attributed to the normalization yield is included
in the statistical uncertainty quoted for each result in this paper.
Differences between data and simulation in the modeling of the $B^0\to
\Dstarm 3\pi$ decay impact the efficiency of the normalization channel
and result in a $2.0\%$ systematic uncertainty on
${\cal{K}}(\Dstarm)$.

The branching fraction for the normalization channel, obtained by
averaging the measurements of Refs.~\cite{LHCb-PAPER-2012-046,TheBABAR:2016vzj,Majumder:2004su}, has an uncertainty of 3.9\%.
A 2.0\% uncertainty arising from the knowledge of the
$B^0\rightarrow D^{*-}\mu^+ \nu_{\mu}$ branching fraction is added in quadrature to
obtain a 4.5\% total uncertainty on ${\cal{R}}(\Dstarm)$ due to
external inputs.

\subsection{Summary of systematic uncertainties}

Table~\ref{tab:systematics} summarizes the systematic uncertainties on
the measurement of the ratio
$\BR(\Bz \to D^{*-}\tau^+\nu_{\tau})/\BR(\Bz \to D^{*-}3\pi)$. The total
uncertainty is $9.1\%$. For ${\cal{R}}(\Dstarm)$, a $4.5\%$
systematic uncertainty due to the knowledge of the external
branching fractions is added.
\begin{table}[h]
  \centering
  \caption{List of the individual systematic uncertainties for the measurement of the
ratio $\BR(\Bz \to D^{*-}\tau^+\nu_{\tau})/\BR(\Bz \to D^{*-}3\pi)$.
}
  \label{tab:systematics}
    \begin{tabular}{lc}
      \hline
      Contribution                          & Value in \% \\
      \hline
      ${\cal{B}}(\decay{\taup}{3\pion\neutb})/{\cal{B}}(\decay{\taup}{3\pion(\piz)\neutb})$                     & $0.7$ \\
      Form factors (template shapes)        & $0.7$ \\
      Form factors (efficiency)                                     & $1.0$ \\
      \tauon polarization effects           & $0.4$ \\
      Other \tauon decays                   & $1.0$ \\
      $B\to D^{**}\tau^+\nu_{\tau}$           & $2.3$ \\
      $B_s^0\to D_s^{**}\tau^+\nu_{\tau}$ feed-down               & $1.5$ \\
    \hline
      $D_s^+ \to 3\pi X$ decay model                            & $2.5$ \\
      $D_s^+$, $D^0$ and $D^+$ template shape                  & $2.9$ \\
      $B\to D^{*-} D_s^+ (X)$ and $B\to D^{*-}D^0(X)$ decay model   & $2.6$ \\
      $\Dstarm 3\pion X$ from \B decays                              & $2.8$ \\
      Combinatorial background (shape + normalization)          & $0.7$ \\
      \hline
      Bias due to empty bins in templates                       & $1.3$ \\
      Size of simulation samples                                & $4.1$ \\
      \hline
      Trigger acceptance                                            & $1.2$ \\
      Trigger efficiency                                            & $1.0$ \\
      Online selection                                              & $2.0$ \\
      Offline selection                                             & $2.0$ \\
      Charged-isolation algorithm                                   & $1.0$ \\
      Particle identification                                       & $1.3$ \\
      Normalization channel                                         & $1.0$ \\
      Signal efficiencies (size of simulation samples)              & $1.7$ \\
      Normalization channel efficiency (size of simulation samples) &
                                                                      $1.6$ \\
      Normalization channel efficiency (modeling of $\Bz\to\Dstarm
      3\pi$) & $2.0$ \\

      \hline
      Total uncertainty                                             & $9.1$ \\
      \hline
    \end{tabular}
\end{table}

\section{Conclusion}
\label{sec:results}

In conclusion, the ratio of branching fractions between the
\decay{\Bz}{\Dstarm\taup\neut} and the
\decay{\Bz}{\Dstarm 3\pi} decays is measured to be
\begin{equation*}
{
\cal{K}}(\Dstarm) =
1.97 \pm 0.13\stat \pm 0.18\syst,
\end{equation*}
\noindent where the first uncertainty is statistical and the second
systematic. Using the branching fraction
${\cal{B}}(\Bz\to\Dstarm 3\pi)=(7.214 \pm 0.28) \times 10^{-3}$ from
the weighted average of the measurements by the LHCb
~\cite{LHCb-PAPER-2012-046}, \babar ~\cite{TheBABAR:2016vzj}, and \belle ~\cite{Majumder:2004su} collaborations,  a value of the absolute branching fraction of the \decay{\Bz}{\Dstarm\taup\neut} decay is
obtained

\begin{equation*}
{
\cal{B}}(\Bz\to\Dstarm\taup\neut)
=\left(1.42\pm 0.094\stat \pm 0.129\syst
\pm0.054\extrn\right)\times 10^{-2},
\end{equation*}
where the third uncertainty originates from the limited knowledge of
the branching fraction of the normalization mode.
The precision of this measurement is comparable to that of the current
world average of Ref.~\cite{PDG2017}.
The first determination of ${\cal{R}}(\Dstarm)$ performed by using three-prong
\tauon decays is obtained by using the measured branching fraction of ${\cal{B}}(\Bz\to\Dstarm\mup\neum) = (4.88\pm 0.10)
\times 10^{-2}$ from Ref.~\cite{HFAG}.  The result
\begin{equation*}
{
\cal{R}}(D^{*-}) = 0.291 \pm 0.019 \stat
    \pm 0.026 \syst \pm 0.013 \extrn
\end{equation*}
is one of the most precise
single measurements performed so far. It is $1.1$
standard deviations higher  than the SM prediction
($0.252 \pm 0.003$) of Ref.~\cite{Fajfer:2012vx}, and consistent
with previous determinations. This R(\Dstar) measurement, being  proportional to   
${\cal{B}}(\Bz\to\Dstarm 3\pi)$, and inversely proportional to  ${\cal{B}}(\Bz\to\Dstarm\mup\neum)$, will need to be rescaled accordingly when more precise values of these inputs are made available in the future. An average
of this measurement with the \lhcb result using
\mbox{\decay{\taup}{\mup\neum\neutb}} decays~\cite{LHCb-PAPER-2015-025}, accounting for small
correlations due to form factors, \tauon polarization and $D^{**}\taup\neut$
feed-down, gives a value of
${\cal{R}}(\Dstarm) = 0.310 \pm 0.0155 \stat \pm 0.0219 \syst$
, consistent
with the world average and $2.2$ standard deviations above the SM prediction.
The overall status of ${\cal{R}}(\D)$ and
  ${\cal{R}}(\Dstar)$ measurements is reported in Ref.~\cite{HFAG}. After inclusion of this result, the combined discrepancy of ${\cal{R}}(\D)$ and
  ${\cal{R}}(\Dstar)$ determinations with the SM prediction is 4.1$\sigma$.

The novel technique presented in this paper, allowing the reconstruction and selection of
semitauonic decays with  $\tau^+ \to\ 3\pion (\pi^0)\neutb$ transitions, can be applied to all the other
semitauonic decays, such as those of \Bp, \Bs, \Bc and \Lb. This technique also
allows isolation of large signal samples with  high purity, which can be
used to measure angular distributions and other observables proposed in the literature
to discriminate between SM and new physics contributions.
The inclusion of further data collected by \lhcb at
 $\sqs=13\tev$ will result in an overall uncertainty on
  ${\cal{R}}(\Dstarm)$ using this technique comparable to that of the current world average.

\section*{Acknowledgements}
%
%
\noindent We express our gratitude to our colleagues in the CERN
accelerator departments for the excellent performance of the LHC. We
thank the technical and administrative staff at the LHCb
institutes. We acknowledge support from CERN and from the national
agencies: CAPES, CNPq, FAPERJ and FINEP (Brazil); MOST and NSFC
(China); CNRS/IN2P3 (France); BMBF, DFG and MPG (Germany); INFN
(Italy); NWO (The Netherlands); MNiSW and NCN (Poland); MEN/IFA
(Romania); MinES and FASO (Russia); MinECo (Spain); SNSF and SER
(Switzerland); NASU (Ukraine); STFC (United Kingdom); NSF (USA).  We
acknowledge the computing resources that are provided by CERN, IN2P3
(France), KIT and DESY (Germany), INFN (Italy), SURF (The
Netherlands), PIC (Spain), GridPP (United Kingdom), RRCKI and Yandex
LLC (Russia), CSCS (Switzerland), IFIN-HH (Romania), CBPF (Brazil),
PL-GRID (Poland) and OSC (USA). We are indebted to the communities
behind the multiple open-source software packages on which we depend.
Individual groups or members have received support from AvH Foundation
(Germany), EPLANET, Marie Sk\l{}odowska-Curie Actions and ERC
(European Union), ANR, Labex P2IO, ENIGMASS and OCEVU, and R\'{e}gion
Auvergne-Rh\^{o}ne-Alpes (France), RFBR and Yandex LLC (Russia), GVA,
XuntaGal and GENCAT (Spain), Herchel Smith Fund, the Royal Society,
the English-Speaking Union and the Leverhulme Trust (United Kingdom).



\addcontentsline{toc}{section}{References}
\setboolean{inbibliography}{true}
\bibliographystyle{LHCb}
\bibliography{main,LHCb-PAPER,LHCb-CONF,LHCb-DP,LHCb-TDR}

\newpage



\newpage
\centerline{\large\bf LHCb collaboration}
\begin{flushleft}
\small
R.~Aaij$^{40}$,
B.~Adeva$^{39}$,
M.~Adinolfi$^{48}$,
Z.~Ajaltouni$^{5}$,
S.~Akar$^{59}$,
J.~Albrecht$^{10}$,
F.~Alessio$^{40}$,
M.~Alexander$^{53}$,
A.~Alfonso~Albero$^{38}$,
S.~Ali$^{43}$
% ...

{\footnotesize \it
% ...
$ ^{5}$Clermont Universit{\'e}, Universit{\'e} Blaise Pascal, CNRS/IN2P3, LPC, Clermont-Ferrand, France\\
% ...
$ ^{10}$Fakult{\"a}t Physik, Technische Universit{\"a}t Dortmund, Dortmund, Germany\\
% ...
$ ^{38}$ICCUB, Universitat de Barcelona, Barcelona, Spain\\
$ ^{39}$Universidad de Santiago de Compostela, Santiago de Compostela, Spain\\
$ ^{40}$European Organization for Nuclear Research (CERN), Geneva, Switzerland\\
% ...
$ ^{43}$Nikhef National Institute for Subatomic Physics, Amsterdam, The Netherlands\\
% ...
$ ^{48}$H.H. Wills Physics Laboratory, University of Bristol, Bristol, United Kingdom\\
% ...
$ ^{53}$School of Physics and Astronomy, University of Glasgow, Glasgow, United Kingdom\\
% ...
$ ^{59}$University of Cincinnati, Cincinnati, OH, United States\\
% ...
}
\end{flushleft}

\end{document}